%% file: main.tex
\documentclass[12pt, titlepages]{report}
\usepackage[dvipsnames]{xcolor}

\usepackage{amsmath,amssymb}
\usepackage{appendix}
\usepackage{braket}
\usepackage{bm}
\usepackage{eucal}
\usepackage{geometry}
\usepackage{graphicx}
\usepackage{hyperref}
\usepackage{listings}
\usepackage{ifthen}
\usepackage{svg}
\usepackage{tocloft}
\usepackage{tikz}
\usetikzlibrary{decorations.pathreplacing, arrows.meta, positioning, calc,shapes.geometric}

\newboolean{ShowComments}
\setboolean{ShowComments}{false}  
\ifthenelse{\boolean{ShowComments}}%
	{
		\newcommand{\ColorComment}[3]{%
				{\colorbox{#1}{\color{White}   \textsf{\textbf{#2}}} \textcolor{#1}{#3}}}

	}%
	{
		\newcommand{\ColorComment}[3]{}

	}%

\definecolor{cocoricolor}{RGB}{255,127,8}

\makeatletter
\@addtoreset{equation}{section}

\makeatother

\definecolor{base}{gray}{0} 
\definecolor{comment}{rgb}{0.52,0.60,0.00} 
\definecolor{string}{rgb}{0.83,0.21,0.51} 
\definecolor{keyword1}{rgb}{0.15,0.55,0.82} 
\definecolor{keyword2}{rgb}{0.80,0.29,0.09} 
\definecolor{keyword3}{rgb}{0.71,0.54,0.00} 
\definecolor{keyword4}{RGB}{230,133,94}

\colorlet{punct}{red!60!black}
\definecolor{background}{HTML}{EEEEEE}
\definecolor{delim}{RGB}{20,105,176}
\colorlet{numb}{magenta!60!black}

\lstdefinelanguage{RuLa}{
  morekeywords = [1]{ 
  let, rule, import, if, else, match, otherwise, watch, cond, act, ruleset, default, watch, OR, for, in, while, ->, get, set, promote, as, get, 
  },
  morekeywords = [2]{ 
  Message, Result, Repeater, Qubit, 
  },
  morekeywords = [3]{ 
      update, free, transfer, res, recv, meas, 
  },
  morekeywords = [4]{ 
  true, false, u_int, vec, int, str, float, bool, 
  },
  morecomment = [l]{//},
  morecomment = [s]{/*}{*/},
  morestring = [b]{"},
}

\definecolor{peg_base}{gray}{0} 
\definecolor{peg_comment}{rgb}{0.52,0.60,0.00} 
\definecolor{peg_string}{rgb}{0.83,0.21,0.51} 
\definecolor{peg_keyword1}{rgb}{0.15,0.55,0.82} 
\definecolor{peg_keyword2}{rgb}{0.80,0.29,0.09} 

\lstdefinelanguage{PEG}{
  morekeywords = [1]{ 
  SOI, EOI, COMMENT, ASCII_DIGIT, ASCII_ALPHA, ANY, WHITESPACE, ASCII_NONZERO_DIGIT,
  },
  morekeywords = [2]{ 
  aa, rula, program, stmt, interface_def, let_stmt, brace_stmt, ident_typed, ident, expr, import_stmt, fn_def_expr, if_stmt, if_block, for_stmt, match_stmt, ruleset_stmt, rule_stmt, cond_expr, act_expr, comp_expr, braket_expr, variable_call_expr, fn_call_expr, term_expr, literal_expr, ident_list, else_if_stmt, promote_stmt, else_stmt, set_stmt, send_stmt, paren_expr, pattern, promotable, generator, argument_def, fn_call_args, callable, struct_name, tuple, match_arm, match_condition, match_action, ruleset_config, rules, rule_idents, monitor_expr, rule_contents, satisfiable, awaitable, watch_expr, int, float, bool, true_lit, false_lit, string, raw_string, send_expr, for_multi_block, for_generator, series, repeater_call, repeater_ident, ret_type_annotation, typedef_lit, maybe, rule_call_expr, set_expr, get_expr, promote_expr, vector, repeaters, inner_term, terms, op, plus, minus, asterisk, slash, percent, caret, comparable, comp_op, cond_clauses, res_assign, number, binary, bin_num, hex, hex_num, unicord, vector_type, integer_type, unsigned_integer_type, float_type, boolean_type, string_type, qubit_type, repeater_type, message_type, result_type, literal_list, rule_import, 
  },
  morecomment = [l]{//},
  morecomment = [s]{/*}{*/},
  morestring = [b]{"},
  morestring = [b]{'},
}

\colorlet{punct}{red!60!black}
\definecolor{background}{HTML}{EEEEEE}
\definecolor{delim}{RGB}{20,105,176}
\colorlet{numb}{magenta!60!black}

\lstdefinelanguage{json}{
    basicstyle = \fontsize{10}{2}\ttfamily\scriptsize,
    numbers=left,
    stepnumber=1,
    numbersep=8pt,
    showstringspaces=false,
    breaklines=true,
    frame = tbrl,
    backgroundcolor=\color{background},
    literate=
     *{0}{{{\color{numb}0}}}{1}
      {1}{{{\color{numb}1}}}{1}
      {2}{{{\color{numb}2}}}{1}
      {3}{{{\color{numb}3}}}{1}
      {4}{{{\color{numb}4}}}{1}
      {5}{{{\color{numb}5}}}{1}
      {6}{{{\color{numb}6}}}{1}
      {7}{{{\color{numb}7}}}{1}
      {8}{{{\color{numb}8}}}{1}
      {9}{{{\color{numb}9}}}{1}
      {:}{{{\color{punct}{:}}}}{1}
      {,}{{{\color{punct}{,}}}}{1}
      {\{}{{{\color{delim}{\{}}}}{1}
      {\}}{{{\color{delim}{\}}}}}{1}
      {[}{{{\color{delim}{[}}}}{1}
      {]}{{{\color{delim}{]}}}}{1},
}

\begin{document}
\input{Chapters/Title}

\tableofcontents
\newpage
\listoffigures
\listoftables

\clearpage

\input{Chapters/1.Introduction/1.0.Introduction}

\input{Chapters/2.Preliminaries/2.0.Preliminaries}

\input{Chapters/3.ProblemDefinition/3.0.ProblemDefinition}

\input{Chapters/4.Proposal/4.0.Proposal}


\input{Chapters/5.Evaluation/5.0.Evaluation.tex}

\input{Chapters/6.Conclusion/6.0.Conclusion}



\bibliographystyle{unsrt}
\bibliography{references}

\appendix
\input{Chapters/B.Appendix}
\end{document}

%% file: Chapters/Title.tex
\begin{titlepage}
   \begin{center}
   \large
    Master's Thesis (Academic Year 2022)\\
    \vspace*{1cm}
    \huge
    \textbf{RuLa: A Programming Language for RuleSet-based Quantum Repeaters}\\
    \Large
    \vspace{9.5cm}
    A thesis presented for the degree of\\
    Master of Media and Governance\\
    \end{center}
    \begin{flushleft}
        \Large
        \textbf{Keio University}\\
        \large
        \textbf{Graduate School of Media and Governance}\\
    \end{flushleft}
    \begin{flushright}
    \Large
    \textbf{Ryosuke Satoh}
    \end{flushright}
\end{titlepage}

\newpage

\begin{titlepage}
   \begin{center}
    \large
    Abstract of Master’s Thesis Academic Year 2022\\
    \vspace*{1cm}
    \Large
    \textbf{RuLa: A Programming Language for RuleSet-based Quantum Repeaters}\\
    \end{center}
    \Large
    \vspace{1cm}
    \normalsize
    
    \vspace{1cm}
    Quantum Repeaters are one critical technology for scalable quantum networking. One of the key challenges regarding quantum repeaters is their management of how they provide quantum entanglement for distant quantum computers. We focus on the RuleSet architecture, which is a decentralized way to manage repeaters. The RuleSet concept is designed to scale up the management of quantum repeaters for future quantum repeaters, suitable because of its flexibility and asynchronous operation, however, it is still at the conceptual level of definition and it is very hard to define raw RuleSets. In this thesis, we introduce a new programming language, called ``RuLa'', to write the RuleSets in an intuitive and coherent way. The way RuLa defines RuleSet and Rule is very similar to how the Rule and RuleSets are executed, so that the programmer can construct the RuleSets the way they want repeaters to execute them. We provide some examples of how the RuleSets are defined in RuLa and what is the output of the compilation. We also discussed future use cases and applications of this language.

    \underline{4. Domain Specific Language}, 
    \vspace{1cm}
    \begin{flushleft}
        \Large
        \textbf{Keio University}\\
        \large
        \textbf{Graduate School of Media and Governance}\\
    \end{flushleft}
    \begin{flushright}
    \Large
    \textbf{Ryosuke Satoh}
    \end{flushright}
\end{titlepage}

%% file: Chapters/1.Introduction/1.0.Introduction.tex
\chapter{Introduction}
\input{Chapters/1.Introduction/1.1.Background}
\input{Chapters/1.Introduction/1.2.ResearchCongribution}
\input{Chapters/1.Introduction/1.3.ThesisStructure}

%% file: Chapters/1.Introduction/1.1.Background.tex
\section{Background}
Quantum technology has been one of the biggest technological challenges in recent decades. Quantum computers have been built and are even available to researchers and students around the world through the cloud system~\cite{cross2018ibm}. Recently, Google claimed quantum supremacy, which demonstrated that quantum computers experimentally exceed classical computers in terms of computational power~\cite{arute2019quantum} (In another paper, quantum computing simulation with more than fifty qubits~\cite{zhou2020limits} has been demonstrated). Quantum communication is also a very active research field that has also gained research interest over the years because of its uniqueness and a wide variety of applications.  

Quantum communication allows us to connect multiple quantum computers or small quantum processors to exchange quantum states using quantum teleporation~\cite{bennett1993teleporting}, enhance quantum computational power~\cite{yimsiriwattana2004distributed}, and securely transfer classical information through a public quantum channel using quantum key distribution~\cite{ekert1992quantum}. In addition to these, we can use quantum communication to improve the sensitivity of sensor networks~\cite{giovannetti2011advances}. 

However, there is a huge gap between the current status of quantum networking development and the quantum networks that are capable of running those applications. Current state-of-the-art experiments have shown entanglement generation in a few hops of quantum devices~\cite{humphreys2018deterministic, hensen2016loophole}. The biggest challenge is to keep the entangled qubits in a nearly perfect state for a long time. Researchers have put a great deal of effort into hardware development to improve the coherence time of quantum memories~\cite{cho2016highly, wang2017single}, to reduce the errors and losses in fiber.

In addition to the quality of the hardware, the architecture and protocols of quantum networks are also important to develop truly useful quantum networks~\cite{ kozlowski2020designing, van2021quantum}. RuleSet protocol~\cite{matsuo2019quantum, van2021quantum} is one of the research projects that define how to manage quantum repeaters, which is one of the most important components in quantum networking. This approach supports scalable asynchronous operations for quantum repeaters, and it is easy to distribute the generated RuleSet to intermediate quantum repeaters. 
However, one of the biggest difficulties is creating a large number of RuleSets, especially when there are complex Rules inside the RuleSet. 

In this thesis, we approach this problem by making a new programming language to define the RuleSet and the corresponding systems to generate a large number of RuleSets. 

%% file: Chapters/1.Introduction/1.2.ResearchCongribution.tex
\section{Research Contribution}
The main research contribution of this project is to develop a new programming language called RuLa that makes it easier to write and execute RuleSets. 
The key idea for scalable RuleSet generation is to provide an abstract way to define multiple rules with flexible configurations. In addition, since RuleSet does not allow any condition operators such as \verb|if~else|, Rula introduces a higher-level condition that is automatically translated into several different condition clauses in different rules. RuLa provides a way to modularly define the Rules that allows us to improve the visibility and maintainability of the RuleSet.

We assume two types of users who would find this project beneficial. 
One is for users who want to try a new quantum networking protocol in the formulation of the RuleSet. Since RuleSet is fairly general for describing any operation in quantum networking, it is possible to implement a new protocol in a RuleSet format. RuLa allows this user to accelerate the development cycle by providing a user interface to generate RuleSets. 

The second user is the network manager who actually manages the quantum network in the future. This could be a human manager who decides how the repeaters work or even a computer program that automatically deals with requests from other quantum computers. Once the way of generation of the RuleSet is defined in RuLa, it is possible to generate an arbitrary number of RuleSets by only changing its configurations.

%% file: Chapters/1.Introduction/1.3.ThesisStructure.tex
\section{Thesis Structure}
This thesis is structured as follows. Chapter~\ref{chap:preliminaries} provides basic preliminary knowledge on quantum information processing, such as the mathematical representation of qubits, quantum circuits, and entanglement. It also introduces the current status of quantum networking development and important protocols for quantum communications.
The problems to be tackled in this thesis are defined in Chapter~\ref{chap:prob_def}. The actual contents of RuleSet and how the RuleSet is described are also explained in this chapter. In Chapter~\ref{chap:proposal}, actual proposals including language design decisions and how the compiler works are explained with its requirements and assumptions. Finally, we evaluate our proposals in terms of their feasibility, use cases, and scalability in Chapter~\ref{chap:evaluation} and conclude this thesis in Chapter~\ref{chap:conclusion}.

%% file: Chapters/2.Preliminaries/2.0.Preliminaries.tex
\chapter{Preliminaries}
\label{chap:preliminaries}
This chapter covers preliminary knowledge of quantum computing and networking from an overview of the history of quantum information processing and fundamental calculation to specific types of operation such as entanglement swapping and purification, which are very important protocols in quantum communications.  In addition to that, this chapter introduces the concept of domain-specific languages.
\input{Chapters/2.Preliminaries/2.1.HistoryOfQuantum}

\input{Chapters/2.Preliminaries/2.2.Qubit}
\input{Chapters/2.Preliminaries/2.3.QuantumInformationProcessing}
\input{Chapters/2.Preliminaries/2.4.Entanglement}
\input{Chapters/2.Preliminaries/2.5.QuantumCommunication}
\input{Chapters/2.Preliminaries/2.6.Tomography}
\input{Chapters/2.Preliminaries/2.7.EntanglementPurification}
\input{Chapters/2.Preliminaries/2.8.EntanglementSwapping}
\input{Chapters/2.Preliminaries/2.9.QuantumErrorCorrection}
\input{Chapters/2.Preliminaries/2.10.DomainSpecificLanguage}

%% file: Chapters/2.Preliminaries/2.1.HistoryOfQuantum.tex
\section{History of Quantum Technology}
Quantum technology in the 21st century is more than just a theory in research papers. Real quantum computers and networks have been deployed and operated by companies,  universities, and national labs. Researchers in several fields in academia, leading companies, and startups are getting involved in this world competition for developing large quantum computers and networks to see the next generation of computation. 

The term ``quantum computer'' was proposed by Richard Feynman in a talk at a conference in 1981~\cite{feynman1982simulating}. His original motivation for building a quantum computer was to simulate quantum systems. In general, simulating a large quantum system is a very hard task for our laptop computers, and he thought it was more natural to simulate such a complex system with quantum computers. 

As Feynman proposed, simulating a quantum system is a promising application of quantum computing~\cite{brown2010using}. Quantum computing allows us to simulate complex quantum systems more precisely and faster. Researchers demonstrated the calculation of the energy surface of molecular hydrogen in 2016 with a real superconducting quantum device~\cite{o2016scalable, kandala2017hardware}.

A qubit is the heart of a quantum computer that can be implemented by different types of physical systems. One of the prominent examples is the superconducting qubit proposed by Nakamura et al. in~\cite{nakamura1999coherent}. There are also qubits with photons~\cite{kok2006linear, kok2007linear} and trapped ions~\cite{schindler2013quantum}.

Current quantum computers that are called NISQ (Noisy Intermediate Scale Quantum) devices do not have error correction capability and do not scale up for more complex problems and algorithms~\cite{preskill2018quantum}. However, computational power is reaching the edge of what classical computers cannot do. 
In 2019, Google claimed that its quantum processor "Sycamore" experimentally outperformed classical computers and achieved quantum supremacy~\cite{arute2019quantum} (However, the result has been still controversial because the simulation of more than fifty qubits was also demonstrated~\cite{zhou2020limits}). Quantum computational advantage has also been demonstrated by using photonic quantum computers with solving gaussian boson sampling on top of their quantum computers~\cite{zhong2020quantum}.

In quantum communication, the first loophole-free CHSH experiment has been demonstrated, which shows the feasibility of generating genuine entangled states over a distance~\cite{hensen2015loophole}. There is also a demonstration of Quantum Key Distribution (QKD) with satelite~\cite{liao2017satellite}. This experiment shows that it is possible to actually apply QKD protocols over the globe.

%% file: Chapters/2.Preliminaries/2.2.Qubit.tex
\section{Qubit}
\label{pre:qubit}
The qubit is a quantum analogue of tge bit in classical computers and is a fundamental unit of quantum information processing. Unlike a classical bit representing '0' or '1' deterministically, a qubit could be a superposition of '0' and '1'. 

\subsection{Notation}
Dirac's bra-ket notation invented by Paul Dirac~\cite{dirac1939new} is a powerful representation of the quantum state. A quantum state can be represented by a vector in a Hilbert space $\mathcal{H}$.
As the simplest example, the zero (one) state can be represented as $\ket{0}$ ($\ket{1}$).
\begin{align}
\label{eq:ket}
\ket{0} = 
\begin{bmatrix}
1 \\  
0  
\end{bmatrix},
\hspace{1cm} 
\ket{1} = 
\begin{bmatrix} 
0 \\ 
1
\end{bmatrix}.
\end{align}

The superposition of the quantum states can be represented as linear combinations of the computational basis states.
\begin{align}
    \ket{\psi} = \alpha \ket{0} + \beta \ket{1}
\end{align}
where $\alpha$ and $\beta$ are complex coefficients ($\alpha, \beta \in \mathbb{C}$) that satisfy the relationship $|\alpha|^2 + |\beta|^2 = 1$.
For example, when $\alpha = \frac{1}{\sqrt{2}}$, $\beta = \frac{1}{\sqrt{2}}$, the quantum state is
\begin{align}
    \ket{\psi} = \frac{1}{\sqrt{2}}\ket{0} + \frac{1}{\sqrt{2}}\ket{1}
\end{align}
which is an equal superposition of the basis states $\ket{0}$ and $\ket{1}$ ($|\alpha|^2 = \frac{1}{2}$, $|\beta|^2 = \frac{1}{2}$).

If there are $m$ bases in a state, the superposition of all basis states is as follows.
\begin{align}
    \ket{\psi} = a_0 \ket{0} + a_1 \ket{1} + ... + a_{m-1} \ket{m-1}
\end{align}
The sum of the squared coefficients must also be 1 ($|a_0|^2 + |a_1|^2 + ... + |a_{m-1}|^2 = 1$). 

It is simply possible to extend this representation to the multi-qubit quantum state. The multi-qubit quantum state can be constructed by tensor products $otimes$ of qubit states.
\begin{align}
    \ket{0}\otimes\ket{0} = \ket{00} =
\begin{bmatrix} 
1 \\ 0 \\ 0 \\ 0
\end{bmatrix}
\end{align}
The resulting Hilbert space is just a tensor product of two Hilbert spaces ($\mathcal{H} = \mathcal{H}_1 \otimes \mathcal{H}_2$).
When there are $n$ qubits in the system and all the qubit states are $\ket{0}$, the quantum state $\ket{\psi}$ is
\begin{align}
    \ket{\psi} = \ket{0}^{\otimes n} = \ket{00..0}.
\end{align}
An $n$ qubit state can be represented as a vector with $2^n$ elements.

We can also simply represent the inner product and the outer product of quantum states with the bra-ket notation. Let $\ket{\psi}$ and $\ket{\phi}$ be quantum states $\ket{\psi} = \{a_1, a_2, ..., a_n\}$ and $\ket{\phi} = \{b_1, b_2, ..., b_n\}$ where $a_k$ and $b_k$ are complex probability amplitudes ($a_k, b_k\in \mathbb{C}$). 
The inner product of two quantum states is
\begin{align}
    \bra{\psi}\ket{\phi} = a_1 b_1 + a_2 b_2 + ... + a_n b_n.
\end{align}
Their outer product is
\begin{align}
    \ket{\psi}\bra{\phi} = 
    \begin{bmatrix}
        a_1\\
        a_2\\
        \vdots\\
        a_n
    \end{bmatrix}
    \begin{bmatrix}
        b_1 b_2 \cdots b_n
    \end{bmatrix} = 
    \begin{bmatrix}
    a_1 b_1 & a_2 b_1 & \cdots & a_n b_1\\
    a_1 b_2 & \ddots  &       & \\
    \vdots & & &\\
    a_1 b_n &&& a_n b_n
    \end{bmatrix}.
\end{align}

The density matrix is one good way to represent a mixed state to introduce noise into a quantum state. If a qubit is in pure state $\ket{\psi}$, the density matrix $\rho$ of $\ket{\psi}$ is
\begin{align}
    \rho = \ket{\psi}\bra{\psi}.
\end{align}
When the target quantum state is noisy, we probably have unwanted quantum states in some probability.
\begin{align}
    \rho' = \sum_{i=0}^n p_i\ket{\psi_i}\bra{\psi_i}.
\end{align}
In the simplest example, if a qubit state $\ket{0}$ flips with 20$\%$ chance, the density matrix is as follows:
\begin{align}
    \rho &= 0.8 \rho_{\ket{0}} + 0.2 \rho_{\ket{1}} \\
    &= 0.8 \ket{0}\bra{0} + 0.2 \ket{1} \bra{1} \\
    &= \begin{bmatrix}
    0.8 & 0\\
    0 & 0.2
    \end{bmatrix}
\end{align}
where $\rho_{\ket{0}}$ ($\rho_{\ket{1}}$) is the density matrix of pure state $\ket{0}$ ($\ket{1}$).

\subsection{Bloch Sphere}
\begin{figure}[h]
    \centering
    \includegraphics[width=0.5\linewidth]{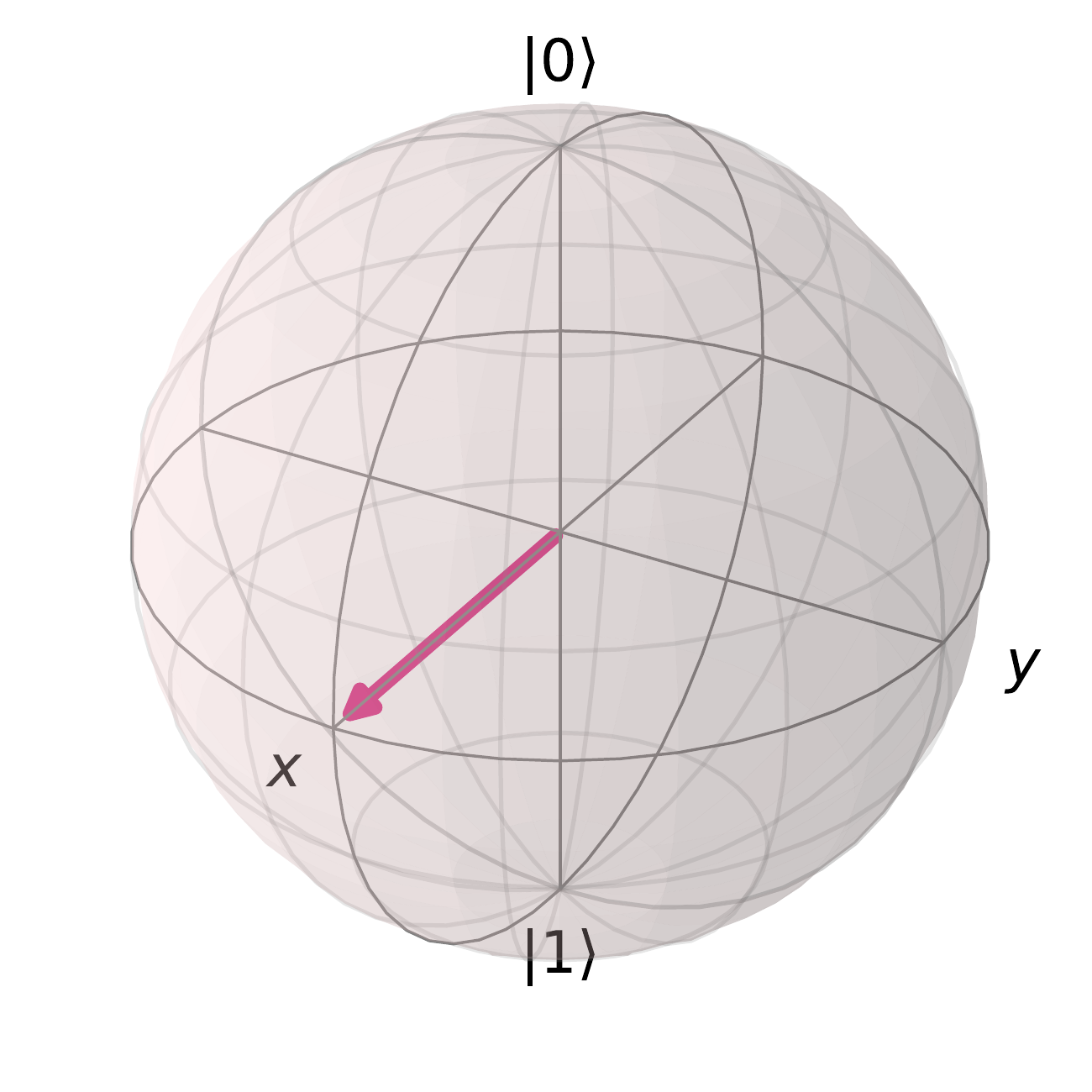}
    \caption[Bloch Sphere Representation]{Bloch sphere representation for one qubit state. Image taken from~\cite{qiskit_text}.}
    \label{fig:bloch_sphere}
\end{figure}

A Bloch sphere shown in Figure~\ref{fig:bloch_sphere} is a representation of one qubit state.
A vector inside the Bloch sphere represents the current quantum state, and the north pole represents the $\ket{0}$ state. The surface of the Bloch sphere is the of possible quantum states that one pure qubit can have. 
The points of each axis correspond to the following quantum states.
\begin{alignat*}{2}
    \ket{X^+} &= \frac{\ket{0} + \ket{1}}{\sqrt{2}},
    &&\ket{X^-} = \frac{\ket{0} - \ket{1}}{\sqrt{2}}\\
    \ket{Y^+} &= \frac{\ket{0} + i\ket{1}}{\sqrt{2}},
    &&\ket{Y^-} = \frac{\ket{0} - i\ket{1}}{\sqrt{2}}\\
    \ket{Z^+} &= \ket{0},
    &&\ket{Z^-} = \ket{1}
\end{alignat*}
where the superscript $+ (-)$ represents the positive (negative) directions in the Bloch sphere.

\subsection{Implementation}
Superconducting qubits are one promising candidate for physical qubits (e.g. They come in several types from Cooper-pair box~\cite{nakamura1999coherent}, Transmon~\cite{koch2007charge}), which many companies pursue due to its possibility of fault-tolerant quantum computing. Trapped ions~\cite{friis2018observation} and linear optical qubits~\cite{knill2001scheme} are also said to be good candidates for scalable quantum computing.

However, any quantum system that satisfies DiVincenzo's criteria proposed in~\cite{divincenzo2000physical} can be a candidate for a qubit. 
There are five criteria for the quantum computing qubit and two criteria for the quantum communication qubit.
A qubit needs to have the following characteristics.
\begin{enumerate}
    \item Scalable physical system with well-characterized qubit
    \item Capability of initialization to fiducial state
    \item Longer decoherence time than its gate operation time
    \item Capability of operating a universal set of gates
    \item Measurable a specific qubit
\end{enumerate}
In addition to those, the qubits for quantum communication need to satisfy the following criteria.
\begin{enumerate}
\setcounter{enumi}{5}
    \item Capable of converting a flying qubit to a stationary qubit
    \item Transmittable a flying qubit between specified locations.
\end{enumerate}
In terms of criteria 6 and 7, the trapped ion, the Nitrogen-vacancy center in diamond quantum memory~\cite{childress2013diamond} are good candidates for quantum communication. However, there is also research work that tries to communicate between superconducting qubits via photonic conversions~\cite{leung2019deterministic}.

%% file: Chapters/2.Preliminaries/2.3.QuantumInformationProcessing.tex
\section{Quantum Information Processing}
\label{pre:quantum_com}
A qubit as introduced in Section~\ref{pre:qubit} is nothing more than an information unit that represents a quantum state. To compute some algorithms and calculations on top of it, we need to manipulate those qubits to interpret a problem. The physical controls for these qubits are different in different physical implementations. However, operations can be written as a form of a unitary matrix. 

\subsection{Quantum Gates}
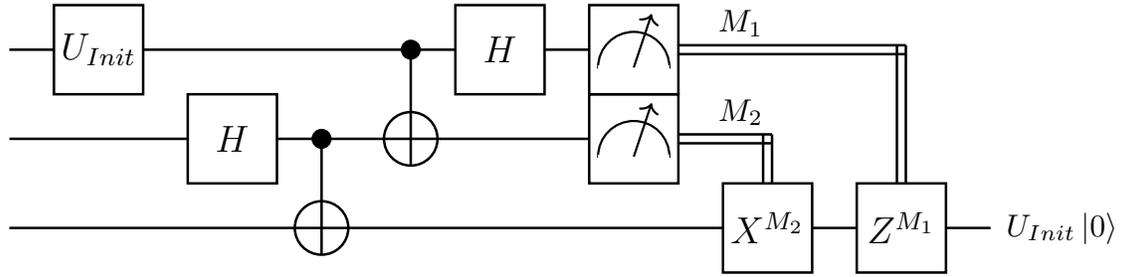
\begin{figure}[ht]
\resizebox{\textwidth}{!}{
\input{Chapters/2.Preliminaries/diagram/2.4.z.SimpleQuantumCircuit}}
    \caption[Example Quantum Circuit]{An example quantum circuit known as quantum teleportation protocol. The lines underneath the boxes represent the qubits. Each box represents a single qubit quantum gate and operations covering two qubits are two-qubit quantum gates. The flow of operation goes from left to right and the top two qubits are measured at the end of the circuit. }
    \label{fig:example_qc}
\end{figure}

Quantum Circuit is one convenient representation of a set of quantum operations such as the one shown in Figure~\ref{fig:example_qc}. 

The horizontal lines represent qubits, and operations are performed from left to right. The boxes are the quantum gate that corresponds to some operation on a single qubit. There are also two-qubit gates which are represented as a combination of line and circle. The black dot is the control qubit and the other edge of the vertical line is the target qubit. At the end of the circuit, usually we measure the qubit represented, as a meter mark, and get some classical outputs.

\subsection{Hadamard Gate}
Hadamard gate is a fundamental quantum operation that creates the superposition of a single qubit.

\begin{figure}[ht]
\resizebox{\textwidth}{!}{
\input{Chapters/2.Preliminaries/diagram/2.3.2.Hadamard}}
    \caption[Hadamard Gate Operation]{Circuit representation of Hadamard gate}
    \label{fig:h_gate}
\end{figure}
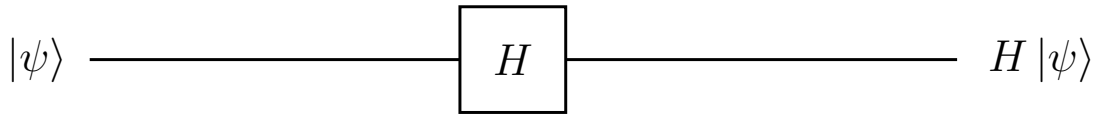
Hadamard gate can be written in matrix form as
\begin{align}
    H = \frac{1}{\sqrt{2}}
    \begin{bmatrix}
    1 & 1\\
    1 & -1
    \end{bmatrix}.
\end{align}
When the initial state is $\ket{0}(\ket{1})$ and the H gate is applied to it, the resulting state is,
\begin{align}
    H\ket{0} &= \ket{+} = \frac{1}{\sqrt{2}}\ket{0} + \frac{1}{\sqrt{2}}\ket{1}\\
    H\ket{1} &= \ket{-} = \frac{1}{\sqrt{2}}\ket{0} - \frac{1}{\sqrt{2}}\ket{1}
\end{align}
which is known as $\ket{+} (\ket{-})$ one of the equal superposition states in a single qubit state.

\subsection{Pauli Gates}
Pauli gates (X, Y, Z) are also important quantum operations that correspond to rotations about the corresponding axis of the Bloch sphere.
The matrices of Pauli X, Y, and Z are
\begin{align}
\label{eq:pauli_matrix}
    X = \begin{bmatrix}
    0 & 1\\
    1 & 0
    \end{bmatrix},
    Y = \begin{bmatrix}
    0 & -i\\
    i & 0
    \end{bmatrix},
    Z = \begin{bmatrix}
    1 & 0\\
    0 & -1
    \end{bmatrix}.
\end{align}
Those operations are the $\pi$ rotation over the corresponding bases. Pauli X gate flips $\ket{0}$ to $\ket{1}$ vice versa, and the Z gate applies $\theta = \pi$ phase rotation to $\ket{\psi} = \frac{\ket{0} + e^{i\theta}\ket{1}}{\sqrt{2}}$. When the original state is $\ket{1}$, the Pauli Z operation gives us $Z\ket{1} = -\ket{1}$ which introduces a global phase.

\subsection{Controlled NOT (X) Gate}
Controlled NOT gate, CNOT gate, in short, is a quantum operation over two different qubits.
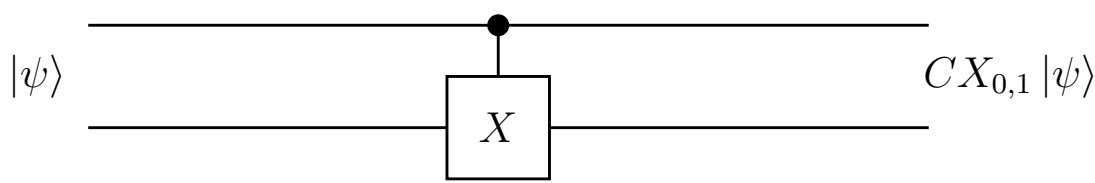
\begin{figure}[h]
\resizebox{\textwidth}{!}{
\input{Chapters/2.Preliminaries/diagram/2.3.4.CNOTgate}}
    \caption[CNOT Gate Operation]{Circuit representation of CNOT (CX) gate. This notation is equivalent to the gate with a circle and a line with a black dot used in Figure~\ref{fig:example_qc}.}
    \label{fig:cnot_gate_qc}
\end{figure}
Figure~\ref{fig:cnot_gate_qc} shows the gate representation of the CNOT gate. When the first (top) qubit is $\ket{1}$, the CNOT gate applies the X gate to the second (bottom) qubit. 
For example, when the initial state is $\ket{\psi} = \ket{10}$, the state becomes $CX\ket{\psi} = \ket{11}$. 
CNOT gate is represented by a four-by-four matrix because it covers two qubits.
\begin{align}
    CX = \begin{bmatrix}
    1 & 0 & 0 & 0\\
    0 & 1 & 0 & 0\\
    0 & 0 & 0 & 1\\
    0 & 0 & 1 & 0
    \end{bmatrix}
\end{align}

\subsection{Controlled Z Gate}
Controlled Z (CZ) gate is similar to the Controlled NOT gate, but the target operation is not the Pauli X gate, but the Pauli Z gate.
\begin{figure}[h]
\resizebox{\textwidth}{!}{
\input{Chapters/2.Preliminaries/diagram/2.3.5.CZgate}}
    \caption[CZ Gate Operation]{CZ gate}
    \label{fig:cz_gate_qc}
\end{figure}
When the first qubit is $\ket{1}$, this gate applies the Z gate to the second qubit. 
The CNOT gate and the CZ gate are interchangeable since Pauli X and Z are interchangeable over the Hadamard gate.

\subsection{Measurement}
To extract information from the quantum state, we need to measure the qubit and observe it as a classical result.
Let $\{M_m\}$ be a collection of measurement operators where $m$ is the corresponding computational basis state, and $\ket{\psi}$ be the measured quantum state. The probability $p$ of observing a basis state $m$ is
\begin{align}
p(m) = \bra{\psi} M ^\dag_m M _m \ket{\psi}.
\end{align}
For example, when we measure $\ket{0}$ on the two computational bases $\ket{0}\bra{0}$ and $\ket{1}\bra{1}$,
\begin{align}
    p(\ket{0}) = \braket{0|0}\braket{0|0}\braket{0|0}= 1, \\
    p(\ket{1}) = \braket{0|1}\braket{1|1}\braket{1|0}= 0.
\end{align}
The probability of measuring $\ket{0}$ is 1 and $\ket{1}$ is 0. However, when we measure $\ket{0}$ on different bases, for example $\ket{+}\bra{+}$ and $\ket{-}\bra{-}$,
\begin{align}
    p(\ket{+}) = \braket{0|+}\braket{+|+}\braket{+|0}= \frac{1}{2}, \\
    p(\ket{-}) = \braket{0|-}\braket{-|-}\braket{-|0}= \frac{1}{2}.
\end{align}

It is also possible to construct a measurement operation for more than a single qubit. One frequently used basis in quantum communication is the Bell measurement, which measures in a Bell basis (see details on the Bell state in Section~\ref{sec:bell_measurement}).

\subsection{Noise}
When a qubit is placed in a completely closed and perfect system, its state could always be perfect, and that qubit always tells us the actual quantum state. However, the reality is different, and its state could be far from the ideal state. A quantum state can interact with the outer world and become a noisy quantum state. 

There are multiple types of noisy channels in the quantum world. 

\textbf{Bit flip channel}: One of the simplest types of errors is the bit flip error, which can also occur in classical bits. 
Suppose that the bit-flip error occurs with probability $p$. The noisy quantum state $\rho'$ can be described as
\begin{align}
    \rho' &= (1-p)\ket{\psi}\bra{\psi} + pX\ket{\psi}\bra{\psi}X\\
        &= (1-p)\rho + p X\rho X
\end{align}
where $\ket{\psi}$ is the ideal state without noise.

\textbf{Phase flip channel}: Similar to the bit flip error, we can also think of the phase flip error by introducing Z errors in the quantum state.
\begin{align}
    \rho'= (1-p)\ket{\psi}\bra{\psi} + pZ\ket{\psi}\bra{\psi}Z
\end{align}

\textbf{Depolarizing channel}: The depolarizing channel is an important error type in the quantum world, in which a qubit becomes a completely mixed state with some probability.
\begin{align}
\label{eq:depolarizing}
    \rho' = (1-p)\rho + \frac{p}{d}I
\end{align}
where $I$ is the identity matrix and $d$ is the dimension of the quantum system. More generally, $\frac{I}{d}$ satisfies the following equation for arbitrary $\rho$.
\begin{align}
    \frac{I}{d} = \frac{\rho + X\rho X + Y \rho Y + Z \rho Z}{2d}
\end{align}
where $X$, $Y$ and $Z$ are Pauli matrices defined in Equation~\ref{eq:pauli_matrix}
Thus, Equation~\ref{eq:depolarizing} can be rewritten as
\begin{align}
    \rho' = \left(1-\frac{2d-1}{2d}p\right)\rho + \frac{p}{2d}(X\rho X + Y \rho Y + Z \rho Z)
\end{align}

\subsection{Fidelity}
Fidelity is a metric that measures the distance between two quantum states. The fidelity $F$ between two quantum states $\rho$ and $\sigma$ is defined as
\begin{align}
    F(\rho, \sigma) = \text{tr}\sqrt{\sqrt{\rho}\sigma\sqrt{\rho}}
\end{align}
where $\text{tr}$ is trace of the resulting matrix.
It is also possible to measure the fidelity between the pure state $\ket{\psi}$ and an arbitrary quantum state $\rho$ as
\begin{align}
    F(\ket{\psi}, \rho) = \sqrt{\bra{\psi}\rho\ket{\psi}}.
\end{align}

For example, if we measure the fidelity between $\ket{0}$ and $\ket{1}$,
\begin{align}
F(\ket{0}, \rho_{\ket{1}}) = \sqrt{\bra{0}\ket{1}\bra{1}\ket{0}} = 0.
\end{align}
The fidelity between $\ket{0}$ and $\ket{+}$ is,
\begin{align}
F(\ket{0}, \rho_{\ket{+}}) = \sqrt{\bra{0}\ket{+}\bra{+}\ket{0}} = \frac{1}{2}
\end{align}
where $\rho_{\ket{\psi}}$ is the density matrix of a pure state $\ket{\psi}$.

%% file: Chapters/2.Preliminaries/diagram/2.4.z.SimpleQuantumCircuit.tex
\begin{tikzpicture}[semic/.style args={#1,#2}{semicircle,minimum width=#1,draw,anchor=arc end,rotate=#2},outer sep=0pt,line width=.7pt]  
    
    \draw[thick] (1,0)--node[midway,above]{}(12,0);
    \draw[thick] (1,1)--node[midway,above]{}(8,1);
    \draw[thick] (1,2)--node[midway,above]{}(8.2,2);
    
    \draw[thick, fill=white] (1.5,2.5) rectangle (2.5,1.5) node [pos=.5]{$U_{Init}$};
    
    \draw[thick, fill=white] (3,1.5) rectangle (4,0.5) node [pos=.5]{$H$};
    \draw[fill=black] (4.5, 1) circle(0.1cm);
    \draw[thick] (4.5,1)--node[midway,above]{}(4.5,-0.3);
    \draw[thick, fill=none] (4.5, 0) circle(0.3cm);
    
    \draw[fill=black] (5.5, 2) circle(0.1cm);
    \draw[thick] (5.5,2)--node[midway,above]{}(5.5,0.7);
    \draw[thick, fill=none] (5.5, 1) circle(0.3cm);
    \draw[thick, fill=white] (6,2.5) rectangle (7,1.5) node [pos=.5]{$H$};
    
    \draw[thick] (8,0.95)--node[midway,above]{}(9.55,0.95);
    \draw[thick] (8,1.05)--node[midway,above]{}(9.55,1.05);
    \draw[thick] (8,1.95)--node[midway,above]{}(11.05,1.95);
    \draw[thick] (8,2.05)--node[midway,above]{}(11.05,2.05);
    
    \draw[thick] (9.45,1.05)--node[midway,above]{}(9.45,0);
    \draw[thick] (9.55,1.05)--node[midway,above]{}(9.55,0);
    \draw[thick] (10.95,2.05)--node[midway,above]{}(10.95,0);
    \draw[thick] (11.05,2.05)--node[midway,above]{}(11.05,0);

    \draw[thick, fill=white] (7.5,0.5) rectangle (8.5,1.5) node [pos=.5]{};
    \node [semic={0.8cm,0}]    at (7.6,0.8){};
    \draw[draw=none, color=white, fill=white] (7.6,0.7) rectangle (8.4,0.82) node [pos=.5]{};
    \draw[arrows=->](8, 0.8)--(8.2,1.4);
    \draw[thick, fill=white] (7.5,1.5) rectangle (8.5,2.5) node [pos=.5]{};
    \node [semic={0.8cm,0}]    at (7.6,1.8){};
    \draw[draw=none, color=white, fill=white] (7.6,1.7) rectangle (8.4,1.82) node [pos=.5]{};
    \draw[arrows=->](8, 1.8)--(8.2,2.4);
    
    \draw[thick, fill=white] (9,-0.5) rectangle (10,0.5) node [pos=.5]{$X^{M_2}$};
    \draw[thick, fill=white] (10.5,-0.5) rectangle (11.5,0.5) node [pos=.5]{$Z^{M_1}$};
    
    \node[draw=none] at (9.2,2.3) {\footnotesize $M_1$};
    \node[draw=none] at (9.2,1.3) {\footnotesize $M_2$};
    \node[draw=none] at (12.8,0) {\footnotesize $U_{Init}\ket{0}$};


\end{tikzpicture}

%% file: Chapters/2.Preliminaries/diagram/2.3.2.Hadamard.tex
\begin{tikzpicture}
    \node[draw=none] at (0.5,0) {$\ket{\psi}$};
    \draw[thick] (1,0)--node[midway,above]{}(9.2,0);
    \draw[thick, fill=white] (4.5,-0.5) rectangle (5.5,0.5) node [pos=.5]{$H$};
    \node[draw=none] at (10,0) {$H\ket{\psi}$};
\end{tikzpicture}

%% file: Chapters/2.Preliminaries/diagram/2.3.4.CNOTgate.tex
\begin{tikzpicture}
    \node[draw=none] at (0.5, 0.5) {$\ket{\psi}$};
    \draw[thick] (1,0)--node[midway,above]{}(9.2,0);
    \draw[thick] (1,1)--node[midway,above]{}(9.2,1);
    \draw[fill=black] (5, 1) circle(0.1cm);
    \draw[thick] (5,1)--node[midway,above]{}(5,0);
    \draw[thick, fill=white] (4.5,-0.5) rectangle (5.5,0.5) node [pos=.5]{$X$};
    \node[draw=none] at (10,0.5) {$CX_{0, 1}\ket{\psi}$};
\end{tikzpicture}

%% file: Chapters/2.Preliminaries/diagram/2.3.5.CZgate.tex
\begin{tikzpicture}
    \node[draw=none] at (0.5, 0.5) {$\ket{\psi}$};
    \draw[thick] (1,0)--node[midway,above]{}(9.2,0);
    \draw[thick] (1,1)--node[midway,above]{}(9.2,1);
    \draw[fill=black] (5, 1) circle(0.1cm);
    \draw[thick] (5,1)--node[midway,above]{}(5,0);
    \draw[thick, fill=white] (4.5,-0.5) rectangle (5.5,0.5) node [pos=.5]{$Z$};
    \node[draw=none] at (10,0.5) {$CZ_{0, 1}\ket{\psi}$};
\end{tikzpicture}

%% file: Chapters/2.Preliminaries/2.4.Entanglement.tex
\section{Entanglement}
Entanglement is a fundamental phenomenon in quantum systems that shows a stronger correlation between two quantum systems. This correlation cannot be mimicked by the classical correlation. 

\subsection{Bell State}
\label{sec:bell_state}
A fundamental example of entangled states is given by the four Bell states, which can be
\begin{align}
    \ket{\Phi^+} &= \frac{\ket{00} + \ket{11}}{\sqrt{2}} \label{eq:phip} \\
    \ket{\Phi^-} &= \frac{\ket{00} - \ket{11}}{\sqrt{2}} \\
    \ket{\Psi^+} &= \frac{\ket{01} + \ket{10}}{\sqrt{2}} \\
    \ket{\Psi^-} &= \frac{\ket{01} - \ket{10}}{\sqrt{2}}.
\end{align}
The first and the second qubits are strongly correlated and the state cannot be determined separately when they are measured. For example, when the first qubit in \ref{eq:phip} is $\ket{0}(\ket{1})$, the second qubit is measured as $\ket{0}(\ket{1})$ respectively.

\subsection{Multipartite Entangled State}
It is also possible to think that more than two quantum bits are entangled. A particular example of a three-qubits entangled state is given by the GHZ (Greenberger-Horne-Zeilinger) state~\cite{greenberger1989going}.
\begin{align}
    \ket{GHZ} = \frac{\ket{000} + \ket{111}}{\sqrt{2}}.
\end{align}
In general, an $n$ qubit GHZ state is 
\begin{align}
    \ket{GHZ} = \frac{\ket{00...0}+ \ket{11...1}}{\sqrt{2}}.
\end{align}

Another example of a multipartite entangled state is a graph state. Suppose that we have a graph $G = (V, E)$ that is composed of a set of vertices $\{v_i\}\in V$ and edges $\{v_i, v_j\}\in E$. The qubits in a system are corresponding to the vertices of the target graph and the entanglement between two qubits are edges in the graph. 
The following equation shows how the graph state is created.
\begin{align}
    \ket{G} = \bigotimes_{\{v_i, v_j\}\in E} CZ_{\{v_i, v_j\}}\ket{+}^n
\end{align}
All the qubits are initialized into superposition state $\ket{+}$, and CZ gates are applied according to the edges of the target graph.

\subsection{Bell Measurement}
\label{sec:bell_measurement}
Let $\ket{\psi}$ be the Bell state $\ket{\Phi^+}$ shown in Equation~\ref{eq:phip}.
If we measure this Bell state on the following Bell basis, 
\begin{align}
    &M_1 = \ket{\Phi^+}\bra{\Phi^+}, M_2 = \ket{\Phi^-}\bra{\Phi^-},\\
    &M_3 = \ket{\Psi^+}\bra{\Psi^+}, M_4 = \ket{\Psi^-}\bra{\Psi^-}.
\end{align}
The measurement result of $\ket{\psi}$ can be written as follows.
\begin{align}
    &p(\ket{\Phi^+}) = \bra{\psi}M_1^\dag M_1\ket{\psi} = 1\\
    &p(\ket{\Phi^-}) = \bra{\psi}M_2^\dag M_2\ket{\psi} = 0\\
    &p(\ket{\Psi^+}) = \bra{\psi}M_3^\dag M_3\ket{\psi} = 0\\
    &p(\ket{\Psi^-}) = \bra{\psi}M_4^\dag M_4\ket{\psi} = 0
\end{align}
Measurement in the Bell basis is very important in quantum communication. It is used to perform the entanglement swapping explained in Section~\ref{sec:entanglement_swapping}.

\subsection{CHSH Experiment}
In \cite{bell2004speakable}, John Stewart Bell stated the upper limit of correlations in hidden local variables. However, this Bell's theorem was revealed to be not applicable to quantum physics. In \cite{bell1964einstein}, Bell explained the violation of his previous theorem. 
To prove the existence of the entanglement, the extended inequality called the CHSH inequality is supposed to be satisfied.

Suppose that there are two people, Alice and Bob, who select two numbers from either +1 or -1 at randomly without coordination. Let $A$ and $a$ be the values that Alice chose, and $B$ and $b$ be the values that Bob chose.

If we think of the following quantity,
\begin{align}
\label{eq:chsh_classical}
    AB + Ab + aB - ab = (A+a)B + (A-a)b.
\end{align}
where either $(A+a)a$ or $(A-a)b$ becomes 0 so that the possible values that~\ref{eq:chsh_classical} can take are $\pm 2$.
The following inequality holds when we consider the expectation values for each combination.
\begin{align}
\label{eq:chsh_e_classical}
    \mathbb{E}(AB) + \mathbb{E}(Ab) + \mathbb{E}(aB) - \mathbb{E}(ab) \leq 2
\end{align}

However, when it comes to the quantum system, Equation~\ref{eq:chsh_e_classical} no longer holds.

Alice and Bob prepare two measurement basis. Let $A$ and $B$ be the measurement basis of $Z$ and $X$. Bob chooses $\frac{-Z -X}{\sqrt{2}}$ and $\frac{Z-X}{\sqrt{2}}$ as $a$ and $b$ which are diagonal to Alice's basis. 

If Alice and Bob share one of the Bell states $\ket{\psi} = \frac{\ket{01} - \ket{10}}{\sqrt{2}}$, and measure it on the corresponding basis, the maximum expectation values of the measurement results are the following.

\begin{align}
    \langle AB \rangle &= \langle Ab \rangle = \langle aB \rangle = \frac{1}{\sqrt{2}}\\
    \langle ab \rangle &= -\frac{1}{\sqrt{2}}
\end{align}
where $\langle \cdot \rangle$ is the average observable value.
Thus, the total expectation value is
\begin{align}
    \langle AB \rangle + \langle Ab \rangle + \langle aB \rangle - \langle ab \rangle \leq 2\sqrt{2}.
\end{align}

The generation of entanglement has been demonstrated experimentally and~\cite{hensen2015loophole} is the first demonstration without any loopholes. 

%% file: Chapters/2.Preliminaries/2.5.QuantumCommunication.tex
\section{Quantum Communication}
Quantum computing has promising applications such as quantum simulation, factoring, data search in databases, and optimization. However, quantum communication allows us to scale up and think of more complex applications. 
Shor's algorithm requires 20 million noisy qubits to factor in 2048-bit RSA as reported in~\cite{gidney2021factor}. Preparing such a large number of qubits is not realistic if we assume only the capacity of a single quantum processor. Quantum communication plays an important role here in making quantum computing scalable. The simplest case is to connect several quantum processors to each other by preparing a bus, as we do on our laptops. In~\cite{meter2008arithmetic}, the authors proposed a method to achieve arithmetic on distributed quantum processors that might allow us to implement Shor's algorithm in a distributed manner. 

\subsection{Quantum Networking Stack}
There are proposed layering models for quantum communications that are modeled based on the roles in the network~\cite{kozlowski2020designing, wehner2018quantum,wu2021sequence}.
\begin{figure}[ht]
    \centering
    \includegraphics[width=\linewidth]{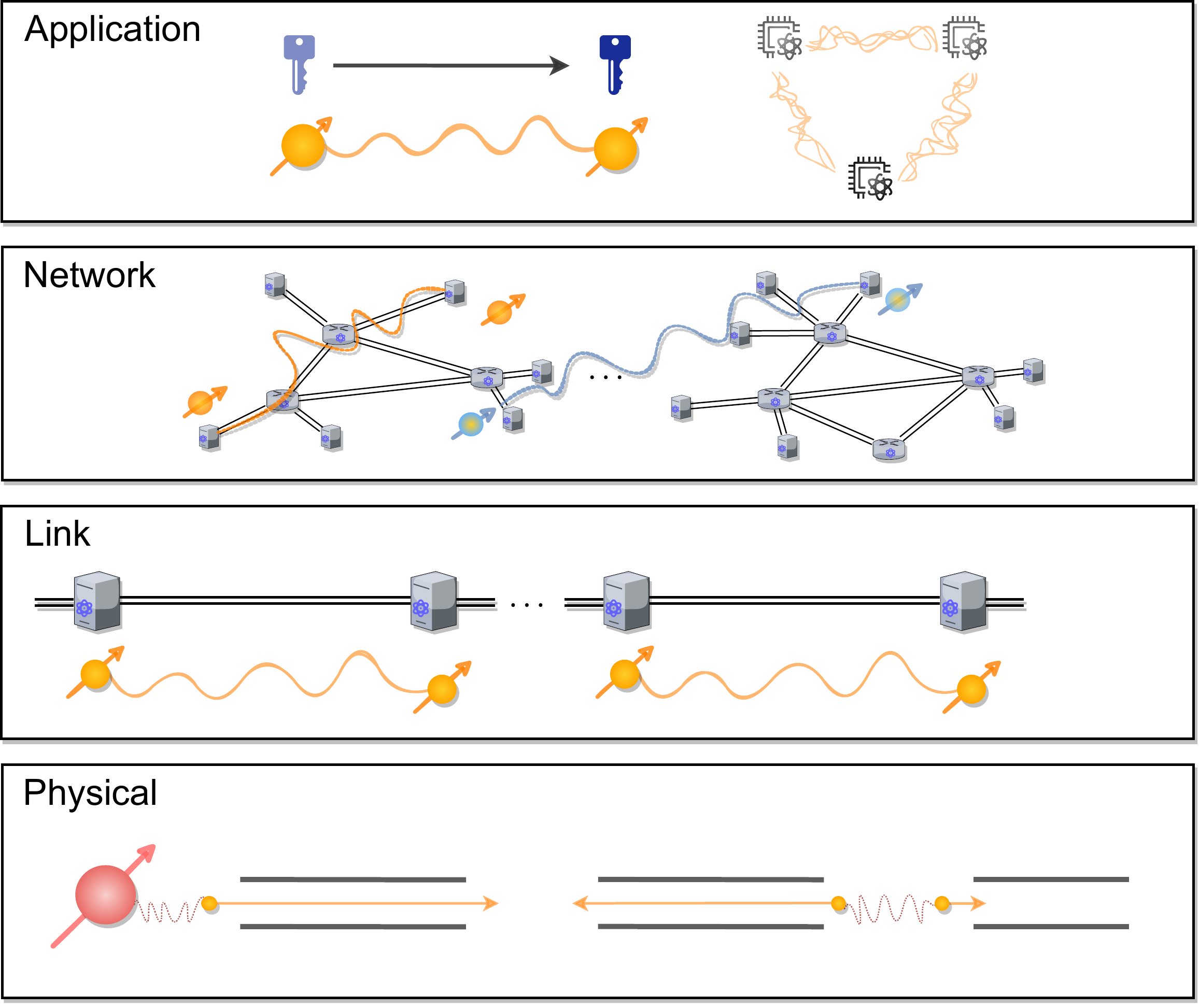}
    \caption[Quantum Networking Layering Model]{Quantum Layering model}
    \label{fig:quantum_layer}
\end{figure}

Figure~\ref{fig:quantum_layer} is a proposed layering model for quantum networking there are roughly four or five layers from the physical layer up to the application layer. 
\begin{itemize}
    \item \textbf{Application}: The layer responsible for managing end-to-end applications such as distributed quantum computing and quantum key distribution. The application layer consumes generated entanglement in the network layer. 
    \item \textbf{Network}: The network layer is responsible for end-to-end entanglement generation and management to reach the requirements for the applications.
    \item \textbf{Link}: The link layer is in charge of generating entanglement with neighboring nodes for passing that link-level entanglement to the network layer. The link layer is also responsible for managing errors for link-level entanglement. 
    \item \textbf{Physical}: This layer contains complex hardware and firmware that allow us to generate entanglement and manipulate the actual qubit to generate entanglement. 
\end{itemize}

\subsection{Applications}
\label{sec:application}
Many applications of quantum communication have been proposed, and a promising application is distributed quantum computing (DQC), which allows us to scale up the capability of quantum computing by connecting multiple quantum computers with quantum links~\cite{cuomo2020towards, diadamo2021distributed}. 

Quantum key distribution (QKD) is another very prominent example of a quantum networking application~\cite{bennett1985quantum, ekert1992quantum}. In a shared key cryptosystem, one of the biggest challenges is sharing the generated key securely. Public key cryptography solved that problem by asymmetric key generation. One of the most famous asymmetric key generation protocols is Diffie–Hellman key exchange protocol which allows us to generate a secret key from a public key that is shared with anyone in the world. The point is that we do not have to expose the generated secret key in public to share it. The security of this protocol is supported by the complexity of the modular exponentiations. 
After the Diffie-Hellman key exchange was proposed, RSA cryptography was proposed ~\cite{rivest1978method} that allows us to authenticate and sign digital signatures.   However, as Peter Shor revealed in~\cite{shor1999polynomial}, the problem of prime number factoring on which RSA relies for its security became known as a polynomial-time solvable problem using quantum computers. Most of the current cryptosystem relies on RSA or a discrete logarithmic problem, and those are said to be solved by quantum computers, so we have to invent a new method to share our information securely. One way to do this is to develop a completely new cryptographic algorithm that cannot be solved by a quantum computer. This is an ongoing project for which many people are trying to develop a new scheme that seems to be hard to solve with quantum computers. In another way, we use a quantum system as cryptography. A key factor behind this method is called the "No-cloning Theorem"~\cite{wootters1982single}. This theorem prohibits creating a perfect copy of an unknown quantum state. However, this theorem can also be proof of its security because an intermediate attacker cannot clone the original information and spoof the receiver. The security of QKD has been proven in several different articles for different protocols~\cite{lo1999unconditional, shor2000simple}.
Furthermore, to improve the security of QKD, an improved scheme called the decoy-state quantum key distribution has also been developed~\cite{lo2005decoy}.

Quantum metrology~\cite{giovannetti2011advances} and quantum sensor networks~\cite{rubio2020quantum} are also interesting applications of quantum communication. By the use of quantum entanglement, it is possible to increase the sensitivity of measurement for physical systems such as gravitational waves~\cite{schnabel2010quantum}.

\subsection{Quantum Repeater}
\label{sec:repeater}
\begin{figure}[h]
    \centering
    \includegraphics[width=.8\linewidth]{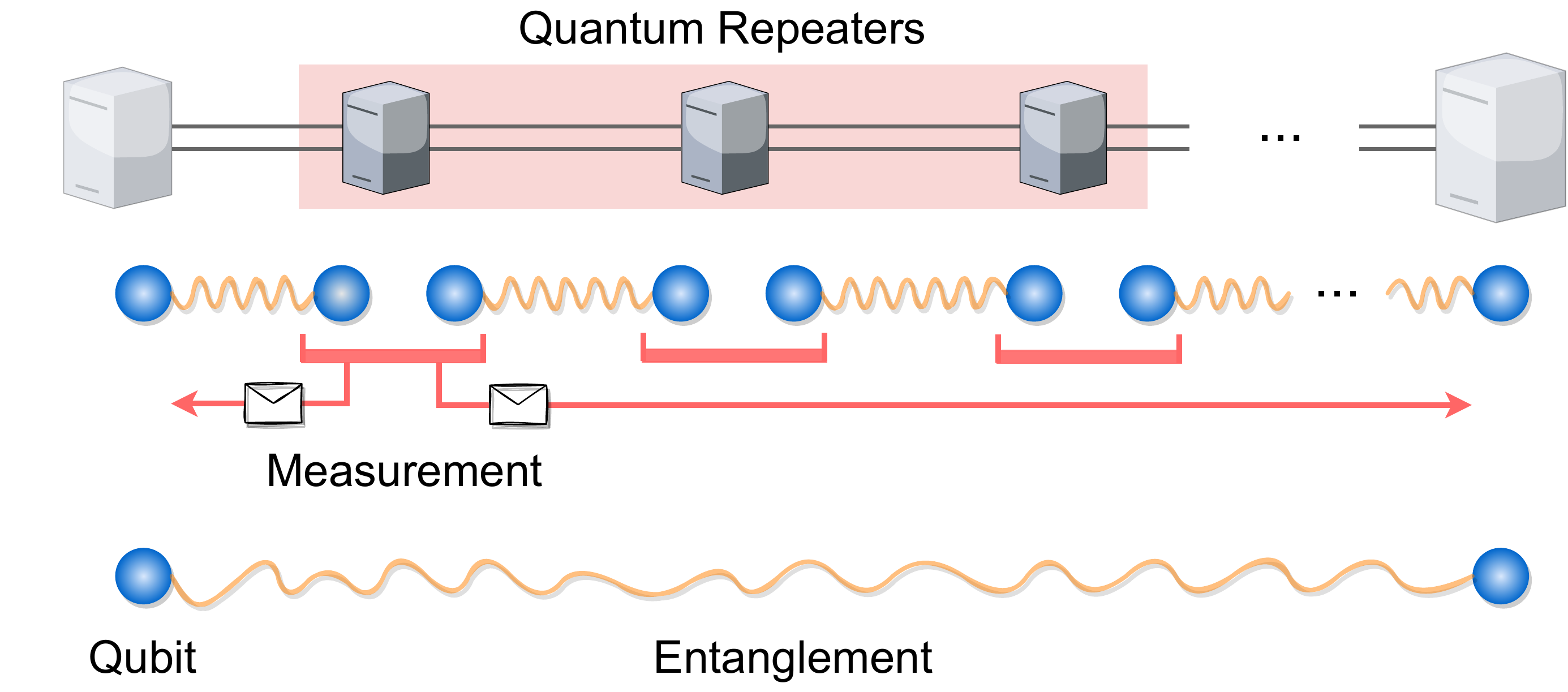}
    \caption[Entanglement Swapping Procedure]{Schematic diagram of entanglement swapping.}
    \label{fig:repeater_network}
\end{figure}

Generating long-range entanglement is a challenging task with noisy qubits and lossy fibers. The quantum repeater plays an important role in generating such a long-distance entanglement by using quantum teleportation techniques. 
Figure~\ref{fig:repeater_network} shows a very simple repeater network in which three quantum repeaters are placed between two end nodes. The distances between quantum repeaters are relatively short, so repeaters can generate link-level entanglement with higher success probability and quality. 

Once two link-level entangled states are ready between neighboring nodes, repeaters measure the corresponding two qubits on a Bell basis.

Quantum repeaters send their measurement results to the appropriate partner nodes, and the partner nodes apply the corresponding quantum gates to the entangled state. 

These processes are called "entanglement swapping" (See Section~\ref{sec:entanglement_swapping} for more details).

However, extending the entanglement is not the only task for quantum repeaters. Quantum repeaters are also responsible for managing generated entanglement. If the generated entanglement is sitting in quantum memory for a long time, it becomes noisy as time goes by. To deal with this decoherence on the entanglement, we might need purification and error correction (see Section~\ref{sec:purification} and See~\ref{sec:error_correction}) that could treat the errors in the entanglement.

Quantum repeaters can be classified into three types by their type of error management~\cite{muralidharan2016optimal}. 
\begin{itemize}
    \item \textbf{1G Repeater} is the first generation repeater that is not capable of quantum error correction. The only operations that this generation's repeater can perform are entanglement swapping and purifications.
    \item \textbf{2G Repeater} is the second generation repeater that is capable of a simple quantum error correction scheme that is tolerated for operational errors such as gate operations and measurement operations. 
    \item \textbf{3G Repeater} is the third generation repeater that has a strong ability to correct not only operational errors but also loss errors. 
\end{itemize}

There are also quantum repeaters that utilize multipartite entanglement such as graph state as a fundamental resource~\cite{azuma2015all, buterakos2017deterministic, russo2018photonic}.
Those repeaters could be more feasible because they do not even assume the existence of quantum memories. 

\subsection{Quantum Router}
\begin{figure}[h]
    \centering
    \includegraphics[width=.8\linewidth]{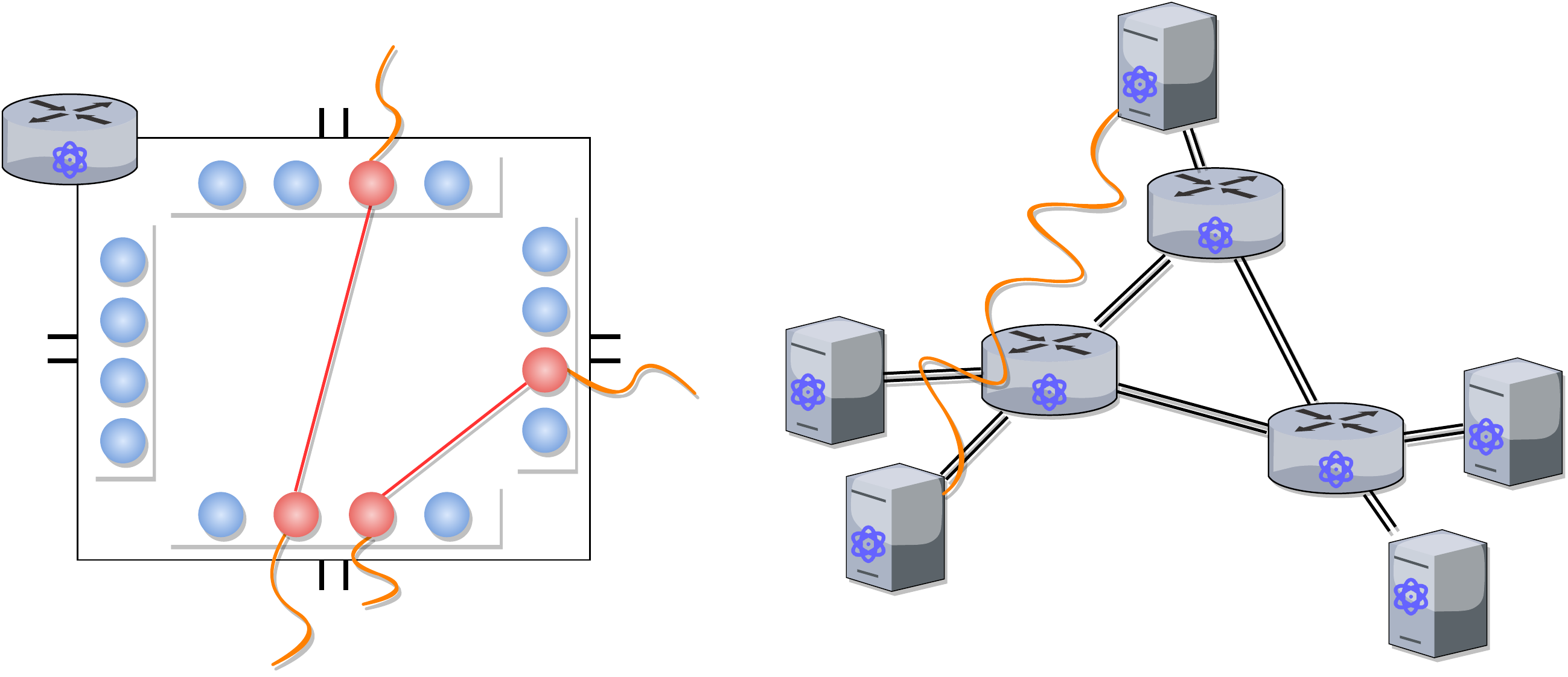}
    \caption[Quantum Router]{Quantum Router that routes quantum entanglement from one port to another port. Quantum router selectively measures qubits based on the routing table.}
    \label{fig:quantum_router}
\end{figure}

A quantum router works in a manner similar to quantum repeaters. However, they have more than two interfaces in general~\cite{lee2022quantum}. This functionality allows us to generate entanglement between multiple different parties in different paths. 
Figure~\ref{fig:quantum_router} is a diagram of how the quantum repeater works.

In terms of the routing scheme, there are several approaches to routing entangled states based on network coding~\cite{hayashi2007quantum, epping2016robust}, virtual quantum link \cite{schoute2016shortcuts}, and local complementation in multipartite entangled state~\cite{hahn2019quantum}.


\subsection{Quantum Memory}
Fundamentally, every operation is performed at the level of the qubit. The term "quantum memory" is used in the same way that qubit is used. However, quantum memory in the context of quantum networking has more requirements. First, quantum memory may require a longer lifetime because it has to wait until all operations are done along with the network. In addition, the candidates for quantum memory might have to be able to emit photons to interact with remote quantum memory. 

Several methods are proposed to build quantum memories~\cite{lvovsky2009optical}. The optical delay line is the easiest way to use long fibers to delay and store the photon inside the fiber~\cite{leong2020long}. However, the information could be lossy and the lifetime of quantum information is very short. Another approach called atomic frequency comb memory has a longer life span~\cite{ma2021one}. 

\subsection{Quantum Link}
A quantum link connects quantum repeaters and routers to communicate quantum information between them. Optical fiber is one of the simplest ways to transfer a quantum bit over distant devices. In general, such fiber is specialized to the wavelength used in quantum communications~\cite{niizeki2020two}. 

There are also free-space quantum link communications that utilize satellite~\cite{yin2017satellite, liao2018satellite}, drones~\cite{schirber2021quantum}.

\begin{figure}[ht]
    \centering
    \includegraphics[width=0.5\linewidth]{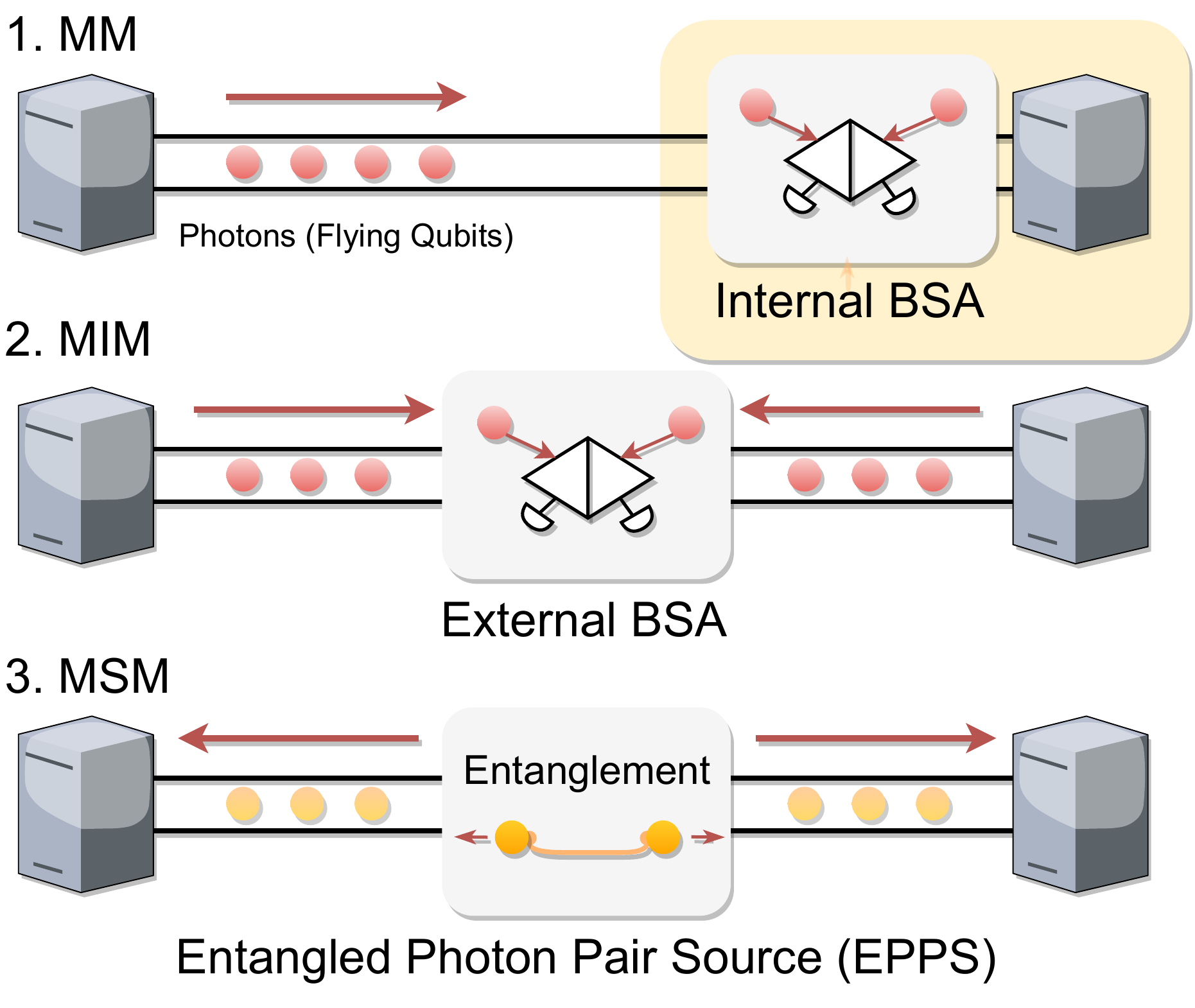}
    \caption[Link Types]{Three different types of quantum link. }
    \label{fig:quantum_link}
\end{figure}

In terms of link architecture, there are three types of links according to their methods to generate the entanglement shown in Figure~\ref{fig:quantum_link}. 

\begin{itemize}
    \item \textbf{MM link} stands for the Memory-Memory link in which two memories placed on two different repeaters are connected to the directory. One of the two quantum repeaters has an internal Bell state analyzer that captures two photons both from the other repeater and the own repeater.
    \item \textbf{MIM link} stands for the Memory-Interferometer-Memory link, in which an external Bell state analyzer is installed in the middle of the link.
    \item \textbf{MSM link} stands for the Memory-Source-Memory link, in which there is an entangled photon source in the middle and sends an entangled pair of photons to both ends. 
\end{itemize}

%% file: Chapters/2.Preliminaries/2.6.Tomography.tex



%% file: Chapters/2.Preliminaries/2.7.EntanglementPurification.tex
\section{Entanglement Purification~(Distillation)}
\label{sec:purification}
\textbf{Entanglement Purification} (also known as \textbf{Entanglement Distillation}) is a process that consumes two or more entangled states to generate higher fidelity quantum entanglement. 
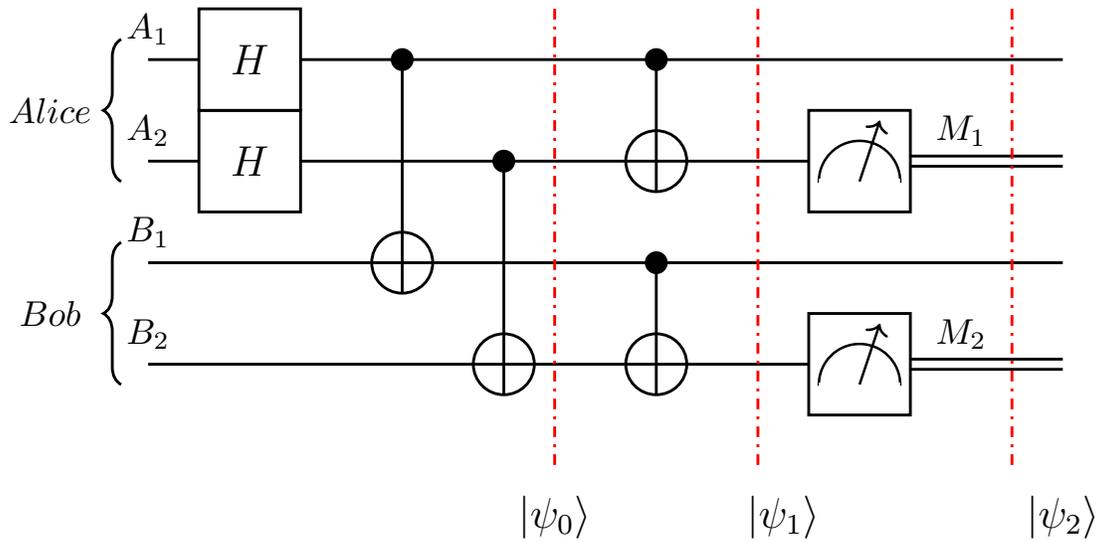
\begin{figure}[ht]
\resizebox{\textwidth}{!}{
\input{Chapters/2.Preliminaries/diagram/2.7.PurificationCircuit}}
    \caption[Purification Circuit]{The circuit representation of a basic quantum purification protocol.}
    \label{fig:purification_circuit}
\end{figure}

There are basically three steps in the simplest purification shown in Figure~\ref{fig:purification_circuit}.
There are two people Alice and Bob and they have both a quantum channel and a classical channel between them. The first step of quantum purification is to prepare two Bell states. In this figure, the target Bell state is the $\ket{\phi^+} = \frac{\ket{00} + \ket{11}}{\sqrt{2}}$ state. 
\begin{align}
    \ket{\psi_0} = \ket{\Phi^+}_{A_1, B_1}\ket{\Phi^+}_{A_2, B_2}
\end{align}
where $A_1, A_2$ are Alice's qubits and $B_1, B_2$ are Bob's qubits.

Second, they apply local CNOT operations on local qubits.
\begin{align}
    \label{eq:purification_no_error}
    \ket{\psi_1} &= CX_{A_1, A_2}CX_{B_1, B_2}\ket{\psi_0}\\
    &= CX_{A_1, A_2}CX_{B_1, B_2}\left(\frac{\ket{00} + \ket{11}}{\sqrt{2}}\right)_{A_1, B_1}
    \left(\frac{\ket{00}+\ket{11}}{\sqrt{2}}\right)_{A_2, B_2}\\
        &= \frac{1}{2}CX_{A_1, A_2}CX_{B_1, B_2}(\ket{0000} + \ket{0011} + \ket{1100} + \ket{1111})_{A_1, B_1, A_2, B_2}\\
    &=\frac{1}{2}CX_{A_1, A_2}CX_{B_1, B_2}(\ket{0000} + \ket{0101} + \ket{1010} + \ket{1111})_{A_1, A_2, B_1, B_2}\\
    &=\frac{1}{2}(\ket{0000} + \ket{0011} + \ket{1111} + \ket{1100})_{A_1, B_1, A_2, B_2}\\
    &=\ket{\Phi^+}_{A_1, B_1}\ket{\Phi^+}_{A_2, B_2}
\end{align}
Finally, Alice and Bob measure their second qubits ($A_2, B_2$) and exchange the measurement results. If the measurement results are the same, the remaining Bell pair can be used as a resource with higher fidelity than that of the input state. Otherwise, the remaining Bell pair is discarded.

In the case where the generated Bell pairs are perfect without any noise, the resulting state is the same as the input state and the parity between these measurement results is the same. 
However, if there is an error in one of the Bell pairs, the state could become $\ket{\Phi^-}$, $\ket{\Psi^-}$, and $\ket{\Psi^-}$.

If we assume that there is a bit-flip error that occurs in one of the Bell pairs, then the resulting state $\ket{\psi_1^{'}}$ is
\begin{align}
    \ket{\psi_1^{'}} &= CX_{A_1, A_2}CX_{B_1, B_2}\ket{\psi_0}\\
    &= CX_{A_1, A_2}CX_{B_1, B_2}
    \left(\frac{\ket{01} + \ket{10}}{\sqrt{2}}\right)_{A_1, B_1}\left(\frac{\ket{00}+\ket{11}}{\sqrt{2}}\right)_{A_2, B_2}\\
    &=\frac{1}{2}CX_{A_1, A_2}CX_{B_1, B_2}(\ket{0100} + \ket{0111} + \ket{1000} + \ket{1011})_{A_1, B_1, A_2, B_2}\\
    &=\frac{1}{2}CX_{A_1, A_2}CX_{B_1, B_2}(\ket{0010} + \ket{0111} + \ket{1000} + \ket{1101})_{A_1, A_2, B_1, B_2}\\
    &=\frac{1}{2}(\ket{0011} + \ket{0110} + \ket{1100} + \ket{1001})_{A_1, A_2, B_1, B_2}\\
    &=\frac{1}{2}(\ket{0101} + \ket{0110} + \ket{1010} + \ket{1001})_{A_1, B_1, A_2, B_2}\\
    &=\ket{\Psi^+}_{A_1, B_1}\ket{\Psi^+}_{A_2, B_2}
\end{align}
In this case, the resulting state is both $\ket{\psi^+}$, which is an undesirable state. When Alice and Bob observe different outputs in such a case ("0" and "1") Alice and Bob can know that the state is not the desired state. If the quantum state is not the desired state, they discard that state and start the process again.

When we assume that $F_{in}$ is the fidelity of the input state, the final output fidelity of the unmeasured Bell pair $F_{out}$ becomes
\begin{align}
    F_{out} = \frac{F_{in}^2}{F_{in}^2 + (1-F_{in}^2)}.
\end{align}
Theoretically, the output fidelity increases to 1.0 by repeating this process infinitely many times if the target noise model is only Pauli X error. If the noise model is more complex, the simple repetition of the purification does not work.

The original entanglement purification protocol was proposed by Bennett et al. in 1996~\cite{bennett1996purification}. Various improved purification schemes with more complex quantum operations have been proposed~\cite{fujii2009entanglement}. 

Initially, the purification protocol was proposed only for bipartite states such as Bell pairs. However, as quantum repeaters and computers that utilize special types of quantum states have been proposed, the demand for quantum purification for a more general state has also been discovered.  \cite{aschauer2005multiparticle, kruszynska2006entanglement} are purification schemes that can be applied to the GHZ state and the graph state, which are extremely important for quantum computing and communication. Furthermore, entanglement purifications for the encoded state have also been studied in~\cite{dur2003entanglement, hostens2006hashing}, their methods can purify the encoded quantum state for fault-tolerant schemes. 

Entanglement purification has been demonstrated on different types of physical devices. Several experiments have been conducted in linear optical devices~\cite{pan2003experimental} and superconducting devices~\cite{yan2022entanglement, zhong2021deterministic}. 

%% file: Chapters/2.Preliminaries/diagram/2.7.PurificationCircuit.tex
\begin{tikzpicture}[semic/.style args={#1,#2}{semicircle,minimum width=#1,draw,anchor=arc end,rotate=#2},outer sep=0pt,line width=.7pt]  
    \draw [thick, decorate,decoration={brace,amplitude=5pt},xshift=1pt,yshift=0pt]
(0.7,1.8) -- (0.7 ,3.2) node [black,midway,xshift=-0.7cm] 
{\footnotesize $Alice$};

    \draw [thick, decorate,decoration={brace,amplitude=5pt},xshift=1pt,yshift=0pt]
(0.7, -0.2) -- (0.7 ,1.2) node [black,midway,xshift=-0.7cm] 
{\footnotesize $Bob$};

\node[draw=none] at (1,3.3) {\footnotesize $A_1$};
\node[draw=none] at (1,2.3) {\footnotesize $A_2$};

\node[draw=none] at (1,1.3) {\footnotesize $B_1$};
\node[draw=none] at (1,0.3) {\footnotesize $B_2$};
    
    \draw[thick] (1,0)--node[midway,above]{}(8,0);
    \draw[thick] (1,1)--node[midway,above]{}(10,1);
    \draw[thick] (1,2)--node[midway,above]{}(8,2);
    \draw[thick] (1,3)--node[midway,above]{}(10,3);
    
    \draw[thick, fill=white] (1.5 ,2.5) rectangle (2.5 ,1.5) node [pos=.5]{$H$};
    
    \draw[thick, fill=white] (1.5, 3.5) rectangle (2.5, 2.5) node [pos=.5]{$H$};

    \draw[fill=black] (4.5, 2) circle(0.1cm);
    \draw[thick] (4.5,2)--node[midway,above]{}(4.5,-0.3);
    \draw[thick, fill=none] (4.5, 0) circle(0.3cm);
    
    \draw[fill=black] (3.5, 3) circle(0.1cm);
    \draw[thick] (3.5,3)--node[midway,above]{}(3.5,0.7);
    \draw[thick, fill=none] (3.5, 1) circle(0.3cm);

    \draw[fill=black] (6, 1) circle(0.1cm);
    \draw[thick] (6,1)--node[midway,above]{}(6,-0.3);
    \draw[thick, fill=none] (6, 0) circle(0.3cm);
    
    \draw[fill=black] (6, 3) circle(0.1cm);
    \draw[thick] (6,3)--node[midway,above]{}(6,1.7);
    \draw[thick, fill=none] (6, 2) circle(0.3cm);
    
    \draw[thick] (8,-0.05)--node[midway,above]{}(10,-0.05);
    \draw[thick] (8,0.05)--node[midway,above]{}(10,0.05);
    \draw[thick] (8,1.95)--node[midway,above]{}(10,1.95);
    \draw[thick] (8,2.05)--node[midway,above]{}(10,2.05);

    \draw[thick, fill=white] (7.5,1.5) rectangle (8.5,2.5) node [pos=.5]{};
    \node [semic={0.8cm,0}]    at (7.6,1.8){};
    \draw[draw=none, color=white, fill=white] (7.6,1.7) rectangle (8.4,1.82) node [pos=.5]{};
    \draw[arrows=->](8, 1.8)--(8.2,2.4);
    
    \draw[thick, fill=white] (7.5,-0.5) rectangle (8.5,0.5) node [pos=.5]{};
    \node [semic={0.8cm,0}]    at (7.6, -0.2){};
    \draw[draw=none, color=white, fill=white] (7.6, -0.3) rectangle (8.4,-0.18) node [pos=.5]{};
    \draw[arrows=->](8, -0.2)--(8.2,0.4);
    
    
    \node[draw=none] at (9,2.3) {\footnotesize $M_1$};
    \node[draw=none] at (9,0.3) {\footnotesize $M_2$};

    \draw[thick,dash dot, color=red] (5.0,3.5)--node[midway,above]{}(5.0,-1);
    \draw[thick,dash dot, color=red] (7,3.5)--node[midway,above]{}(7,-1);
    \draw[thick,dash dot, color=red] (9.5,3.5)--node[midway,above]{}(9.5,-1);
    
    \node[draw=none] at (5.0,-1.5) {$\ket{\psi_0}$};
    \node[draw=none] at (7.25,-1.5) {$\ket{\psi_1}$};
    \node[draw=none] at (10,-1.5) {$\ket{\psi_2}$};
\end{tikzpicture}

%% file: Chapters/2.Preliminaries/2.8.EntanglementSwapping.tex
\section{Entanglement Swapping}
\label{sec:entanglement_swapping}
Entanglement swapping is one of the most primitive operations for quantum repeaters to concatenate shorter entanglements into longer-range entanglements. Entanglement swapping is based on a method called quantum teleportation. 
\subsection{Teleportation}
\begin{figure}
    \centering
    \includegraphics{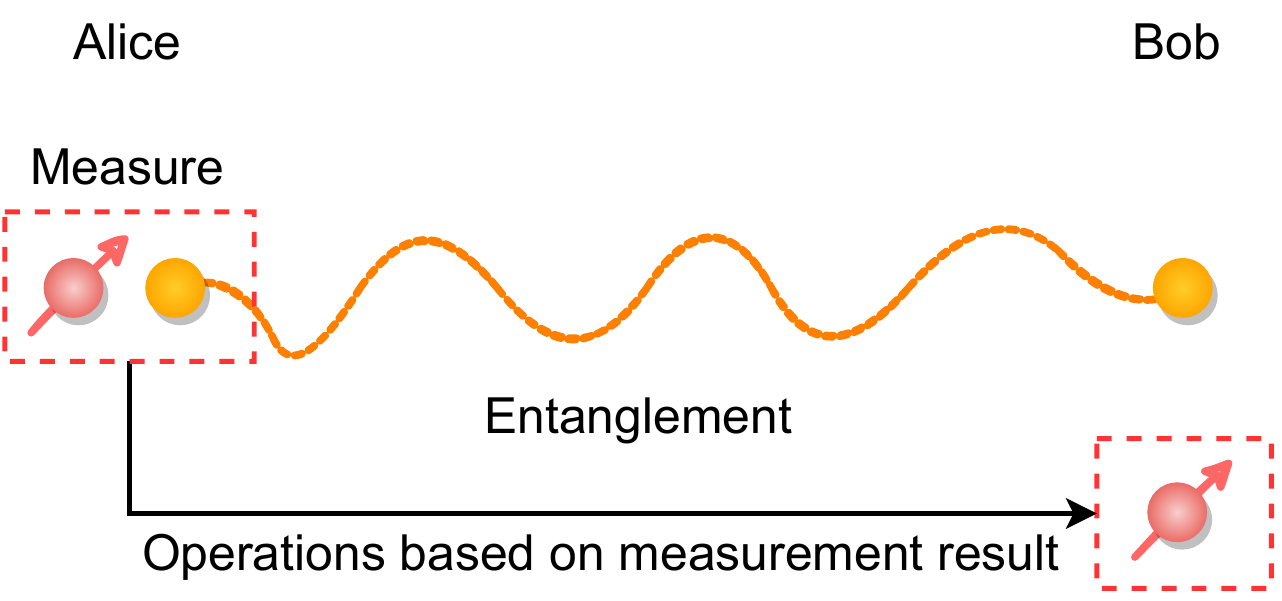}
    \caption[Quantum Teleporation]{Quantum teleportation protocol: Two parties Alice and Bob share the entanglement (e.g. Bell pair) to share a single qubit state. Once Alice gets the state ready to send, Alice measures her qubits and entanglement in the Bell basis and sends their measurement results to Bob. Bob applies the corresponding quantum gates to his quantum state and recovers the original state Alice wished to send.}
    \label{fig:quantum_teleporatation}
\end{figure}

Quantum Teleportation~\cite{bennett1993teleporting} is a method that transfers a quantum state from one place to another by exploiting quantum entanglement. Figure~\ref{fig:quantum_teleporatation} represents the schematic diagram of quantum teleportation. 

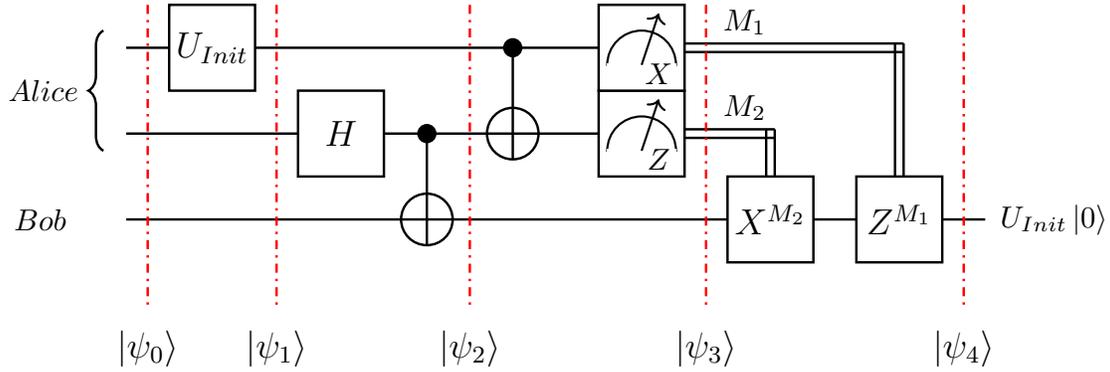
\begin{figure}[ht]
\resizebox{\textwidth}{!}{
\input{Chapters/2.Preliminaries/diagram/2.7.1TeleporationCircuit}}
    \caption[Quantum Teleportation Circuit]{Procedure of quantum teleportation in the circuit diagram. In this diagram, the entanglement between Alice and Bob is established in the process from $\ket{\psi_1}$ to $\ket{\psi_2}$. However, it is also possible to think that using preshared entanglement stored in quantum memories. }
    \label{fig:teleportation_circuit}
\end{figure}

Figure~\ref{fig:teleportation_circuit} shows the quantum teleportation procedure. The top two qubits are Alice's qubits, and the bottom qubit is Bob's qubit. In the first step, we start from all zero states. 

\begin{align}
    \ket{\psi_0} = \ket{00}_{A}\ket{0}_{B}
\end{align}

Alice encodes the information she wants to send on the first qubit with a unitary operation $U_{Init}$.
\begin{align}
    \ket{\psi_1} = U_{Init}\ket{00}_A \ket{0}_B = U_{Init}\ket{\psi_0}
\end{align}
where $U_{Init}$ is only applied to Alice's first qubit.

In the next step, Alice generates a Bell state with Bob. We assume that they share the $\Phi^+$ state in this example.
\begin{align}
    \ket{\psi_2} = U_{Init}\ket{0}\ket{\Phi^+}_{A_1, B_0}
\end{align}
Alice measures her qubits in the Bell basis in the next step. Let $X$ measurement be the measurement operation that is made up of the $H$ gate and the $Z$ measurement. The quantum state just before the measurement can be written as
\begin{align}
    \ket{\psi} = \frac{1}{2}
    [&\ket{00}(\alpha\ket{0} + \beta\ket{1}) + \\
    &\ket{01}(\alpha\ket{1} + \beta\ket{0}) +\\
    &\ket{10}(\alpha\ket{0} - \beta\ket{1}) + \\
    &\ket{11}(\alpha\ket{1} - \beta\ket{0})]
\end{align}
where $U_I\ket{0} = \alpha\ket{0} + \beta\ket{1}$. From this result, based on Alice's measurement results, Bob can get the corresponding results. 
When Alice measures $\ket{00}$, the resulting state in Bob's qubit is also the desired state. However, in the case where $\ket{01}, \ket{10}, \ket{11}$, Bob needs to correct his quantum state by applying the corresponding operations $X$, $Z$, and $XZ$, respectively. At the end of the protocol, the initial Bob's qubit has turned into the state that Alice wanted to send. 

Quantum teleportation has been experimentally demonstrated in~\cite{bouwmeester1997experimental, furusawa1998unconditional, takesue2015quantum, ren2017ground}. 

\subsection{Teleporting Entanglement}
\begin{figure}[ht]
    \centering
    \includegraphics[width=.8\linewidth]{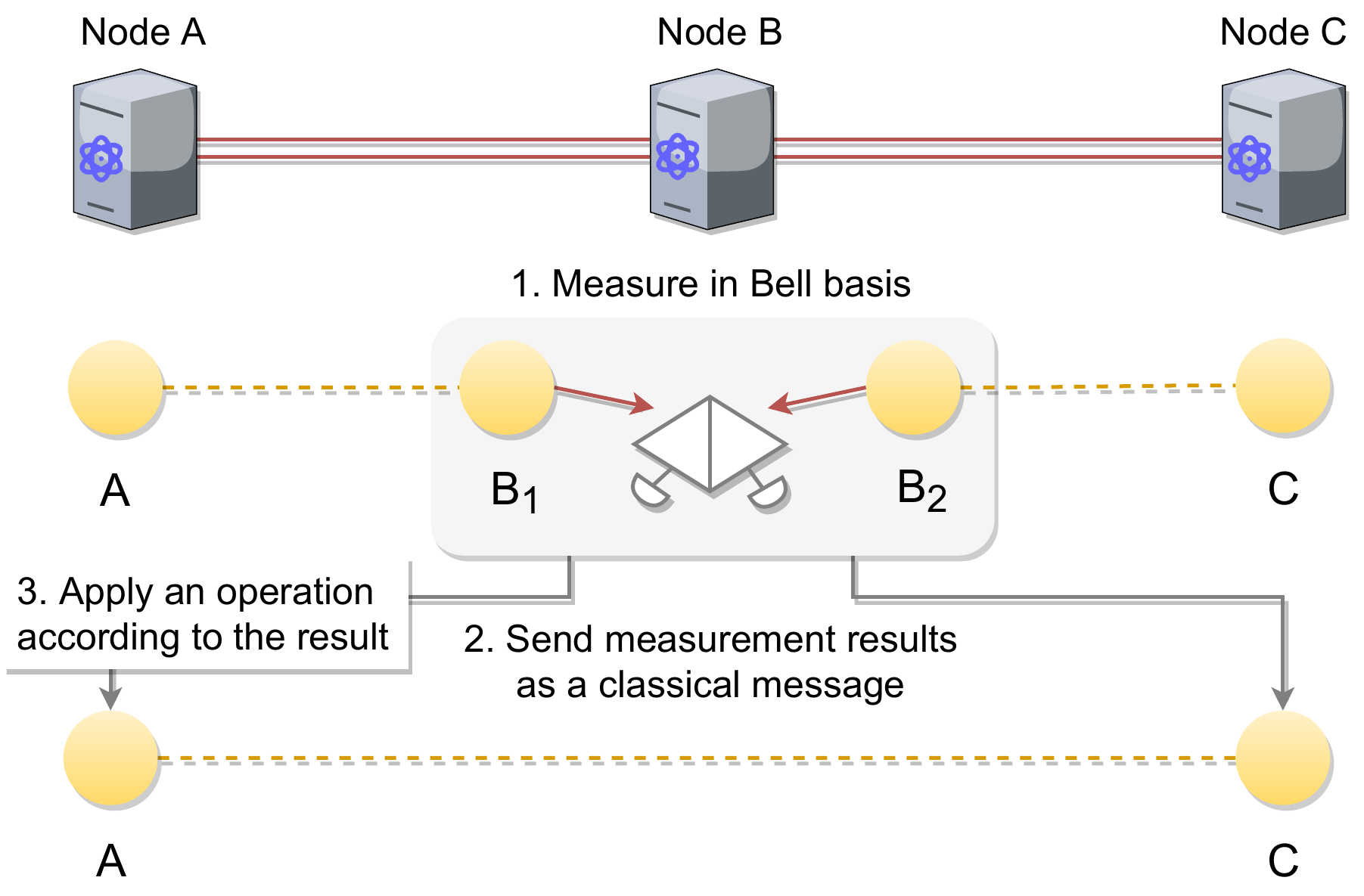}
    \caption[Entanglement Swapping Procedure]{Procedure of entanglement swapping: 1. Measure qubits $B_1$ and $B_2$ which are entangled with $A$ and $C$ in Bell basis. 2. Send the measurement results to the corresponding partner. 3. Apply Pauli correction operations to the proper qubits.}
    \label{fig:entanglement_swapping}
\end{figure}

Quantum teleportation can also teleport entangled states over the entanglement. This technique is called Entanglement Swapping, which is a very important and primitive operation for quantum networking. Figure~\ref{fig:entanglement_swapping} shows the entanglement swapping procedure. 

Suppose that node A and node B, node B and node C share Bell states, respectively. Node B measures two qubits $B_1$ and $B_2$ in the Bell basis and sends the measurement results to the corresponding partner. In the figure, node B sends the results to both node A and node C. However, it is also possible to send both results to a single partner. Once the partner nodes receive the measurement results, they apply the quantum operation corresponding to the result they receive, as we do in quantum teleportation. Finally, the range of entangled states goes from one single hop to the end (from node A to node C). 

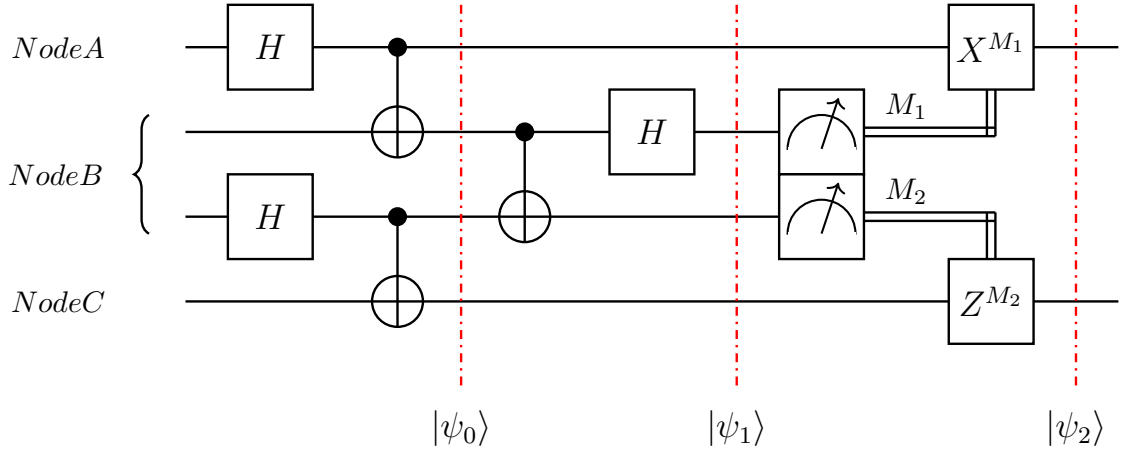
\begin{figure}[ht]
\resizebox{\textwidth}{!}{
\input{Chapters/2.Preliminaries/diagram/2.8.EntanglementSwappingCircuit}}
    \caption[Entanglement Swapping Circuit]{The circuit representation of the entanglement swapping protocol}
    \label{fig:swapping_circuit}
\end{figure}

Figure~\ref{fig:swapping_circuit} shows the circuit representation for the entanglement swapping protocol for the $\ket{\Phi^+}$ state. The first step starts with generating Bell pairs between Node A and Node B and between Node B and Node C.
\begin{align}
\ket{\psi_0} = \left(\frac{\ket{00} + \ket{11}}{\sqrt{2}}\right)_{A, B_1}
\left(\frac{\ket{00}+\ket{11}}{\sqrt{2}}\right)_{B_2, C}
\end{align}

Once both Bell pairs are ready, Node B applies the Bell measurement to qubits. Just before measurement, the quantum state is described as follows.
\begin{align*}
    \ket{\psi_1} &= H_{B_1}CX_{B_1, B_2}\frac{1}{2}(\ket{0000} + \ket{0011} + \ket{1100} + \ket{1111})_{A, B_1, B_2, C}\\
    &= H_{B_1}\frac{1}{2}(\ket{0000} + \ket{0011} + \ket{1110} + \ket{1101})_{A, B_1, B_2, C}\\
    &=\frac{1}{\sqrt{2}}(\ket{0000} + \ket{0100} + \ket{0011} + \ket{0111}\\ 
    &+ \ket{1010} -\ket{1110} + \ket{1001} - \ket{1101})_{A, B_1, B_2, C}\\
\end{align*}
When we change the order of the qubit, $\ket{\psi_1}$ can be written as
\begin{align}
    \ket{\psi_1} = \frac{1}{2\sqrt{2}} &\ket{00}_{B_1, B_2}(\ket{00} + \ket{11})_{A, C}\\ + &\ket{01}_{B_1, B_2}(\ket{01} + \ket{10})_{A, C}\\
    + &\ket{10}_{B_1, B_2}(\ket{00} - \ket{11})_{A, C}\\
    + &\ket{11}_{B_1, B_2}(\ket{01} - \ket{10})_{A, C}.\\
\end{align}
If node B measures $\ket{00}$, the remaining nodes (node A and node C) share $\ket{\Phi^+}$. Otherwise, node A and node C cannot share the target Bell pairs $\ket{\Phi^+}$. However, if node B tells its measurement results to node A and node C, it is possible to revive the original $\ket{\Phi^+}$ state. 

\begin{align}
    \ket{\psi_2} &= X_A\frac{\ket{01} + \ket{10}}{\sqrt{2}}_{A, C} = \ket{\Phi^+}_{A, C}\\
    \ket{\psi_2} &= Z_C\frac{\ket{00} - \ket{11}}{\sqrt{2}}_{A, C} = \ket{\Phi^+}_{A, C}\\
    \ket{\psi_2} &= X_A Z_C \frac{\ket{01} - \ket{10}}{\sqrt{2}}_{A, C}= \ket{\Phi^+}_{A, C}
\end{align}

In any case, we can see that the final state is $\ket{\Phi^+}$, which we intended to create. This process can be extended to any number of hops naturally, so creating a relatively shorter entangled state and concatenating them with entanglement swapping is a reasonable approach to creating a long-range entanglement.

This entanglement swapping protocol has been experimentally demonstrated in several physical systems~\cite{pan1998experimental, ma2012experimental, jin2015highly}.

%% file: Chapters/2.Preliminaries/diagram/2.7.1TeleporationCircuit.tex
\begin{tikzpicture}[semic/.style args={#1,#2}{semicircle,minimum width=#1,draw,anchor=arc end,rotate=#2},outer sep=0pt,line width=.7pt]  
    \draw [thick, decorate,decoration={brace,amplitude=5pt},xshift=1pt,yshift=0pt]
(0.7,0.8) -- (0.7 ,2.2) node [black,midway,xshift=-0.7cm] 
{\footnotesize $Alice$};
    \node[draw=none] at (0,0) {\footnotesize $Bob$};
    
    \draw[thick] (1,0)--node[midway,above]{}(11,0);
    \draw[thick] (1,1)--node[midway,above]{}(7,1);
    \draw[thick] (1,2)--node[midway,above]{}(7,2);
    
    \draw[thick, fill=white] (1.5,2.5) rectangle (2.5,1.5) node [pos=.5]{$U_{Init}$};
    
    \draw[thick, fill=white] (3,1.5) rectangle (4,0.5) node [pos=.5]{$H$};
    \draw[fill=black] (4.5, 1) circle(0.1cm);
    \draw[thick] (4.5,1)--node[midway,above]{}(4.5,-0.3);
    \draw[thick, fill=none] (4.5, 0) circle(0.3cm);
    
    \draw[fill=black] (5.5, 2) circle(0.1cm);
    \draw[thick] (5.5, 2)--node[midway,above]{}(5.5,0.7);
    \draw[thick, fill=none] (5.5, 1) circle(0.3cm);
    
    \draw[thick] (7,0.95)--node[midway,above]{}(8.55,0.95);
    \draw[thick] (7,1.05)--node[midway,above]{}(8.55,1.05);
    \draw[thick] (7,1.95)--node[midway,above]{}(10.05,1.95);
    \draw[thick] (7,2.05)--node[midway,above]{}(10.05,2.05);
    
    \draw[thick] (8.45,1.05)--node[midway,above]{}(8.45,0);
    \draw[thick] (8.55,1.05)--node[midway,above]{}(8.55,0);
    \draw[thick] (9.95,2.05)--node[midway,above]{}(9.95,0);
    \draw[thick] (10.05,2.05)--node[midway,above]{}(10.05,0);

    \draw[thick, fill=white] (6.5,0.5) rectangle (7.5,1.5) node [pos=.5]{};
    \node [semic={0.8cm,0}]    at (6.6,0.8){};
    
    \draw[draw=none, color=white, fill=white] (6.6,0.7) rectangle (7.4,0.82) node [pos=.5]{};
    \draw[arrows=->](7, 0.8)--(7.2,1.4);
    \draw[thick, fill=white] (6.5,1.5) rectangle (7.5,2.5) node [pos=.5]{};
    \node [semic={0.8cm,0}]    at (6.6,1.8){};
    \draw[draw=none, color=white, fill=white] (6.6,1.7) rectangle (7.4,1.82) node [pos=.5]{};
    \draw[arrows=->](7, 1.8)--(7.2,2.4);
    
    \draw[thick, fill=white] (8,-0.5) rectangle (9,0.5) node [pos=.5]{$X^{M_2}$};
    \draw[thick, fill=white] (9.5,-0.5) rectangle (10.5,0.5) node [pos=.5]{$Z^{M_1}$};
    
    \node[draw=none] at (7.2,1.7) {\footnotesize $X$};
    \node[draw=none] at (8.2,2.3) {\footnotesize $M_1$};
    \node[draw=none] at (7.2,0.7) {\footnotesize $Z$};
    \node[draw=none] at (8.2,1.3) {\footnotesize $M_2$};
    \node[draw=none] at (11.8,0) {\footnotesize $U_{Init}\ket{0}$};

    \draw[thick,dash dot, color=red] (1.25,2.5)--node[midway,above]{}(1.25,-1);
    \draw[thick,dash dot, color=red] (2.75,2.5)--node[midway,above]{}(2.75,-1);
    \draw[thick,dash dot, color=red] (5.0,2.5)--node[midway,above]{}(5.0,-1);
    \draw[thick,dash dot, color=red] (7.75,2.5)--node[midway,above]{}(7.75,-1);
    \draw[thick,dash dot, color=red] (10.75,2.5)--node[midway,above]{}(10.75,-1);

    \node[draw=none] at (1.25,-1.5) {$\ket{\psi_0}$};
    \node[draw=none] at (2.75,-1.5) {$\ket{\psi_1}$};
    \node[draw=none] at (5.0,-1.5) {$\ket{\psi_2}$};
    \node[draw=none] at (7.75,-1.5) {$\ket{\psi_3}$};
    \node[draw=none] at (10.75,-1.5) {$\ket{\psi_4}$};
\end{tikzpicture}

%% file: Chapters/2.Preliminaries/diagram/2.8.EntanglementSwappingCircuit.tex
\begin{tikzpicture}[semic/.style args={#1,#2}{semicircle,minimum width=#1,draw,anchor=arc end,rotate=#2},outer sep=0pt,line width=.7pt]  

    \node[draw=none] at (-.5,3) {\footnotesize $Node A$};
    \draw [thick, decorate,decoration={brace,amplitude=5pt},xshift=2pt,yshift=0pt]
        (.5, 0.8) -- (.5 ,2.2) node [black,midway,xshift=-1.1cm] 
{\footnotesize $Node B$};
    \node[draw=none] at (-.5,0) {\footnotesize $Node C$};
    \draw[thick] (1,0)--node[midway,above]{}(12,0);
    \draw[thick] (1,1)--node[midway,above]{}(8,1);
    \draw[thick] (1,2)--node[midway,above]{}(8,2);
    \draw[thick] (1,3)--node[midway,above]{}(12,3);
    
    \draw[thick, fill=white] (1.5, 3.5) rectangle (2.5, 2.5) node [pos=.5]{$H$};
    
    \draw[thick, fill=white] (1.5 ,1.5) rectangle (2.5 ,0.5) node [pos=.5]{$H$};
    
    \draw[fill=black] (3.5, 3) circle(0.1cm);
    \draw[thick] (3.5,3)--node[midway,above]{}(3.5,1.7);
    \draw[thick, fill=none] (3.5, 2) circle(0.3cm);
    
    \draw[fill=black] (3.5, 1) circle(0.1cm);
    \draw[thick] (3.5,1)--node[midway,above]{}(3.5,-0.3);
    \draw[thick, fill=none] (3.5, 0) circle(0.3cm);

    \draw[fill=black] (5, 2) circle(0.1cm);
    \draw[thick] (5,2)--node[midway,above]{}(5,0.7);
    \draw[thick, fill=none] (5, 1) circle(0.3cm);
    
    \draw[thick, fill=white] (6 ,2.5) rectangle (7 ,1.5) node [pos=.5]{$H$};
    
    \draw[thick] (8.5,0.95)--node[midway,above]{}(10.55,0.95);
    \draw[thick] (8.5,1.05)--node[midway,above]{}(10.55,1.05);
    \draw[thick] (8.5,1.95)--node[midway,above]{}(10.55,1.95);
    \draw[thick] (8.5,2.05)--node[midway,above]{}(10.55,2.05);

    \draw[thick] (10.45,1.95)--node[midway,above]{}(10.45,3.05);
    \draw[thick] (10.55,1.95)--node[midway,above]{}(10.55,3.05);
    \draw[thick] (10.45,1.05)--node[midway,above]{}(10.45,0);
    \draw[thick] (10.55,1.05)--node[midway,above]{}(10.55,0);
    
    \draw[thick, fill=white] (8,1.5) rectangle (9,2.5) node [pos=.5]{};
    \node [semic={0.8cm,0}]    at (8.1,1.8){};
    \draw[draw=none, color=white, fill=white] (8.1,1.7) rectangle (8.9,1.82) node [pos=.5]{};
    \draw[arrows=->](8.5, 1.8)--(8.7,2.4);
    
    \draw[thick, fill=white] (8,0.5) rectangle (9,1.5) node [pos=.5]{};
    \node [semic={0.8cm,0}]    at (8.1, 0.8){};
    \draw[draw=none, color=white, fill=white] (8.1, 0.7) rectangle (8.9,0.82) node [pos=.5]{};
    \draw[arrows=->](8.5, 0.8)--(8.7,1.4);

    \draw[thick, fill=white] (10,2.5) rectangle (11, 3.5) node [pos=.5]{$X^{M_1}$};
    \draw[thick, fill=white] (10,-0.5) rectangle (11,0.5) node [pos=.5]{$Z^{M_2}$};
    
    \node[draw=none] at (9.5,2.3) {\footnotesize $M_1$};
    \node[draw=none] at (9.5,1.3) {\footnotesize $M_2$};

    \draw[thick,dash dot, color=red] (4.25,3.5)--node[midway,above]{}(4.25,-1);
    \draw[thick,dash dot, color=red] (7.5,3.5)--node[midway,above]{}(7.5,-1);
    \draw[thick,dash dot, color=red] (11.5,3.5)--node[midway,above]{}(11.5,-1);
    
    \node[draw=none] at (4.25,-1.5) {$\ket{\psi_0}$};
    \node[draw=none] at (7.5,-1.5) {$\ket{\psi_1}$};
    \node[draw=none] at (11.5,-1.5) {$\ket{\psi_2}$};
\end{tikzpicture}

%% file: Chapters/2.Preliminaries/2.9.QuantumErrorCorrection.tex
\section{Quantum Error Correction}
\label{sec:error_correction}
Quantum Error Correction (QEC) is a method that can correct errors in a quantum system by consuming ancillary qubits. 
The capability of the QEC code can be represented using three parameters $[n, k, d]$ where $n$ is the number of bits to use to encode the bit information $k$ and $d$ is the distance from the code. Let $C$ be an error correction code and $x, y$ be the codewords of $C$. The distance of the code is defined as 
\begin{align}
    d(C) = \underset{x, y \in C, x\neq y}{\text{min}} d(x, y)
\end{align}
where $d(x, y)$ is the Hamming distance of two codewords. Furthermore, the Hamming distance $d(x, y)$ can be written as $d(x, y) = \text{wt}(x + y)$ where $\text{wt}(x + y)$ is the Hamming weight of $x+y$ and $+$ is bitwise modulo addition or XOR. Then the code distance $d(C)$ can be described as
\begin{align}
    d(C) = \underset{x\in C, x\neq 0}{\text{min}} \text{wt}(x).
\end{align}
Distance $d$ is related to the number of possible errors that the code can correct. The code $C$ can correct $t$ errors when its code distance is $2t + 1$.

In general, QEC is performed in three steps.
\begin{enumerate}
    \item Encoding: A quantum state is usually encoded to a higher-dimensional Hilbert space by using ancillary qubits. These ancillary qubits are used to detect errors in the quantum state.
    \item Syndrome measurement: By measuring the ancillary qubit, we can know what kind of error is happening in which qubit. 
    \item Decoding with state recovery: Finally, we use the error information taken in the syndrome measurement to correct the error and recover the noise-free quantum state.
\end{enumerate}

\subsection{Repetition Code}
As a very primitive example of quantum error correction code, this section introduces the repetition code. 

If we assume only one of two types of errors (bit-flip error, phase-flip error), we can correct the error with three qubits. 
Let $\ket{0_{L}} (\ket{1_{L}})$ be a logical zero state (one state) encoded as
\begin{align}
    \ket{0_{L}} = \ket{000}, \ket{1_{L}} = \ket{111}.
\end{align}

\begin{figure}[ht]
  \begin{minipage}[b]{0.5\linewidth}
    \centering
    \resizebox{\linewidth}{!}{
\input{Chapters/2.Preliminaries/diagram/2.9.1.RepetitionBit}}
  \end{minipage}
  \begin{minipage}[b]{0.5\linewidth}
    \centering
    \resizebox{\linewidth}{!}{
    \input{Chapters/2.Preliminaries/diagram/2.9.1.RepetitionPhase}}
  \end{minipage}
  \label{fig:repetition_code}
  \caption[Encoding Circuits]{A quantum circuit encodes a single quantum state with repetition code. The left circuit shows the repetition code for the bit flip error and the right circuit represents the repetition code for the phase flip error.}
\end{figure}
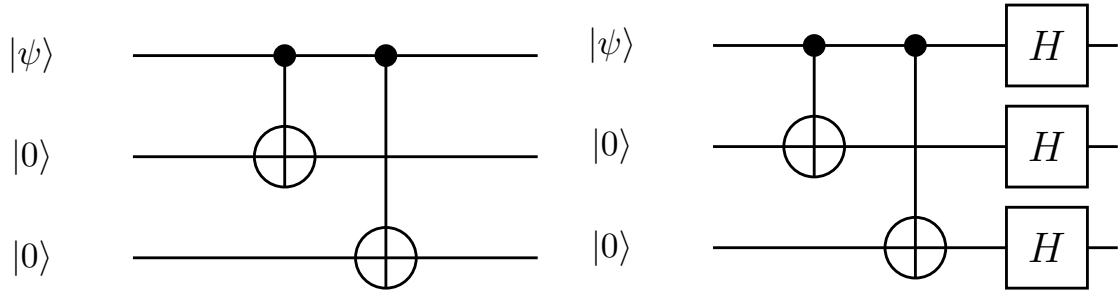

\textbf{1. Encoding}:
Quantum circuits that encode a quantum state $\ket{\psi}$ are shown in Figure~\ref{fig:repetition_code}.

When we assume only one bit-flip error we use the left quantum circuit. If we assume a phase flip error, we encode the logical plus state $\ket{+_L} = \ket{+++}$ and the minus state $\ket{-_L} = \ket{---}$ state.

If the state encoding is done perfectly without error, the quantum state is $\ket{\psi} = \alpha\ket{000} + \beta\ket{111}$.
After encoding, the state passes through a noisy channel that adds errors to the quantum state. In this example, we assume that the quantum state goes through a bit-flip channel.
 
\textbf{2. Syndrome measurement}:
If we limit the number of bit-flip occurrences in this quantum state to one, then we can think of the following four syndromes.
\begin{enumerate}
    \item No error
    \item Bit-flip on the first qubit
    \item Bit-flip on the second qubit
    \item Bit-flip on the third qubit
\end{enumerate}
To detect these four events, we can come up with four types of projection operators.
\begin{align}
    P_1 &= \ket{000}\bra{000} + \ket{111}\bra{111}\\
    P_2 &= \ket{100}\bra{100} + \ket{011}\bra{011}\\
    P_3 &= \ket{010}\bra{010} + \ket{101}\bra{101}\\
    P_4 &= \ket{001}\bra{001} + \ket{110}\bra{110}
\end{align}

Suppose that there is a bit-flip error in the first qubit $\ket{\psi'} = \alpha\ket{100} + \beta\ket{011}$. When we think of the following observable,
\begin{align}
    \bra{\psi'}P_2\ket{\psi'} &= (\alpha^*\bra{100} + \beta^*\bra{011})(\ket{100}\bra{100} + \ket{011}\bra{011})(\alpha\ket{100} + \beta\ket{011})\\
    &= |\alpha|^2 + |\beta|^2 = 1
\end{align}
If we apply a different projection matrix to it, the observable is 0 since they are orthogonal.

\textbf{3. Recovery}:
We use syndrome measurement to tell which qubit has a bit-flip error. If the state is in the case where 0. (No error), no operation is performed. Otherwise, we have to apply another bit-flip operation (X gate) to the corresponding error bit to correct the state. 

However, in this example, we assumed that the bit-flip error occurs on one or fewer of the three qubits in the system. 
Let $p$ be the probability that the bit-flip error occurs in a qubit. The probability that we can correct the error is
\begin{align}
    (1-p)^3 + 3\cdot p(1-p)^2 = 2p^3 -3p^2 + 1.
\end{align}

In any process of error correction, we do not have to know the original state information $\alpha$ and $\beta$.


\subsection{Stabilizer Code}
\label{sec:stabilizer_code}
When we think of a larger quantum state, it is easier to handle a set of operators called stabilizers instead of using state vector representation in general. 
Let $P$ be an arbitrary operator. When the operator satisfies the following relationship, $P$ is called a "stabilizer" for the state $\ket{\psi}$.
\begin{align}
    P\ket{\psi} = \ket{\psi}
\end{align}
This also means that $\ket{\psi}$ is an eigenvector of $P$ with eigenvalue $+1$.
The stabilizer restricts the degree of freedom for the quantum state such as Bell pairs that can be fully characterized by the set of stabilizer operators. However, when there are degrees of freedom, the state cannot be stabilized by a set of stabilizers. 

Suppose that $\ket{\psi}$ is a Bell state $\ket{\psi} = \ket{00} + \ket{11}/\sqrt{2}$, the following equation indicates that $X_0X_1$ and $Z_0Z_1$ are both stabilizers of $\ket{\psi}$.
In other words, $\ket{\psi}$ is stabilized by $X_0X_1$ and $Z_0Z_1$.
\begin{align}
    X_0X_1&\ket{\psi} = \ket{\psi},\\
    Z_0Z_1&\ket{\psi} = \ket{\psi}
\end{align}

As an example, we consider fixing a single bit-flip error with the three-qubit stabilizer code as we did in previous sections.
Suppose that there is a stabilizer generator $g$ that stabilizes $\ket{000}$ and $\ket{111}$.
\begin{align}
    g = \{ Z_1Z_2I_3, I_1Z_2Z_3\}
\end{align}
We can check that these two stabilizer generators stabilize $\ket{000}$ and $\ket{111}$.
\begin{align}
    Z_1Z_2I_3&\ket{000} = \ket{000}, Z_1Z_2I_3\ket{111} = \ket{111}\\
    I_1Z_2Z_3&\ket{000} = \ket{000}, I_1Z_2Z_3\ket{111} = \ket{111}
\end{align}
If we assume that there is a bit flip error in one of the qubits in $\ket{\psi} = \alpha\ket{000} + \beta\ket{111}$, measurement results with these stabilizers are shown in Table~\ref{tab:stabilizer_code_syndrome_meas}.
These measured syndromes are corresponding to the eigenvalues of corresponding eigenvectors (e.g. $Z_1 Z_2 I_3 \ket{100} = (-1) \ket{100}$). 
\begin{table}[]
\centering
\renewcommand{\arraystretch}{1.2}
\caption[Syndrome Measurement]{Correspondence between stabilizers and measured syndromes}
\label{tab:stabilizer_code_syndrome_meas}
\begin{tabular}{|l|c|c|c|c|}
\hline
            & \multicolumn{1}{l|}{\begin{tabular}[c]{@{}l@{}}No bitflip\\ $\ket{000} +  \ket{111}$\end{tabular}} & \multicolumn{1}{l|}{\begin{tabular}[c]{@{}l@{}}Bit flip on the \\ first qubit\\ $\ket{100} + \ket{011}$\end{tabular}} & \multicolumn{1}{l|}{\begin{tabular}[c]{@{}l@{}}Bit flip on the \\ second qubit\\ $\ket{010} +  \ket{101}$\end{tabular}} & \multicolumn{1}{l|}{\begin{tabular}[c]{@{}l@{}}Bit flip on the \\ third qubit\\ $\ket{001} +  \ket{110}$\end{tabular}} \\ \hline\hline
$Z_1Z_2I_3$ & (+1, +1)                                                                                             & (-1, -1)                                                                                                                 & (-1, -1)                                                                                                                  & (+1, +1)                                                                                                                 \\ \hline
$I_1Z_2Z_3$ & (+1, +1)                                                                                             & (+1, +1)                                                                                                                 & (-1, -1)                                                                                                                  & (-1, -1)                                                                                                                 \\ \hline
\end{tabular}
\end{table}
According to the table, if all results are +1, there is no error in the state. Otherwise, there is a bit-flip error in the corresponding position. 

We can think of various stabilizer codes with different stabilizer generators. When we think of the following stabilizer generators $\{g_1, g_2, g_3, g_4, g_5, g_6\}$, we can construct the Steane seven-qubit code~\cite{steane1996multiple}. 
\begin{align}
    g_1 &\equiv IIIXXXX\\
    g_2 &\equiv IXXIIXX\\
    g_3 &\equiv XIXIXIX\\
    g_4 &\equiv IIIZZZZ\\
    g_5 &\equiv IZZIIZZ\\
    g_6 &\equiv ZIZIZIZ
\end{align}

CSS codes, named after three different researchers (Robert Calderbank, Peter Shor, and Andrew Steane), are also a subclass of the stabilizer code. CSS codes utilize classical error correction to correct quantum errors. 

%% file: Chapters/2.Preliminaries/diagram/2.9.1.RepetitionBit.tex
\begin{tikzpicture}[semic/.style args={#1,#2}{semicircle,minimum width=#1,draw,anchor=arc end,rotate=#2},outer sep=0pt,line width=.7pt]  
    \node[draw=none] at (0,2) {\footnotesize $\ket{\psi}$};
    \node[draw=none] at (0,1) {\footnotesize $\ket{0}$};
    \node[draw=none] at (0,0) {\footnotesize $\ket{0}$};
    
    \draw[thick] (1,0)--node[midway,above]{}(5,0);
    \draw[thick] (1,1)--node[midway,above]{}(5,1);
    \draw[thick] (1,2)--node[midway,above]{}(5,2);
    
    
    \draw[fill=black] (3.5, 2) circle(0.1cm);
    \draw[thick] (3.5,2)--node[midway,above]{}(3.5,-0.3);
    \draw[thick, fill=none] (3.5, 0) circle(0.3cm);
    
    \draw[fill=black] (2.5, 2) circle(0.1cm);
    \draw[thick] (2.5, 2)--node[midway,above]{}(2.5, 0.7);
    \draw[thick, fill=none] (2.5, 1) circle(0.3cm);
\end{tikzpicture}

%% file: Chapters/2.Preliminaries/diagram/2.9.1.RepetitionPhase.tex
\begin{tikzpicture}[semic/.style args={#1,#2}{semicircle,minimum width=#1,draw,anchor=arc end,rotate=#2},outer sep=0pt,line width=.7pt]  
    \node[draw=none] at (0,2) {\footnotesize $\ket{\psi}$};
    \node[draw=none] at (0,1) {\footnotesize $\ket{0}$};
    \node[draw=none] at (0,0) {\footnotesize $\ket{0}$};
    
    \draw[thick] (1,0)--node[midway,above]{}(5,0);
    \draw[thick] (1,1)--node[midway,above]{}(5,1);
    \draw[thick] (1,2)--node[midway,above]{}(5,2);
    
    
    \draw[fill=black] (3, 2) circle(0.1cm);
    \draw[thick] (3,2)--node[midway,above]{}(3,-0.3);
    \draw[thick, fill=none] (3, 0) circle(0.3cm);
    
    \draw[fill=black] (2, 2) circle(0.1cm);
    \draw[thick] (2, 2)--node[midway,above]{}(2,0.7);
    \draw[thick, fill=none] (2, 1) circle(0.3cm);
    \draw[thick, fill=white] (3.9,2.4) rectangle (4.7,1.6) node [pos=.5]{$H$};
    \draw[thick, fill=white] (3.9,1.4) rectangle (4.7,0.6) node [pos=.5]{$H$};
    \draw[thick, fill=white] (3.9,0.4) rectangle (4.7,-0.4) node [pos=.5]{$H$};

\end{tikzpicture}

%% file: Chapters/2.Preliminaries/2.10.DomainSpecificLanguage.tex
\section{Domain Specific Language}
In the history of computer science, many programming languages have been developed. In general, programming languages can be used to express a universal set of computations such as C/C++, Rust, Python, Java, and Go. However, some programming languages are created for specific purposes. A prominent example is SQL, which is a query language for database systems. In terms of networking, \cite{bosshart2014p4} is one that specializes in data plane control in software-defined networking. In general, those languages are capable of running any application with useful libraries.

However, there are some advantages to using a domain-specific language. One is to be able to write down the problem more naturally, and sometimes it is easier for users and even nonprogrammers to define the problem with that language. If the language is carefully designed, the syntax and semantics are more natural for users to interpret problems with that language.

Furthermore, if we can design the runtime systems corresponding to that problem domain, it might also be possible to build up more robust and more efficient programming languages.

The types of domain-specific languages can be roughly divided into two. The first DSL is called inner DSL which allows us to write down programs similar to the original language in a highly abstract and easier way. The other is external DSL which has completely different syntax and semantics from other languages.

%% file: Chapters/3.ProblemDefinition/3.0.ProblemDefinition.tex
\chapter{Problem Definition}
\label{chap:prob_def}
This chapter explains and defines the problems we are approaching in this thesis. In the first section~\ref{sec:rulset_based_qnet}, we introduce a new concept called a \textit{RuleSet}-based quantum repeater network, which is a key in this project. Section~\ref{sec:ruleset_def} shows how to define and generate a RuleSet by introducing its schema. In Section~\ref{sec:problem_def}, the problems encountered when we create RuleSets are defined. We finally look at how related works tackle similar problems when they manage quantum repeaters in Section~\ref{sec:related_works}

\input{Chapters/3.ProblemDefinition/3.1.ProblemDomain}
\input{Chapters/3.ProblemDefinition/3.2.RuleSetDefinition}
\input{Chapters/3.ProblemDefinition/3.3.ProblemDefinition}
\input{Chapters/3.ProblemDefinition/3.4.RelatedWorks.tex}

%% file: Chapters/3.ProblemDefinition/3.1.ProblemDomain.tex
\section{RuleSet-based Quantum Repeater Network}
\label{sec:rulset_based_qnet}
The goal of building a quantum network is to transfer an unknown quantum state from one place to another or to share a bipartite or multipartite entangled state between multiple quantum computers. We exploit quantum entanglement as a resource to send a quantum state over the teleportation technique. 
 
In order to achieve long-distance quantum teleportation, we need to deliver entangled states over the distance without losing their fidelity. If the fidelity of the entanglement is small, the reliability of the teleportation drops. In general, when a qubit goes through a long fiber, its noise and loss rate rapidly increase over the fiber length.

As explained in the previous chapter, the quantum repeater is a device that allows us to generate longer entangled states with higher fidelity than direct qubit transfer. Quantum repeaters are placed between the two end nodes, as shown in Figure~\ref{fig:repeater_network}. When it comes to large-scale quantum networking, more and more quantum repeaters are involved in entanglement generation. 

\begin{figure}[h]
    \centering
    \includegraphics[width=\linewidth]{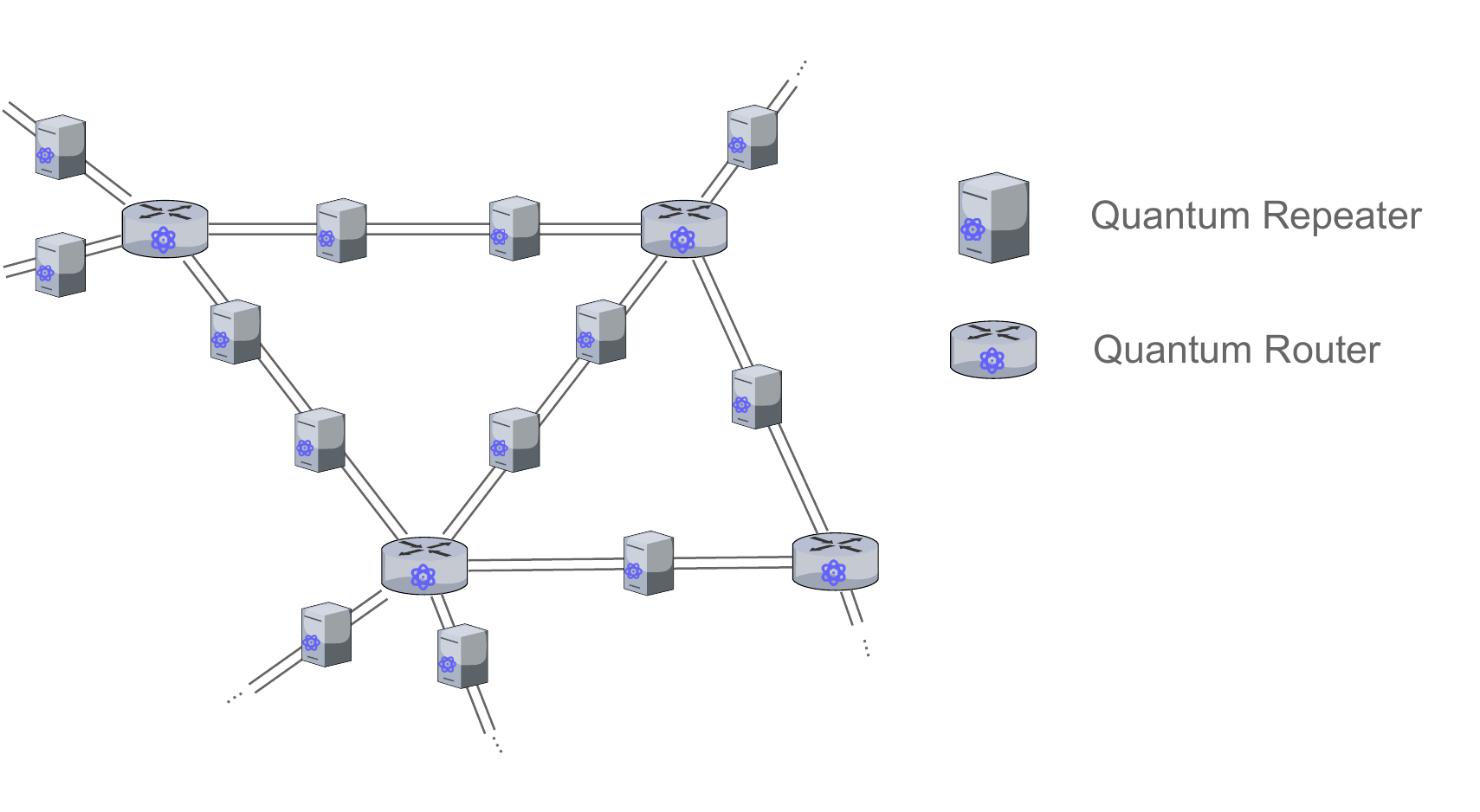}
    \caption[Scale of Quantum repeater Network]{The larger the entire quantum network becomes, the more repeaters are responsible for generating entanglement. }
    \label{fig:larger_repeater_network}
\end{figure}

In Figure~\ref{fig:larger_repeater_network}, the quantum routers and switches are also involved in the network, and quantum repeaters are installed between these components. When this becomes larger and larger in the future, hundreds or thousands of repeaters are working together to generate global entanglement. Furthermore, when we think of internetworking (network of networks), the distance between two networks would be larger. Thus, more and more devices are involved in the entanglement generation process.

Even when we have such a large number of repeaters in the network, these quantum repeaters must be properly managed. There are two strategies to manage these quantum repeaters. One is a purely centralized and synchronized way that allows us to generate some amount of entangled states in specific time cycles. There are several related works that adopted this scheme~\cite{pant2019routing, kozlowski2020designing}. The advantage of this is that we can increase the success rate of entanglement generation. Entanglement generation requires severe timing control so that two photons need to arrive at the Bell state analyzer within a nanosecond to microsecond time difference in current technology~\cite{hensen2015loophole}. However, in large-scale quantum networking, this approach may no longer be useful. One of the biggest challenges of these approaches is aligning the timing over a large number of repeaters. Especially in the case where we assume that two different networks are controlled by two different entities, aligning those clock cycles over the different networks is not an easy task.  

The other is a completely distributed way that limits the level of synchronization as much as possible. RuleSet is a method that allows us to generate entanglement generation asynchronously. 

\begin{figure}[h]
    \centering
    \includegraphics[width=0.8\linewidth]{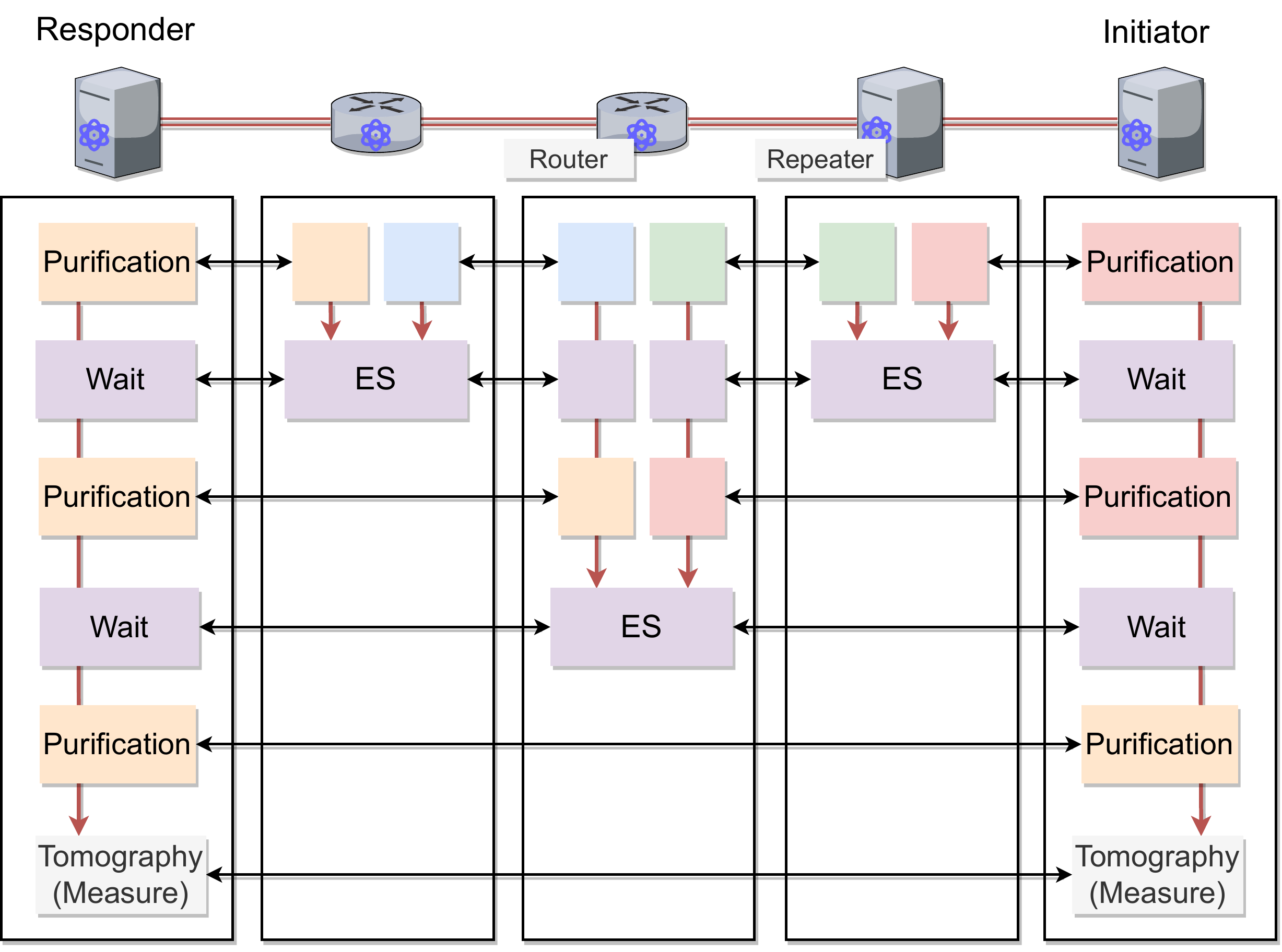}
    \caption[Example RuleSet]{A simplified example of RuleSets: a RuleSet is composed of a set of Rules (e.g. Purification, Entanglement Swapping) that tells each quantum repeater what to do. }
    \label{fig:example_ruleset}
\end{figure}

\textit{RuleSet}~\cite{matsuo2019quantum} is a set of instructions that Repeaters and Routers follow. This RuleSet scheme is also applicable for Quantum Internetworking~\cite{van2021quantum} by translating RuleSet over different networks. 
Figure~\ref{fig:example_ruleset} is an example structure of RuleSet. Each column represents a RuleSet for intermediate quantum devices, and they process their Rules inside the RuleSet from top to bottom. 

The first step in the RuleSet protocol is to distribute RuleSet to intermediate quantum repeaters.
\begin{figure}[h]
    \centering
    \includegraphics[width=.8\linewidth]{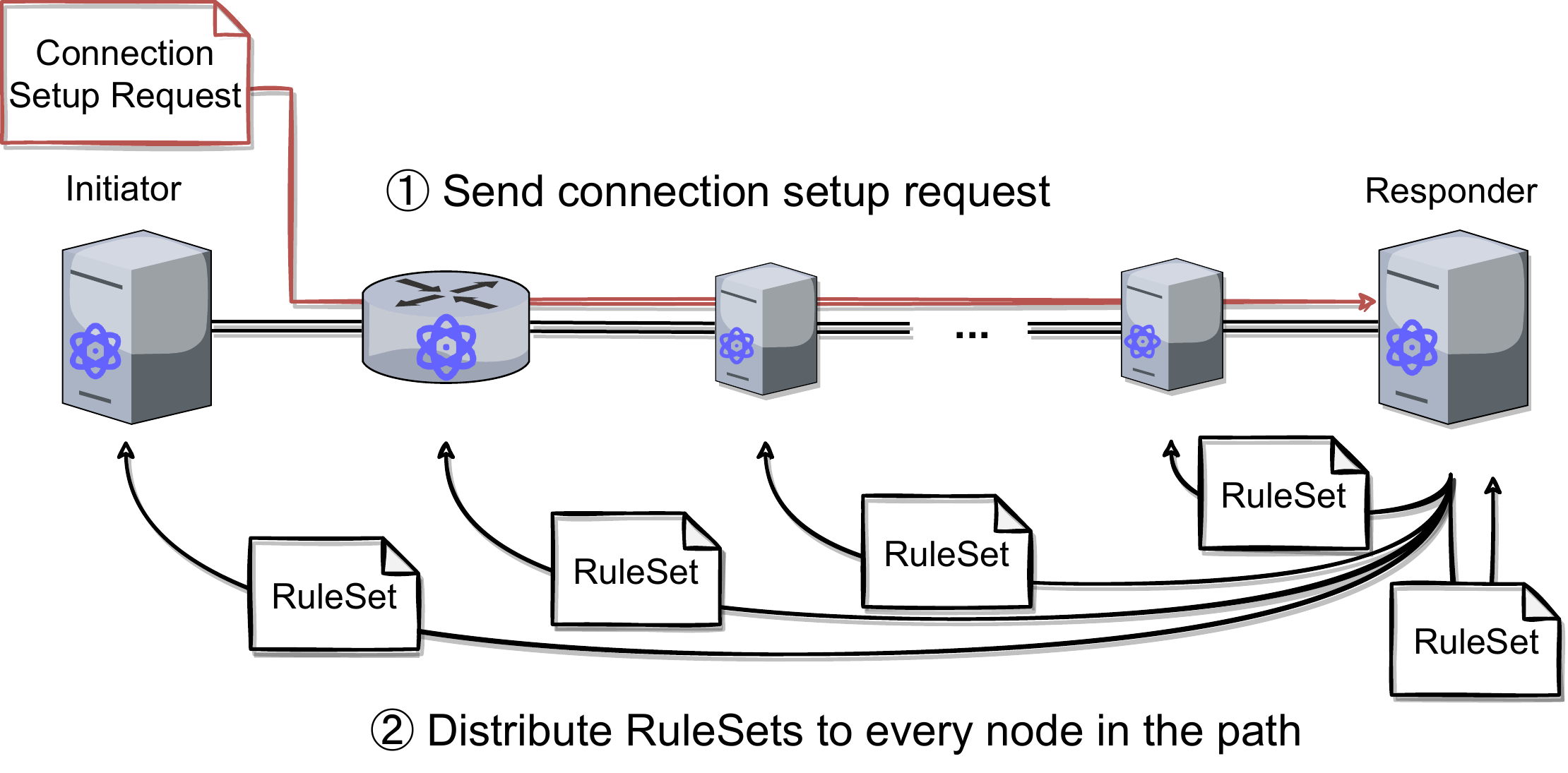}
    \caption[RuleSet Connection Setup]{Connection Setup process in RuleSet}
    \label{fig:connection_setup_request}
\end{figure}

Figure~\ref{fig:connection_setup_request} shows how the initiator sends the request to the responder and the responder returns the RuleSets. The terms "initiator" and "responder" are similar to "source" and "destination" in classical networking. However, in the context of quantum networking, we can theoretically send quantum information in either direction once. Thus, in this paper, we distinguish these terms. In RuleSet protocols, we assume static routing that all routers have their routing table, and once the path is decided, it does not change while the repeaters are generating entanglement. 

The first step in connection setup is to send the request from the initiator to the responder. In this process, the initiator describes the requirements for the entanglement in the request and sends it through the path. Second, the responder receives the request and creates RuleSets for itself and intermediate repeaters. After the RuleSets are prepared, they are sent and executed.

RuleSet is composed of a set of \textit{Rule}s which are instructions for quantum repeaters.
\begin{figure}[h]
    \centering
    \includegraphics[width=\linewidth]{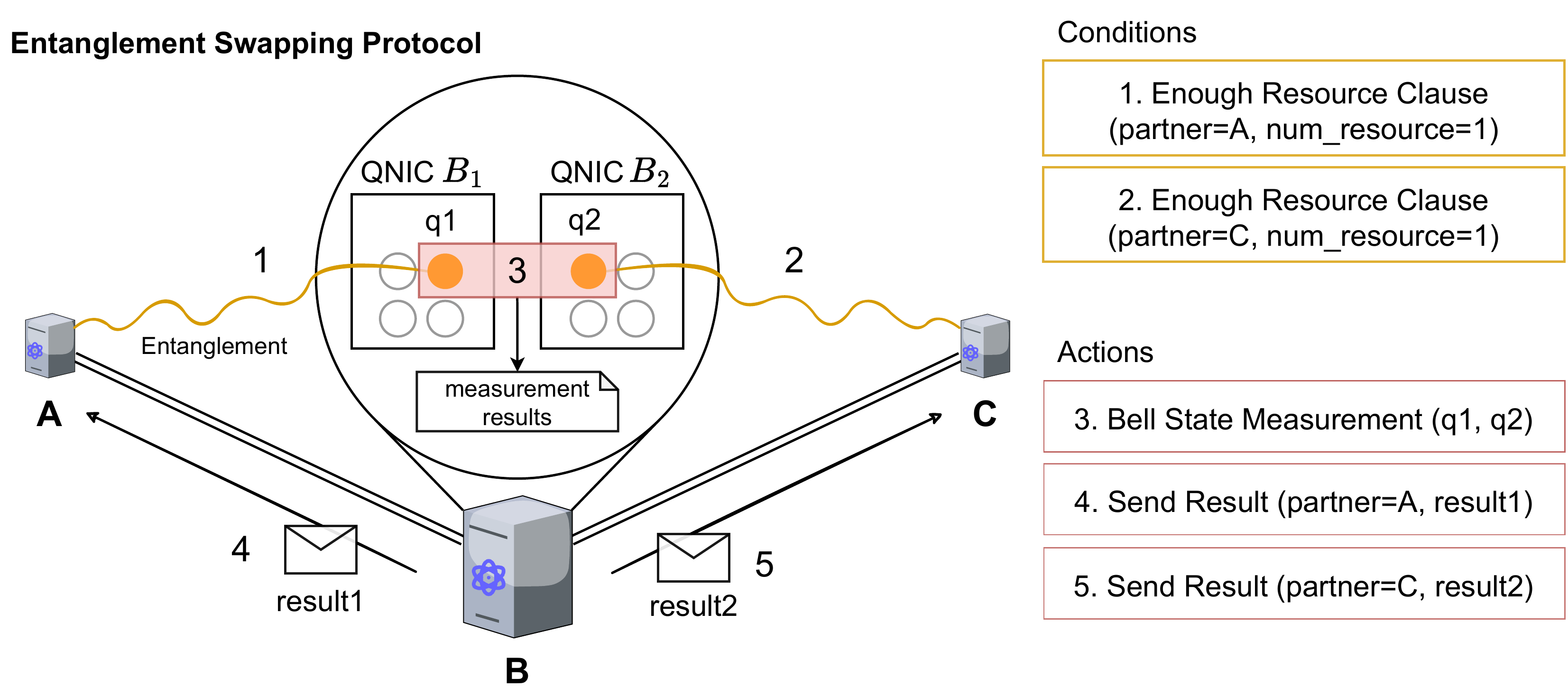}
    \caption[Example RuleSet Construction]{An example entanglement swapping rule structure.}
    \label{fig:rule_contents}
\end{figure}
Figure~\ref{fig:rule_contents} shows an example structure of a single Rule for entanglement swapping. As we showed in Section~\ref{sec:entanglement_swapping}, entanglement swapping consumes two entangled states. In other words, as soon as two entangled states are ready to be used, we can move on to the measurement operation. Thus, there are two conditions called the "Enough Resource Clause", which is satisfied when the quantum repeater has a specified number of entangled states with a specific partner node. Once all the conditions are met, the following actions are taken. In this example, there are three actions. The first action is the "Bell state measurement", which takes two qubits entangled with proper partners and applies the Bell measurement to them. As a result of the measurement, repeaters can extract the results and send them to the partners. 

Each rule has only pure condition clauses and action clauses so there is no way to introduce condition branching in the Rule. However, there is a concept called \textit{stage} in the RuleSet. This stage allows us to divide the conditions into several different rules that can mimic the selective execution of the rules. More details on the stage system are explained in Section~\ref{sec:rule_execution_model}.

There are several predefined conditions and actions introduced in~\cite{van2021quantum}. 
Table~\ref{tab:condition_clauses} is the list of condition clauses.
\begin{table}[h]
\centering
\renewcommand{\arraystretch}{1.2}
\caption[Condition Clauses]{List of primitive condition clauses}
\begin{tabular}{cl}
\hline
\multicolumn{1}{|c|}{Name} & \multicolumn{1}{c|}{Description}                                                     \\ \hline
\hline
\multicolumn{1}{|c|}{Enough Resource (RES)}            & \multicolumn{1}{l|}{Check if there are enough available resources in QNIC} \\ \hline
\multicolumn{1}{|c|}{Comparison (CMP)}                        & \multicolumn{1}{l|}{Compare some properties and check if that is satisfied}          \\ \hline
\multicolumn{1}{|c|}{Timer (TIMER)}                      & \multicolumn{1}{l|}{Check the timer is expired or not}                               \\ \hline
\end{tabular}
\label{tab:condition_clauses}
\end{table}

\textbf{Enough Resource (RES)} is the condition clause that checks if there are enough entangled states with sufficient fidelity. For example, if a quantum repeater A needs two entangled states with fidelity above 0.8 shared with repeater B, the condition clause is described as \verb|RES(1, 0.8, B)| where the arguments correspond to 

\verb|(<the number of resources>|, \verb|<required fidelity>|, \verb|<partner identifier>)|.

\textbf{Comparison (CMP)} checks whether a value is equal, less, or greater than some values. For instance, if we want to check the measurement result as 0 or 1, the condition clause is \verb|CMP(MeasResult, Eq, "0")| where the first argument is the identifier of the classical variable. In this case, since we compare the measurement results, the variable ID is \verb|MeasResult|. The second argument is the operator such as \verb|Eq|("="), \verb|Lt|("<"), \verb|Gt|(">").

\textbf{Timer (TIMER)} Check the timer to be expired. 

Table~\ref{tab:action_clauses} is the list of actions to be performed.
\begin{table}[h]
\centering
\caption[Action Clauses]{List of primitive action clauses}
\renewcommand{\arraystretch}{1.2}
\begin{tabular}{ll}
\hline
\multicolumn{1}{|l|}{Name}      & \multicolumn{1}{l|}{Description}                              \\ \hline\hline
\multicolumn{1}{|l|}{Set Timer (SETTIMER)} & \multicolumn{1}{l|}{Set a timer to count the time}            \\ \hline
\multicolumn{1}{|l|}{Promote (PROMOTE)}   & \multicolumn{1}{l|}{Promote a resource from one rule to next} \\ \hline
\multicolumn{1}{|l|}{Free (FREE)}      & \multicolumn{1}{l|}{Free the qubit no longer used}            \\ \hline
\multicolumn{1}{|l|}{Set (SET)}       & \multicolumn{1}{l|}{Set the RuleSet variable}                 \\ \hline
\multicolumn{1}{|l|}{Measurement (MEAS)}      & \multicolumn{1}{l|}{Measure the qubit}                        \\ \hline
\multicolumn{1}{|l|}{Quantum Circuit (QCIRC)}     & \multicolumn{1}{l|}{Apply quantum gates}                      \\ \hline
\end{tabular}
\label{tab:action_clauses}
\end{table}

\textbf{Set Timer} starts the timer to be expired. (e.g. \verb|TIMER(10)|)

\textbf{Promote (PROMOTE)} move the ownership of the resources to the next Rule or Stage. For example, \verb|PROMOTE(qubit1, Rule2)| moves a qubit which has ID (\verb|qubit1|) to \verb|Rule2| 
 (\verb|PROMOTE(<qubit id>, <Stage or Rule id>)|).

\textbf{Free (FREE)} release the qubit ownership and reinitialize for later use. \verb|FREE(qubit123)| frees the qubit with ID (qubit123) and the resource allocator starts to create link-level entanglement in the next round. 

\textbf{Set (SET)} sets a variable to be shared in Rule, Stage, or RuleSet. For example, to count the number of measurements that occurred in a Rule, \verb|SET(MeasCount)| grabs the current measurement count and updates it.

\textbf{Measurement (MEAS)} measures the qubit and returns the measurement result. \verb|MEAS(qubit1, X)| means that measure a qubit with identifier \verb|qubit1| in X basis (\verb|MEAS(<qubit id>, <measurement basis>)|).

\textbf{Quantum Circuit (QCIRC)} applies a series of gate operations on a qubit. \verb|QCIRC((q1, q2), (("X", q1), ("H", q2)))| is an example usage of \verb|QCIRC| that instructs the system to apply X gate on a qubit with ID \verb|q1| and apply H gate a qubit with ID \verb|q2|. The first argument is the list of qubit identifiers and the second argument is the list of operations.

(\verb|QCIRC((<qubit id, qubit id, ..>), ((<operation>, <qubit id>), ..))|)

In addition to these condition clauses and action clauses, there are also predefined protocol messages to exchange the measurement result and to notify the successful entanglement generation.
Table~\ref{tab:proto_message} is the messages used in RuleSet.
\begin{table}[h]
\centering
\caption[Protocol Messages]{A list of protocol messages}
\renewcommand{\arraystretch}{1.2}
\label{tab:proto_message}
\begin{tabular}{|l|l|}
\hline
Name                                      & \multicolumn{1}{c|}{Description}                                                                                      \\ \hline
Free (FREE)                               & Request to free the used qubit in the partner node                                                                    \\ \hline
Update (UPDATE)                           & \begin{tabular}[c]{@{}l@{}}Request to update the resource (e.g. Pauli correction) \\ in the partner node\end{tabular} \\ \hline
Measurement (MEAS)                        & Tell the measurement results to the partner                                                                           \\ \hline
\multicolumn{1}{|l|}{Transfer (TRANSFER)} & \begin{tabular}[c]{@{}l@{}}Distribute the measurement results \\ and transfer a new resource id.\end{tabular}         \\ \hline
\end{tabular}
\end{table}

%% file: Chapters/3.ProblemDefinition/3.2.RuleSetDefinition.tex
\section{Defining RuleSet}
\label{sec:ruleset_def}

Figure~\ref{fig:example_ruleset} is just a very simplified visualization of the RuleSet. In reality, the RuleSet needs to be defined in some portable format. One possible way to define the RuleSet is the JSON format, which is a very common way of exchanging information between different applications.

Listing~\ref{lst:ruleset_def_json} shows an example definition of the RuleSet.
There are several fields within the RuleSet.
\begin{lstlisting}[caption={Example JSON file for RuleSet definition},captionpos=b, label={lst:ruleset_def_json}, language=json]
{
    "name": "entanglement_swapping",
    "id": 9876543210,
    "owner_addr": 1,
    "stages": [
        {
            "rules": [
                {
                    "name": "swapping",
                    "id": 0,
                    "shared_tag": 0,
                    "qnic_interfaces": {},
                    "condition": {
                        "name": null,
                        "clauses": [
                            {
                                "Cmp": {
                                    "cmp_val": "MeasResult",
                                    "operator": "Eq",
                                    "target_val": {
                                        "MeasResult": "00"
                                    }
                                }
                            }
                        ]
                    },
                    "action": {
                        "name": null,
                        "clauses": [
                            {
                                "QCirc": {
                                    "qgates": [
                                        {
                                            "qubit_identifier": {
                                                "qubit_index": 0
                                            },
                                            "kind": "CxControl"
                                        },
                                        {
                                            "qubit_identifier": {
                                                "qubit_index": 1
                                            },
                                            "kind": "CxTarget"
                                        }
                                    ]
                                }
                            },
                            {
                                "Measure": {
                                    "qubit_identifier": {
                                        "qubit_index": 0
                                    },
                                    "basis": "X"
                                }
                            },
                            {
                                "Measure": {
                                    "qubit_identifier": {
                                        "qubit_index": 1
                                    },
                                    "basis": "Z"
                                }
                            },
                            {
                                "Send": {
                                    "Transfer": {
                                        "partner_addr": 0
                                    }
                                }
                            },
                            {
                                "Send": {
                                    "Transfer": {
                                        "partner_addr": 0
                                    }
                                }
                            }
                        ]
                    },
                    "is_finalized": false
                }
                ...
            ]}
            ...
            ]    
        }
\end{lstlisting}

\begin{itemize}
    \item \verb|ruleset_id|: Since RuleSet is created for the entire connection, RuleSets for one connection share the unique ids called \verb|ruleset_id|. If there are multiple connections at the same time, the specific connection should be identified by this \verb|ruleset_id|.
    \item \verb|num_rules|: This property is used to check the number of rules to be executed inside this RuleSet.
    \item \verb|owner_address|: The address of the owner of this RuleSet. For simplicity, this RuleSet contains integer values such as \verb |owner_address|, however, this field could contain an IP address or any identifier of the specific quantum repeater or router node.
    \item \verb|stages|: The list of stages that contains a set of rules.
\end{itemize}
Rules also contain multiple fields.
\begin{itemize}
    \item \verb|rule_id|: Similarly to \verb |ruleset_id|, each rule also has its own \verb|rule_id|, which can be used to identify, for example, the measurement result generated by a specific rule. In the current RuleSet, \verb|rule_id| is given sequentially.
    \item \verb|shared_tag|: This is an integer value that is shared among several rules. For example, when purification is performed on one quantum repeater, the corresponding pairs have the same shared tags.
    \item \verb|next_rule_id|: The identifier for the next rule. Once this rule is completed, the next rule identified by this value is to be operated.
    \item \verb|condition|: The list of conditions to be satisfied.
    \item \verb|action|: The list of actions to be operated when all conditions are satisfied.
\end{itemize}
These conditions and actions are constructed from their \verb|type| and options such as the number of resources they require and the fidelity they need.

%% file: Chapters/3.ProblemDefinition/3.3.ProblemDefinition.tex
\section{Problem Definition}
\label{sec:problem_def}
As long as a large number of RuleSets are properly generated, the RuleSet protocol might be highly scalable even for quantum internetworking. However, as a user or creator of a RuleSet, it is sometimes difficult to see what is going on in a RuleSet since they are composed of a set of instructions. It is also difficult to check dependencies of Rules such as sending and waiting messages.

In this project, we propose a new domain-specific language to generate such a large number of RuleSets. In addition to that, we provide a new runtime system corresponding to that language.

\begin{figure}[ht]
    \centering
    \includegraphics[width=\textwidth]{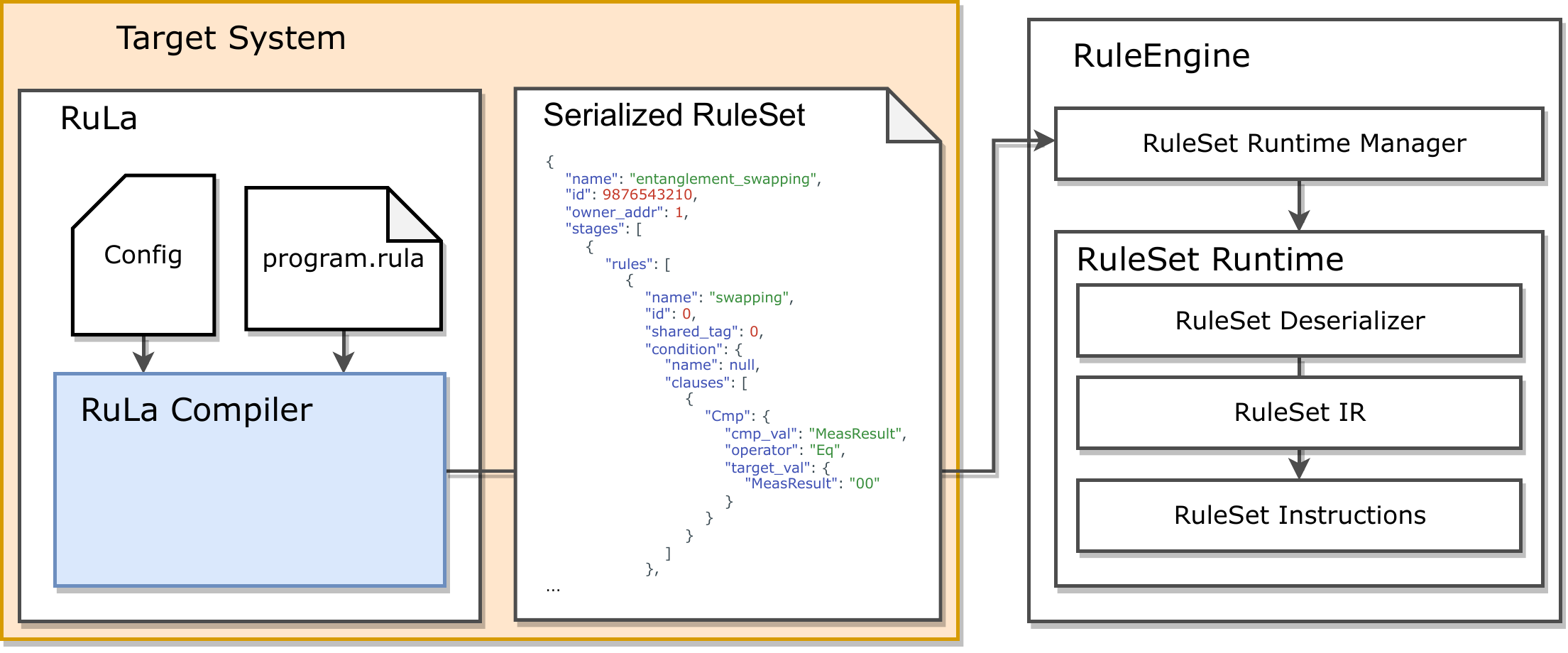}
    \caption[System Overview]{System overview of our proposal.}
    \label{fig:system_overview}
\end{figure}

Figure~\ref{fig:system_overview} gives an overview of our system. Once a network manager for this quantum network or users who want to try their protocol on the RuleSet write a \verb|.rula| program and provide it to the compiler with a configuration, our system provides us a serialized RuleSet that contains a set of instructions. This system allows us to describe RuleSet in more abstract manner.

%% file: Chapters/3.ProblemDefinition/3.4.RelatedWorks.tex
\section{Related Works}
\label{sec:related_works}
\subsection{Quantum Network Architectures}

There are several related works on quantum networking and internet architectures. 
The transition from the current state of the quantum networking to scalable and a future possible quantum network stuck was introduced~\cite{wehner2018quantum}.
A similar layering model was also proposed in a different research project, however, which is more reliable by using GHZ states~\cite{pirker2019quantum}. Higher-level routing implementation with multipartite states is also discussed.

The authors proposed entanglement generation protocols on top of the proposed layering model in~\cite{kozlowski2020designing}. They formalized entanglement tracking protocols and cutoff policies for the lifetime of entanglements. 

The main target of those previous works is small to medium-scale quantum networks especially maintainable by a central node. Our scope in this protocol is further future quantum networking that is fairly less time-sensitive for the sake of longer lifetime quantum memory. 

Our assumption is based on the RuleSet protocol~\cite{van2021quantum} that the main proposal for quantum internet is rewriting RuleSet over the different networks. RuleSets are designed to manage quantum repeaters in an asynchronous and scalable manner.

\subsection{Quantum Programming Language}
To manipulate quantum computers through software, we need a way to describe the instructions for quantum computers. The basic compiler models for quantum computing have been introduced and surveys on programming languages for quantum computers have been conducted~\cite{chong2017programming}. The various aspects such as compiler and debugger have also been discussed~\cite{mosca_et_al:DR:2019:10329}. In the NISQ era, operation imperfection and decoherence are crucial to the computational result. NISQ quantum computer requires noise-aware circuit compilation and transpilation that produces better computational results.  
 
 In terms of the modules for existing programming languages, one of the most popular software development kits for quantum computing is \verb|Qiskit|~\cite{cross2018ibm} from IBM. Google has also been developing its open-source quantum library called \verb|Cirq|~\cite{cirq_developers_2022_6599601}. Both are modules of the Python programming language. They are modules to define quantum circuits and submit quantum circuits as jobs that can be run on real devices or simulators.
\verb|Q#|~\cite{svore2018q} developed by Microsoft is a domain-specific programming language for quantum computing.

There are assembly-like languages to describe low-level gate-level instructions in quantum computing, called OpenQASM~\cite{cross2017open}, which allows us to write a sequence of gate operations and measurement operations. In terms of networking, there is also a similar work that describes quantum networking operations in an assembly-like language called NetQASM~\cite{dahlberg2022netqasm}. The usage of NetQASM is experimentally demonstrated~\cite{pompili2021experimental}.

There is a functional programming language called Quipper that is specialized in the generation of large quantum circuits~\cite{green2013quipper}.
Qwire is also a programming language that defines gate-level quantum circuits~\cite{paykin2017qwire}.
Silq~\cite{bichsel2020silq} is also a high-level programming language that can define a quantum circuit with a small number of lines.
\cite{yuan2022twist} introduces how the entanglement is reasoned in the computer program with a new programming language called twist.

There is also a research direction that tries to integrate classical networking technology into quantum networking. The authors are trying to use the software-defined network (SDN) technique~\cite{humble2018software}. There is also research work that is trying to integrate the manipulation of the data plane of existing p4 data into quantum networking~\cite{kozlowski2020p4}.

%% file: Chapters/4.Proposal/4.0.Proposal.tex
\chapter{Proposal}
\label{chap:proposal}
This chapter introduces our proposal programming language that allows us to write down RuleSet in more abstract way. In Section~\ref{sec:requirements}, the required features for this language are listed. The assumptions we made are discussed in Section~\ref{sec:assumptions}. The design principles of RuLa are introduced in Section~\ref{sec:design_principle} and we dived into the detailed grammar and semantics in Section~\ref{sec:language_design}. Section~\ref{sec:compiler} covers how the RuLa program is compiled into lower-level RuleSet instructions.
\input{Chapters/4.Proposal/4.1.Requirements}
\input{Chapters/4.Proposal/4.2.Assumptions}
\input{Chapters/4.Proposal/4.3.DesignPrinciple}

\input{Chapters/4.Proposal/4.4.LanguageDesign}
\input{Chapters/4.Proposal/4.5.Compiler}

\input{Chapters/4.Proposal/4.6.Example.tex}

%% file: Chapters/4.Proposal/4.1.Requirements.tex
\section{Requirements}
\label{sec:requirements}
In the current RuleSet generation process, there are three possible difficulties when there are a number of Rules and RuleSet.
\begin{enumerate}
    \item \textbf{Visibility}: It becomes difficult to grasp what is going on in a Rule.
    \item \textbf{Consistency}: It is hard to resolve the dependencies of two dependent Rules.
    \item \textbf{Expressibility}: Multiple different Rules are needed to describe multi-condition operations.
\end{enumerate}
From these possible difficulties in the current RuleSets, the requirements for RuLa are introduced as follows.

\textbf{1. Visibility enhancement}: 
The RuleSet and each Rule that is contained in a RuleSet is composed of a set of instructions shown in Table~\ref{tab:condition_clauses}, ~\ref{tab:action_clauses} and ~\ref{tab:proto_message}. Resulting RuleSets are generally produced in a serialized format such as JSON. It is hard to see the relationship between the repeaters and the flow of the operation. RuLa must enhance the visibility of the RuleSet.

\textbf{2. Automatic consistency support}: 
When Rule A sends a message to Rule B, Rule B has to wait for the message from Rule A. In this case, Rule B has a dependency on Rule A. The RuLa compiler should support automatic consistency resolution to reduce the potential inconsistent Rule errors.

\textbf{3.Support multi-conditioning}:
The current RuleSet does not support multi-conditioning in a Rule such as ``if condition A is satisfied, do action A. If condition B is satisfied, do action B". Instead, we need to create two different Rules with two different conditions in a single stage. RuLa should support the functionalities that translate that multi-level conditioning into a set of different Rules in a stage.

In addition to the requirements above, RuLa needs to properly cover all the instructions listed in Table~\ref{tab:condition_clauses}, \ref{tab:action_clauses}, and \ref{tab:proto_message}. 

%% file: Chapters/4.Proposal/4.2.Assumptions.tex
\section{Assumptions}
\label{sec:assumptions}

Since quantum computing and networking are still early-stage technologies, there is no strong consensus on the general architecture. In this section, we explain the assumptions that we made to design this language.

\subsection{Quantum Recursive Network Architecture}
One of the fundamental assumptions we made is the assumption of quantum network architecture. Our primary target architecture is the RuleSet-based Quantum Recursive Network Architecture (QRNA)~\cite{van2021quantum, van2011recursive}. 
The network is divided into multiple layers and is recursively managed.

\begin{figure}[h]
    \centering
    \includegraphics[width=.8\linewidth]{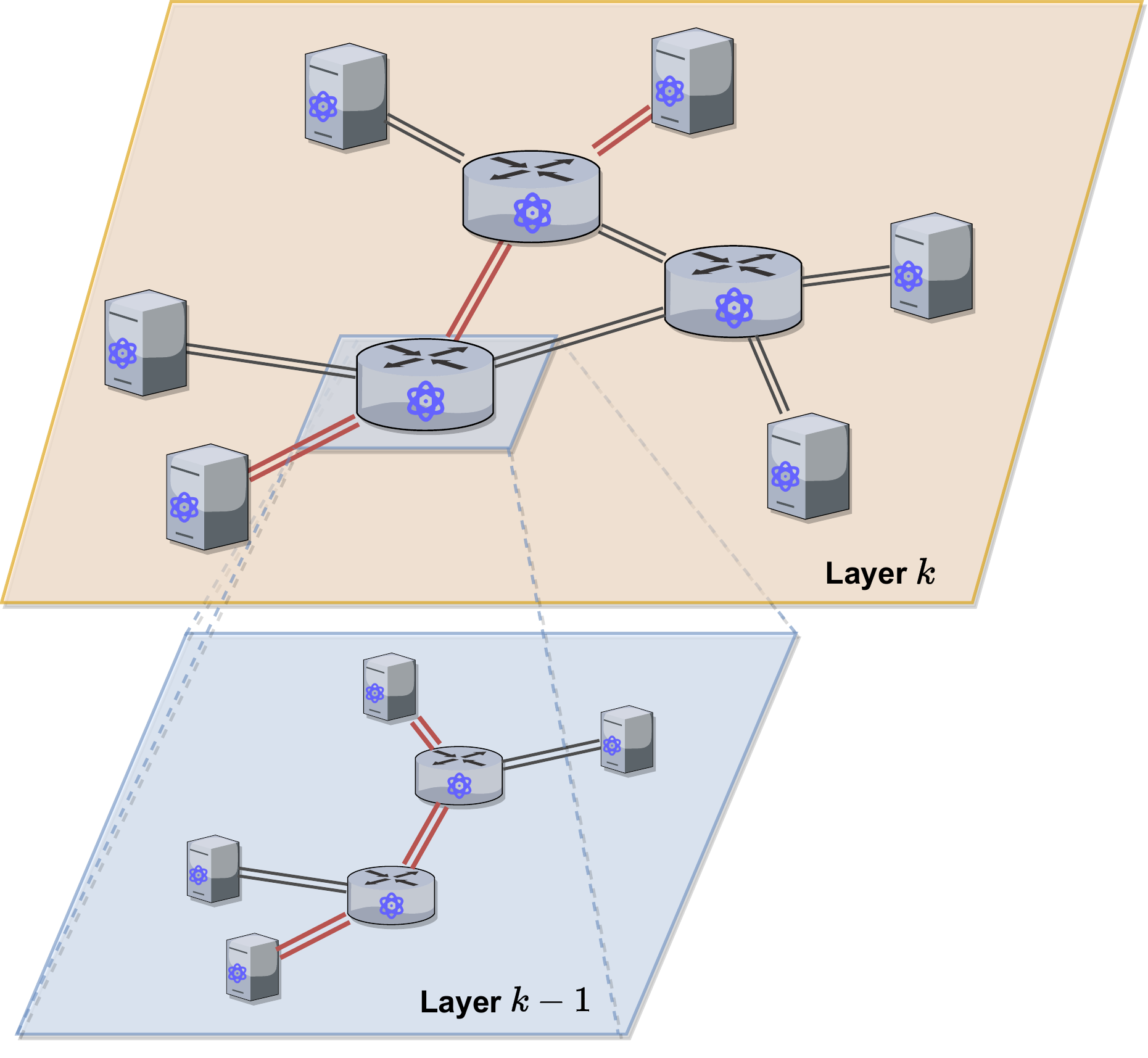}
    \caption[Recursive Network Architecture]{Recursive Network Architecture}
    \label{fig:recursive_network_architecture}
\end{figure}
Figure~\ref{fig:recursive_network_architecture} shows the basic idea of QRNA. The network in layer $k-1$ is virtualized in the higher layer $k$. The advantage of this architecture is that higher-layer quantum repeaters are not involved in lower-layer resource management. This allows us to have independent and scalable resource management in the different layers.

\subsection{Quantum Node Architecture}
\begin{figure}[h]
    \centering
    \includegraphics[width=\linewidth]{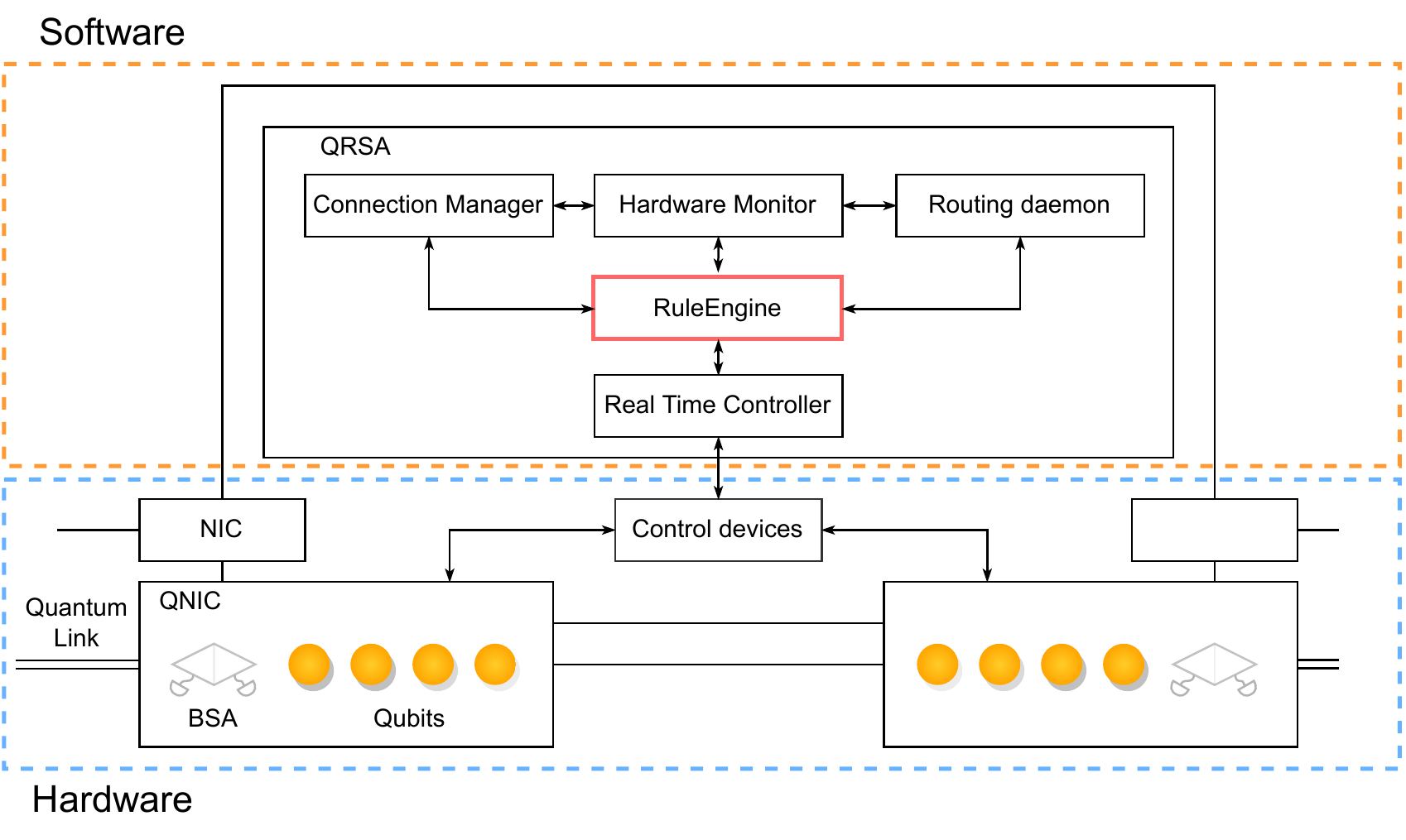}
    \caption[Quantum Node Architecture]{Quantum Node Architecture}
    \label{fig:node_architecture}
\end{figure}

We also need some assumptions on the hardware and software architecture for quantum repeaters. A proposal of the repeater architecture is in Quantum Internet Research Group (QIRG) work in the Internet Research Task Force (IRTF)~\cite{irtf-qirg-principles-11}. This RFC is still a draft version, but we believe that the fundamental concept of quantum networking will not change drastically in the future. In addition, we believe that the architecture based on RuleSet \cite{matsuo2019quantum, van2021quantum} is a good candidate for future quantum networking and internetworking.
Based on these architectural proposals, Figure~\ref{fig:node_architecture} is the basic architecture that we assume in this project.

\subsection{QRSA (Quantum Repeater Software Architecture)}
In terms of software architecture, we assume the Quantum Repeater Software Architecture (QRSA) is the fundamental software architecture for quantum repeaters. QRSA is composed of the following components.
\begin{itemize}
    \item \textbf{Connection Manager}: a software component that is responsible for handling requests and responses for connection setup and creating RuleSet for intermediate quantum repeaters. 
    \item \textbf{Hardware Monitor}: a system that monitors the status of the hardware (e.g. the fidelity of the link, memory). The hardware monitor may provide that information to choose the path for routing and multiplexing.
    \item \textbf{Routing Daemon}: routing daemon is a daemon process that constantly updates the routing table for successful routing. 
    \item \textbf{Real Time Controller}: a system that manipulates the hardware and receives feedback. When the Rule Engine makes a request to apply quantum gates to a qubit, Rule Engine issues a request to Real-Time Controller.
    \item \textbf{Rule Engine}: a critical component that handles RuleSet. Rule Engine is also responsible for managing the resources available for the Rules. Rule Engine receives RuleSet from Connection Manager and converts provided RuleSets into an executable format. The converted RuleSet instructions are executed one by one.
\end{itemize}

\subsection{QNIC (Quantum Network Interface Card)}
Classical computers and servers have one or more Network Interface Card (NIC) to communicate over the network or the Internet. In quantum networking, we assume the existence of the Quantum Network Interface Card (QNIC) which is a quantum analogue of a NIC. Inside the QNIC, there are solid quantum memories that can store quantum information. In this project, we assumed that arbitrary operations among the different QNICs in a single node can be performed without any overhead.


%% file: Chapters/4.Proposal/4.3.DesignPrinciple.tex
\section{Design Principle}
\label{sec:design_principle}
The design of syntax and grammar should describe the way to execute the execution of the RuleSet execution model. The resulting program should also be able to include device-specific configurations to support the flexibility of RuleSet definitions. 

\subsection{Target Users}
The target users for this language are 1. people who want to build up quantum communication protocols in the RuleSet scheme and 2. computer systems that are responsible for RuleSet generation.  

Our primary expectation is the one who wants to try a new quantum communication on a simulator or test devices that supports RuleSet processes. Thus, the grammar and syntax of RuLa should be human-readable for people to write it easily. In addition, we also expect a computer process that is responsible for RuleSet generation such as some controller devices for the quantum network could also use this language to build RuleSets for controlled devices.

\subsection{Rule Execution Model}
\label{sec:rule_execution_model}
The rule execution process is divided into five steps. 
\begin{figure}[ht]
    \centering
    \includegraphics[width=\textwidth]{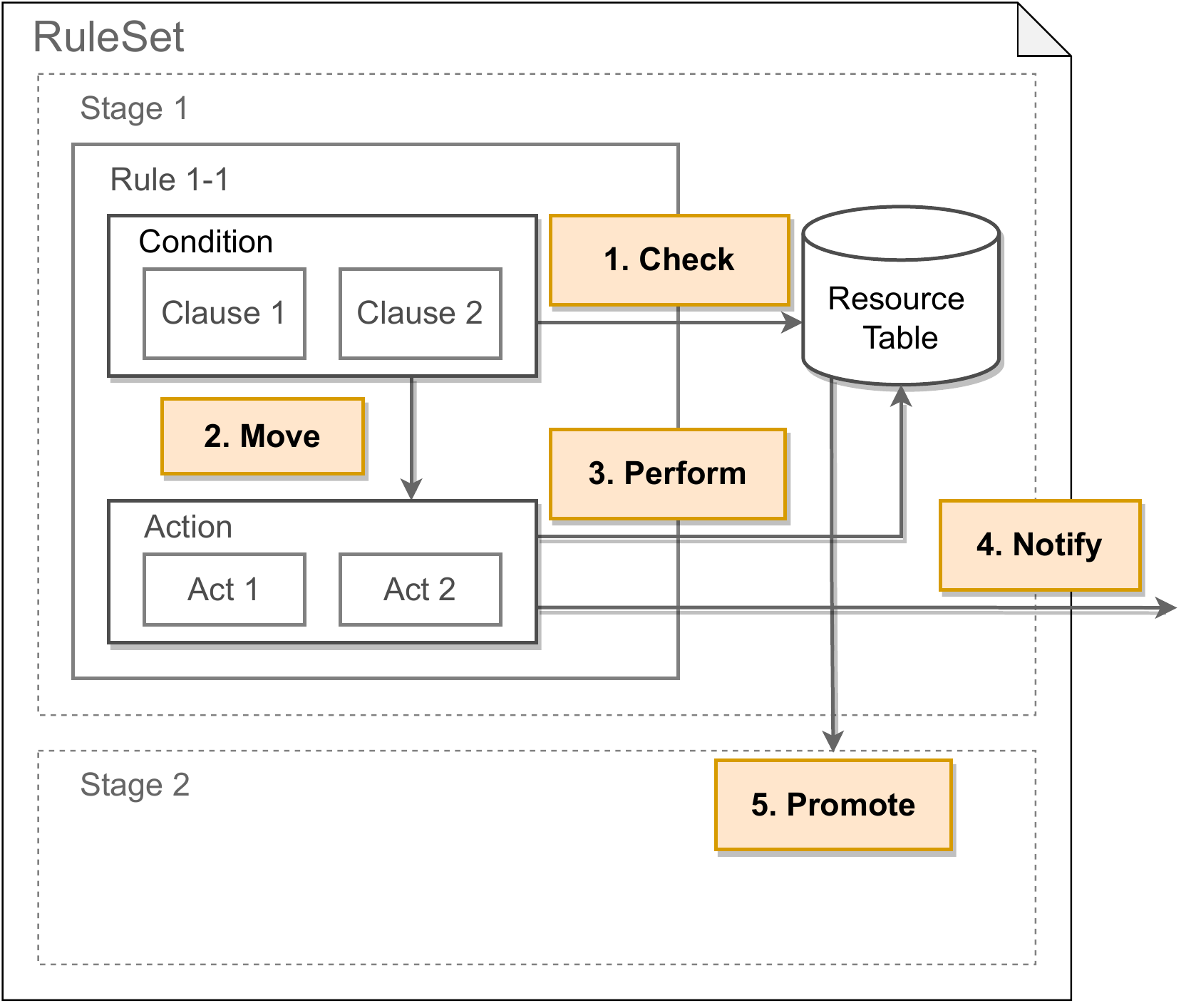}
    \caption[Rule Execution Steps]{Steps for Rule execution}
    \label{fig:rule_execution}
\end{figure}
Figure~\ref{fig:rule_execution} shows how a single rule is executed. 
\begin{enumerate}
    \item \textbf{Check}: Once at a certain time, the condition checks whether the clauses are satisfied or not. The condition checks the table of assigned Bell pairs if their status is ready to use and if the fidelity is high enough, as well as some constant variables.
    \item \textbf{Satisfy}: If all condition clauses are satisfied, the rule calls for the corresponding actions. 
    \item \textbf{Perform}: The action clauses are performed one by one on the target qubit. This might include some classical operations, such as sending measurement results to the other repeaters.
    \item \textbf{Notify}: After every operation is completed, the results of the actions need to be notified to the resource manager inside the RuleEngine and the corresponding rules running on the other quantum repeaters. 
    \item \textbf{Promote}: If there are resources that are moved from the previous rule to the next rule, the resource manager updates the table of the resources, and the corresponding resources are promoted. 
\end{enumerate}
From these items, RuLa must describe 1. What are the conditions to be satisfied, 2. How these conditions are met 3. What operations are performed, 4. How to tell the result of actions to the other rules, 5. How to transfer updated resources from one rule to another.

\subsection{Repeater Identifications}
In order to identify which repeater communicates with which other repeaters, RuLa should provide a way to specify the target repeaters.
However, the RuLa program itself must be independent of the number of repeaters in the path. 
RuLa should offer the ways to identify the repeaters as follows.
\begin{enumerate}
    \item Identify a repeater with a specific index.
    \item Identify a repeater relatively with the number of hops from the target repeaters.
    \item Identify a repeater dynamically with the index variables.
\end{enumerate}
Those three ways of identifying a repeater value allow us to specify a repeater and its partner repeaters without knowing the number of actual repeaters beforehand.

\subsection{Compilation RuLa program}
In terms of the timing that a RuLa program is compiled, the RuLa program is compiled anytime if there are RuLa programs with a user who wants to try a new protocol. In the case that a computer program generates a RuLa program for actual entanglement generation, a RuLa program is compiled when the connection setup request arrives at the responder node. The responder node compiles the RuLa program with the given repeater information and compiles a RuLa program with it.

%% file: Chapters/4.Proposal/4.4.LanguageDesign.tex
\section{Language}
\label{sec:language_design}
In this section, we explain the detailed language design and specifications of RuLa. The full specification is in Appendix~\ref{chap:full_specification}.

\subsection{Grammar Definition}
\label{sec:grammar_def}
We chose Parsing Grammer Expression (PEG)~\cite{ford2004parsing} to define the grammar. PEG can define grammar rules without any ambiguity. 

Let $E_1, E_2$ be defined expressions. Expressions follow the rules listed below in PEG.
\begin{itemize}
    \item $E_1 E_2$: $E2$ follows immediately after $E_1$.
    \item $E_1/E_2$: If $E_1$ is not evaluated, $E_2$ is evaluated.  
    \item $E_1$?: $E_1$ can be omitted
    \item $E_1$*: Zero or more $E_1$
    \item $E_1$+: One or more $E_1$
\end{itemize}

It is also possible to define a new rule with $\leftarrow$.
\begin{align}
    \text{NewRule} \leftarrow \text{Rule1}/\text{Rule2}
\end{align}
where \verb|NewRule| is supposed to be either \verb|Rule1| or \verb|Rule2|.

To define the grammar, we used \verb|pest|\cite{pest_rust} which is a grammar rule generator based on the PEG expression.
The way to define grammar in \verb|pest| is quite similar to how we define grammar in a pure PEG expression; however, there is some syntax sugar to make grammar definition easier.
As an extension of PEG, \verb|pest| supports the following rules for convenience.
\begin{itemize}
    \item \^~"keyword": Case sensitive
    \item !~"keyword": Prohibit "keyword"
\end{itemize}

\subsection{Terminal Characters}
The termination characters are defined as \verb|ASCII_DIGIT| and \verb|ASCII_ALPHA|.
\begin{lstlisting}[language=PEG]
ASCII_ALPHA <- 'a' / 'b' / ... / 'z' / 'A' / 'B' / ... / 'Z'
ASCII_NONZERO_DIGIT <- '1'/'2'/'3'/'4'/'5'/'6'/'7'/'8'/'9'
ASCII_DIGIT <- ASCII_NONZERO_DIGIT / '0'
\end{lstlisting}
There is also a rule \verb|ANY| that includes symbols such as \verb|@|, \verb|*| as well.

\subsection{Supported Data Types}
In RuLa, there are five primitive data types that would be required for RuleSet and five composite data types to simplify data handling.

\vskip\baselineskip
\noindent\textbf{Integer}: The integer value is made up of one or more than one digit of numbers.
\begin{lstlisting}
int <- (ASCII_NONZERO_DIGIT ASCII_DIGIT+ / ASCII_DIGIT) !ASCII_ALPHA
\end{lstlisting}
Internally, it is converted to a 64-bit signed integer value
($-9223372036854775808$ to $9223372036854775807$). 
In the variable definition, the integer value must be annotated with \verb|int|.

\vskip\baselineskip
\noindent\textbf{Unsigned Integer}: If the value is defined as an unsigned integer value, that value can be between $0$ to $2^{64}-1$. ($0$ to $18446744073709551615$)
The unsigned integer variable must be annotated by \verb|u_int| when it is defined.

\vskip\baselineskip
\noindent\textbf{Floating Point}: The floating point value consists of an integer part, a floating point, and fraction values. 
\begin{lstlisting}
float <- int "." int
\end{lstlisting}
The floating-point value is converted to a 64-bit floating value internally. 

($-1.7976931348623157e308$ to $1.7976931348623157e308$).

The floating-point variable must be annotated with \verb|float| in the variable definition.

\vskip\baselineskip
\noindent\textbf{Boolean}: The Boolean value can be \verb|true| or \verb|false|.
\begin{lstlisting}
bool <- true_lit / false_lit 
true_lit <- "true" 
false_lit <- "false" 
\end{lstlisting}

The Boolean value must be annotated with \verb|bool|.

\vskip\baselineskip
\noindent\textbf{String}: The string value is represented as a value quoted by double quotation (\verb|"|).
\begin{lstlisting}
string <- "\"" raw_string* "\"" 
raw_string <- (!( "\\" | "\"" ) ANY)+
\end{lstlisting}
Some characters \verb|\| and \verb|"| cannot be included in the quotation.

\vskip\baselineskip
\textbf{Vector}: A vector value contains multiple data inside, but the type of data needs to be the same among all data.
\begin{lstlisting}
vector <- "[" literal_expr? ("," literal_expr)* ","? "]"
\end{lstlisting}

\vskip\baselineskip
\textbf{Composite Data Types}: For convenience, RuLa natively supports the following composite data types.
\begin{itemize}
   \item \textbf{Qubit} (\verb|Qubit|): An alias type of the target qubit. A qubit value stores an index value that is valid in the rule definition.
   \item \textbf{Repeater} (\verb|Repeater|): The type of quantum repeater that contains several methods to specify the partner quantum repeaters.
   \item \textbf{Message} (\verb|Message|): The message type value stores the message contents from the other repeaters.
   \item \textbf{Result} (\verb|Result|): The result type value stores the measurement result created by the measurement operations. 
   \item \textbf{Time} (\verb|Time|): To check if the timer has expired or not, RuLa provides Time as a primitive data structure to represent the time and check the timer expirations. In the current version, this \verb|Time| type is not available.
\end{itemize}

\subsection{Variable Definition}

\begin{lstlisting}[label={lst:var_def_peg}, language=PEG]
let_stmt <- ^"let" (ident_typed/
            "(" ident_typed ("," ident_typed)*")")"="expr
\end{lstlisting}

\verb|let| statement allows us to define a variable as shown in Listing~\ref{lst:var_def}.
\begin{lstlisting}[caption={variable definition},captionpos=b, label={lst:var_def}, language=RuLa]
// Define x as 10
let int_vec: vec[int] = [1, 2, 3, 4]
let integer_val: int = -10
let uinteger_val: u_int = 19
let floating_val: float = 0.85
let bool_val: bool = true
let string_val: str = "hello world"
\end{lstlisting}

Variable identifiers must be properly annotated in the \verb|let| statement.
\begin{lstlisting}[caption={Definition of identifier and typed identifier}, label={lst:identifier_peg}, language=PEG, captionpos=b]
ident_typed <- ident ":" typedef_lit
ident <- ASCII_ALPHA ( ASCII_ALPHA / ASCII_DIGIT / '_' )*
typedef_lit <- vector_type/integer_type/unsigned_integer_type/float_type/boolean_type/string_type/qubit_type/repeater_type/message_type/result_type
\end{lstlisting}

\subsection{Importing}
To keep the system's modularity, RuLa allows us to import rules and functions defined in other files. It is possible to get functions and structures from other places with \verb|import| as we show in Listing~\ref{lst:import_def}.

\begin{lstlisting}[caption={import functions and structs},captionpos=b, label={lst:import_def}, language=RuLa]
// Import a function bsm()  from the operations library
import std::operation::bsm
import std::operation::{x, y}
\end{lstlisting}
Types and functions are specified using delimiters (\verb|::|).
At the end of the path, it is also possible to set a list of identifiers that represent functions and structures.

\begin{lstlisting}[language=PEG]
import_stmt <- ^"import" ident ("::" ident)* ( "::" "{"ident_list"}")?  !"::"
rule_annotation <- "(rule)"
ident_list <- ( ident / ident_typed )  ( "," ( ident / ident_typed ) )*
\end{lstlisting}

\subsection{Rule definition}
\label{sec:rule_def}
\begin{lstlisting}[language=PEG, caption = {Rule definition}, captionpos=b, label={lst:rule_def_peg}]
rule_stmt <- ^"rule" ident "<" repeater_ident ">" argument_def "{" rule_contents "}"
ret_type_annotation <- typedef_lit maybe? / "(" (typedef_lit maybe?) 
                    ("," typedef_lit maybe?)* ")"
maybe <- "?"
argument_def <- "("( ( ident_typed / ident )?  
                ( "," ( ident_typed / ident ) )* ) ")"
rule_contents <- (let_stmt)* cond_expr "=>" act_expr (stmt)* 
\end{lstlisting}
Rule expression defines a rule that describes how the target repeater interacts with its qubits and other repeaters. Rule definition starts with \verb|"rule"| and \verb|ident| for the name of the rule. The identifier for the repeater \verb|repeater_ident| follows that with angle brackets (\verb|<|, \verb|>|) and a set of arguments within parenthesis (\verb|(|, \verb|)|).

To distinguish an ordinary value and a repeater value, the repeater value has a special identifier starting with \verb|#|. 
\begin{lstlisting}[language=PEG]
repeater_ident <- "#" ident
\end{lstlisting}
The target repeater can be referred to with the given identifier within the rule. 
A rule may promote a qubit value to the next rule. The \verb|ret_type_annotation| is required when the rule has \verb|promote| expression (see Section~\ref{sec:promote_stmt} in details).

The contents of the rule start with let statements that can define variables used within the rules.
The conditions \verb|cond_expr| and actions \verb|act_expr| follow them.
\begin{figure}[ht]
    \centering
    \includegraphics[width=\linewidth]{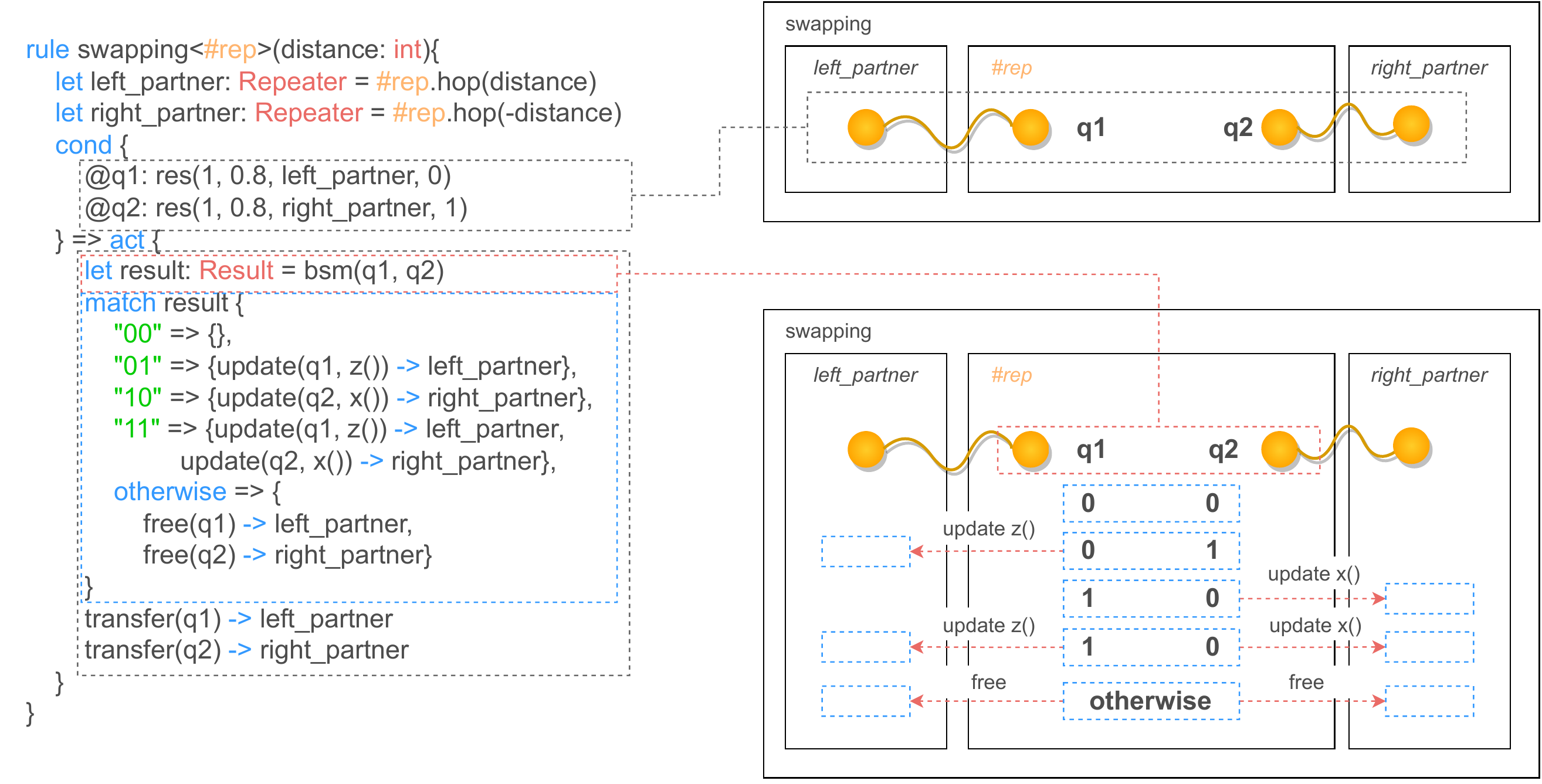}
    \caption[Rule Definition]{Rule definition follows Rule execution model}
    \label{fig:rule_def_model}
\end{figure}

\subsection{Condition Clauses}
\begin{lstlisting}[language=PEG, caption={PEG expressions for condition expression}, captionpos=b, label={lst:cond_def_peg}]
cond_expr <- ^"cond" "{" cond_clauses* "}"
cond_clauses <- res_assign / fn_call_expr / variable_call_expr 
res_assign <- "@" ident ":" fn_call_expr
\end{lstlisting}

Condition expression (\verb|cond|) contains multiple condition clauses to be evaluated. 
\begin{lstlisting}[caption={condition expressions}, captionpos=b, label={lst:cond_def}, language=RuLa, escapechar=|]
    cond {
        @q1: res(1, 0.8, left_partner, 0)  
        @q2: res(1, 0.8, right_partner, 1)
    }
\end{lstlisting}

\subsection{Action Clauses}
\begin{lstlisting}
act_expr <- ^"act" "{" stmt* "}"
\end{lstlisting}
\verb|act| expression describes the corresponding actions to be performed when the condition is met. Listing~\ref{lst:act_def} is an example that applies Bell State Measurement to \verb|q1| and \verb|q2| which are defined in the condition clause.
\begin{lstlisting}[caption={action expression}, captionpos=b, label={lst:act_def}, language=RuLa]
    // ...
    => act {
        let result: Result = bsm(q1, q2)
    }
\end{lstlisting}

\subsection{Repeater Definition}
\begin{figure}[h]
    \centering
    \includegraphics[width=\linewidth]{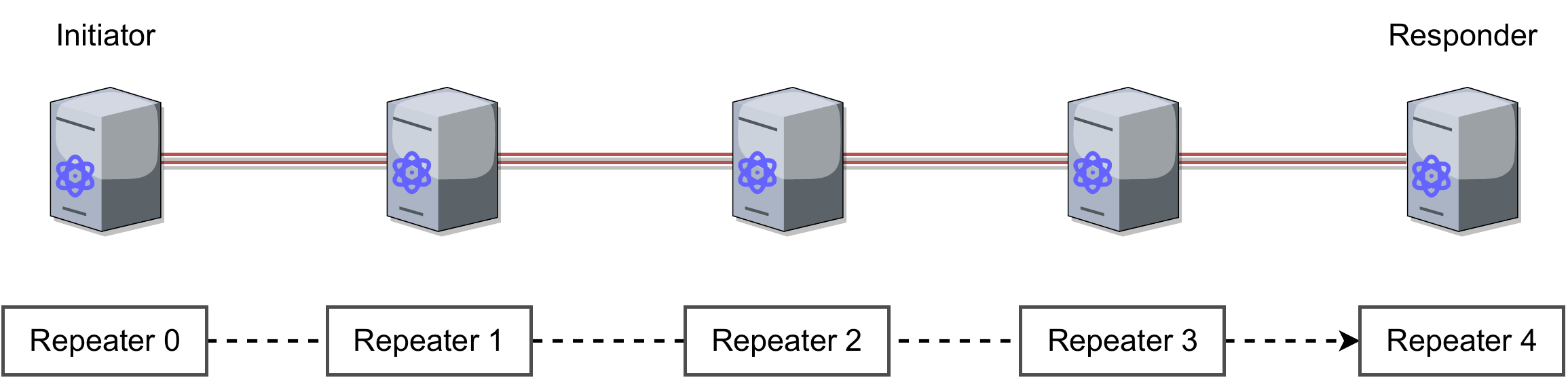}
    \caption[Repeater Indexing]{The repeaters are indexed from the initiator to the responder.}
    \label{fig:repeater_indexing}
\end{figure}

It is possible to specify each individual repeater node in the path in the RuLa program. The very first line of the RuLa program starts with \verb|#repeaters: vec[Repeater]| that contains the repeaters in the path. Figure~\ref{fig:repeater_indexing} shows how repeaters are indexed in the \verb|#repeaters|. 

Each repeater can be specified by the index of \verb|#repeaters| value. 
\begin{lstlisting}
repeater_call <- "#repeaters" "(" ( term_expr / ident / int ) ")"
\end{lstlisting}
For example, the first repeater can be called \verb|#repeaters(0)|. It is also possible to specify with some parameters (e.g. \verb|#repeaters(i+1)|) for dynamic repeater selection.

\subsection{Rule Call Expression}
Calling the defined rule is done similarly to the function call, but the rule call takes an additional argument of \verb|Repeater| value.
\begin{lstlisting}
rule_call_expr <- ident "<" repeater_call ">" 
                "(" fn_call_args? ("," fn_call_args)* ")"
\end{lstlisting} 
For example, the following expression prepares the \verb|swapping| rule for the first repeater.
\begin{lstlisting}
swapping<#repeaters(0)>(1)
\end{lstlisting}
The rule is evaluated with a given repeater value. 

\subsection{RuleSet Definition}
\begin{lstlisting}[language=PEG]
ruleset_stmt <- ^"ruleset" ident "{" stmt* "}"
\end{lstlisting}

\begin{lstlisting}[language=RuLa]
ruleset entanglement_swapping{
    for i in 1..#repeaters.len()-1{
        if (i % 2 == 1){
            swapping<#repeaters(i)>(1)
        }
    }
}
\end{lstlisting}

\subsection{Conditioning}
RuLa supports two types of conditioning to easily construct complex rules.

\vskip\baselineskip
\noindent\textbf{If Statement}

\begin{minipage}{\linewidth}
\begin{lstlisting}[language=PEG, caption={PEG expression for if expression}, label={lst:if_expr_peg}, captionpos=b] 
if_stmt <- ^"if"  "(" if_block ")" "{" stmt* "}" 
            else_if_stmt* else_stmt?
if_block <- get_expr / comp_expr / literal_expr 
else_if_stmt <- ^"else" ^"if" "(" if_block ")" "{" stmt* "}" 
else_stmt <- ^"else" "{" stmt* "}"
\end{lstlisting}
\end{minipage}

\verb|if| statement can take multiple \verb|else if| statement and one \verb|else| statement at the end.
\begin{lstlisting}[caption={if expression},captionpos=b, label={lst:if_def}, language=RuLa]
if(result == "0"){
    // do something here
    ...
}else if (result == "1"){
    // do something else here
}else{
    // if none of the above is executed, do something here
}
\end{lstlisting}
\verb|if_block| needs to return a Boolean value.
\verb|if| statement works differently whether it is defined in the \verb|rule| statement or \verb|ruleset| statement.

\vskip\baselineskip
\noindent\textbf{Match Statement}
\begin{lstlisting}[caption={PEG expression for match expression}, captionpos=b, language=PEG]
match_stmt <- ^"match" expr "{" (match_arm  ",")*
                (^"otherwise" "=>" match_action)? "}"
match_arm <- match_condition "=>" match_action 
match_condition <- satisfiable 
satisfiable <- literal_expr
match_action <- "{" stmt? ("," stmt)* "}"
\end{lstlisting}

Another way to define conditions is by using the expression \verb|match|. In the \verb|match| statement, the conditions are evaluated from top to bottom. If one of these conditions is satisfied, the corresponding actions following after \verb|=>| are executed. We can also define operations with \verb|otherwise| in the case where all conditions are not met.

\begin{lstlisting}[caption={match expression},captionpos=b, label={lst:match_def}, language=RuLa]
match result {
 "0" => {/* do actions corresponding to result == "0" */ },
 "1" => {/* do actions corresponding to result == "1" */},
otherwise => {/* do actions if both the condition doesn't match */}
 }
\end{lstlisting}

These conditioning expressions have a special ability for RuleSet generation. RuleSet originally does not allow any conditional operations in condition clauses. For example, quantum repeaters have to apply selective operations based on the Bell measurement results at other nodes. In RuleSet, those operations are achieved by preparing multiple rules in a stage.

\begin{figure}[ht]
    \centering
    \includegraphics[width=\textwidth]{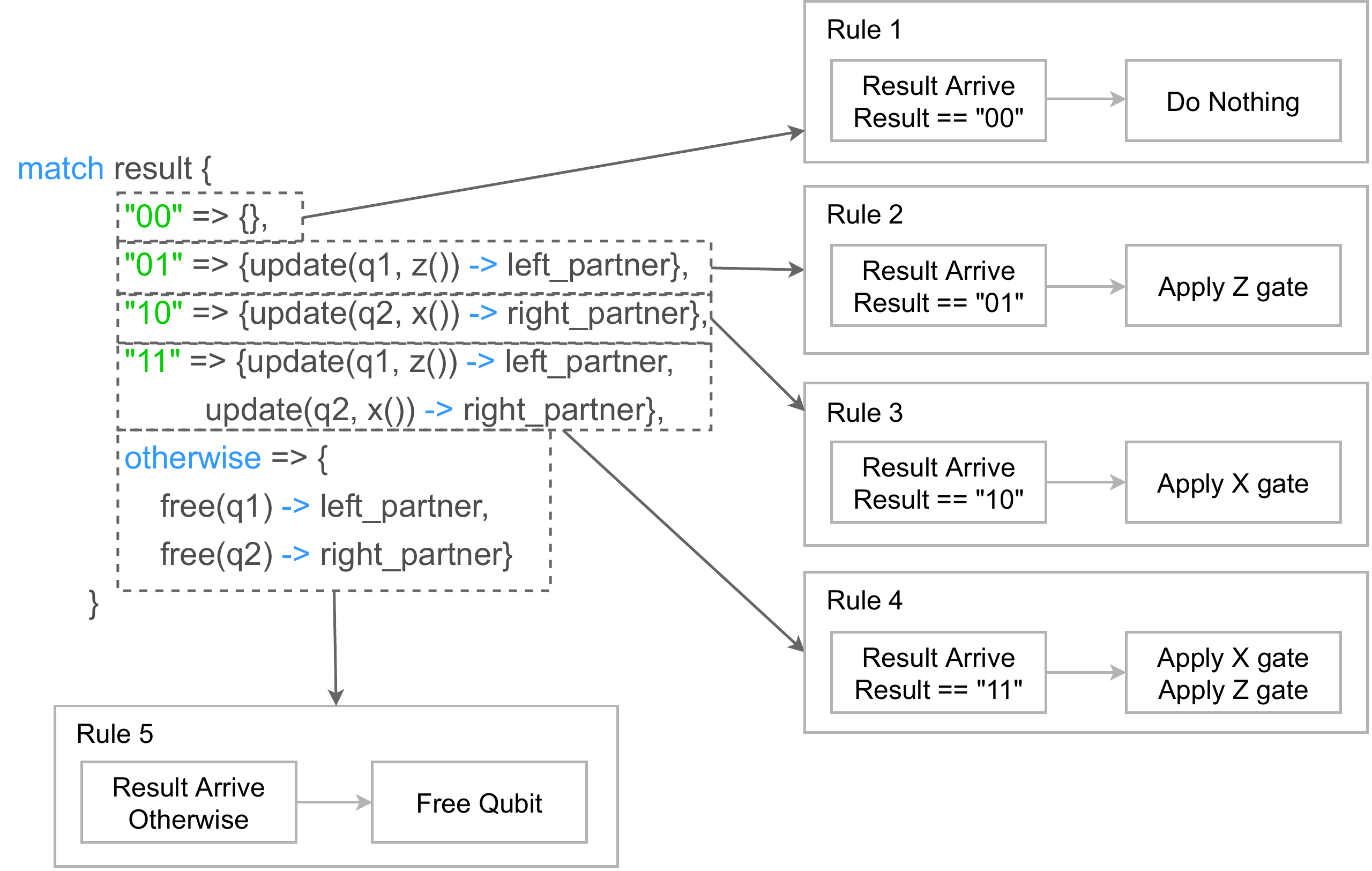}
    \caption[Match Expression in Rule Generation]{Auto expansion in a match expression.}
    \label{fig:auto_expantion}
\end{figure}

When multiple factors are considered in the conditions, the number of rules grows. In RuLa, these conditions are automatically expanded to combinations of rules, even if there are multiple conditions. An example is shown in Figure~\ref{fig:auto_expantion}

\subsection{Loop}
\begin{lstlisting}[language=PEG]
for_stmt <- ^"for" ( ident / for_multi_block ) 
            "in" for_generator "{" stmt* "}"
for_multi_block <- "(" ident_list ")"
for_generator <- series / expr
series <- int ".." expr
\end{lstlisting}
Looping is also an important functionality for most programming languages. We can define loop operations with \verb|for| expression.

Listing~\ref{lst:for_expr} is an example \verb|for| loop expression with a generated function. In this example, the numbers $1, 2, 3, 4, 5$ are substituted, and the statements inside the curly braces are evaluated. 

\begin{minipage}{\linewidth}
\begin{lstlisting}[language=RuLa, caption={Example for expression}, captionpos=b, label={lst:for_expr}]
for i in 1..5 {
    // do something here
    ...
}    
\end{lstlisting}
\end{minipage}

\subsection{Function Call Expression}
To call the functions defined in the other file or another place, the following function call expression can be used. 
\begin{lstlisting}
fn_call_expr <- (repeater_ident / ident)  "("fn_call_args? ("," fn_call_args)*")"
fn_call_args <- fn_call_expr / variable_call_expr / term_expr / literal_expr 
\end{lstlisting}
As an argument for the function, the four types of expression can be in parentheses.

\subsection{Send Statement}
In quantum repeater networks, message exchange often appears in the process. Especially shown in Table~\ref{tab:proto_message}, there are four types of defined protocol messages used to transfer a certain instruction and measurement results to the other quantum repeaters. 

RuLa supports this message transfer with the \verb|->| symbol.
\begin{lstlisting}
send_stmt <- fn_call_expr "->" expr
\end{lstlisting}
The function call goes to the left-hand side of the \verb|->| symbol. This function call needs to be one of four types of functions.
\begin{itemize}
    \item \verb|update(<resource id>, <pauli_correction>)|: Apply the Pauli correction operation to the resource with \verb|<resource id>|.
    \item \verb|free(<resource id>)|: Frees the resource whose resource id is \verb|<resource id>|.
    \item \verb|meas(<resource id>, <measurement_result>)|: Send the measurement result corresponding to the \verb|<resource id>|.
    \item \verb|transfer(<resource id>)|: Notify of successful entanglement swapping and transfer a new resource id.
\end{itemize}
The right-hand side of the send expression can be any expression but needs to be a value with \verb|Repeater| type.

This \verb|#repeaters| vector is resolved by the configuration given in the compilation (see the details in~\ref{sec:compiler}).

\subsection{Promoting Values}
\label{sec:promote_stmt}
\begin{lstlisting}[language=PEG]
promote_stmt <- ^"promote" promotable ("," promotable)*
promotable <- comp_expr / term_expr / vector / tuple / variable_call_expr / literal_expr
\end{lstlisting}
Promoting entangled resources is one important operation in the Rule definition. Right now, the only value we can promote is \verb|Qubit| type values. 

When a Rule contains \verb|promote| statement, that rule requires type annotation.
\begin{lstlisting}[language=RuLa, caption={Example type annotation}, label={lst:type_annotation}, captionpos=b]
rule test_rule<#rep>() -> Qubit?{
    cond {
        // ... definition of qubit
    } => act {
        // ...
        if (result == "0"){
            promote qubit
        }else{
            free(qubit)
        }
    }
}
\end{lstlisting}
Listing~\ref{lst:type_annotation} shows an example type annotation when the rule contains a \verb|promote| statement. In the action clauses \verb|act{...}| there is a \verb|if| statement that could promote a qubit value.


%% file: Chapters/4.Proposal/4.5.Compiler.tex
\section{Compiler}
\label{sec:compiler}
\begin{figure}[ht]
    \centering
    \includegraphics[width=\linewidth]{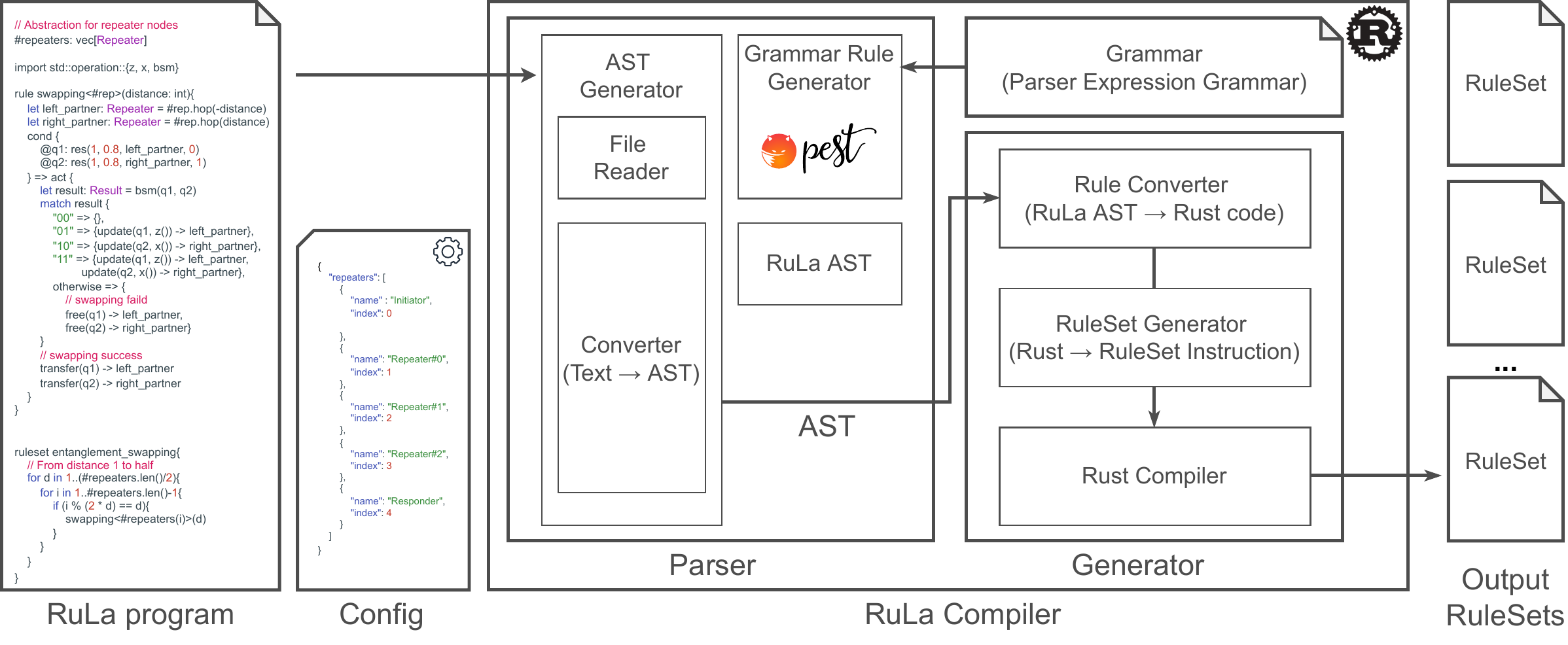}
    \caption[RuLa Compiler Design]{RuLa Compiler Design}
    \label{fig:compiler_design}
\end{figure}

The RuLa program needs to be translated into basic RuleSet instructions. Figure~\ref{fig:compiler_design} is an overview of the RuLa compiler architecture. 

\subsection{Input/Output}
The first input for this compiler is the source code written in RuLa. The file index of the RuLa program should be \verb|.rula|, and the input program is passed to Parser. The second input is a configuration file that contains a set of properties of quantum repeaters. 
Listing~\ref{lst:repeater_config} is an example configuration of repeaters.

\begin{minipage}{\linewidth}
\begin{lstlisting}[captionpos=b, caption={An example configuration file in JSON format}, language=json, label={lst:repeater_config}]
{
    "repeaters": [
        {
            "name" : "#1",
            "address": 0

        },
        {
            "name": "#2",
            "address": 1
        },
        {
            "name": "#3",
            "address": 2
        },
        ...
    ]
}
\end{lstlisting}
\end{minipage}
The configuration file is converted into a vector of \verb|Repeater| values that provide some useful functions to specify the partners. As an example, it is possible to select the partner quantum repeaters with the number of hops generated by this configuration. 
\begin{figure}
    \centering
    \includegraphics[width=\textwidth]{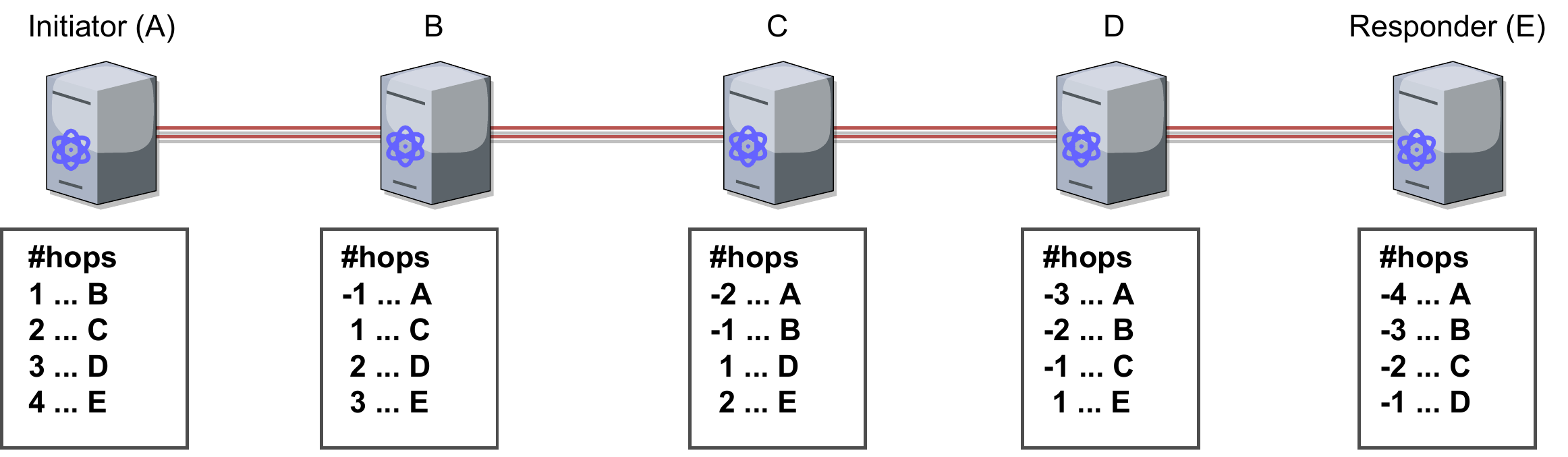}
    \caption[Repeater Config Translation]{Automatically generated the number of hops between the quantum repeaters.}
    \label{fig:repeater_config_resolve}
\end{figure}
For example, if the repeater B in Figure~\ref{fig:repeater_config_resolve} wants to specify the repeater A as a partner node, it is possible to do it with
\begin{lstlisting}[language=RuLa]
let repeater_A: Repeater = #rep.hop(-1)
\end{lstlisting}
where \verb|#rep| is the variable used in the Rule to specify the target repeater (in this example, the target repeater is repeater B).

The output files are RuleSet which is serialized in JSON format. If the configuration file contains properties of five repeaters, the number of output files is also five and each RuleSet is corresponding to each repeater.

\subsection{Parser}
There are two important components in the RuLa compiler. One is a parser and the other is a code generator. The parser generates Abstract Syntax Tree (AST) from an input RuLa program. AST nodes are defined based on the grammar introduced in Section~\ref{sec:grammar_def}.
\begin{figure}
    \centering
    \includegraphics[width=\linewidth]{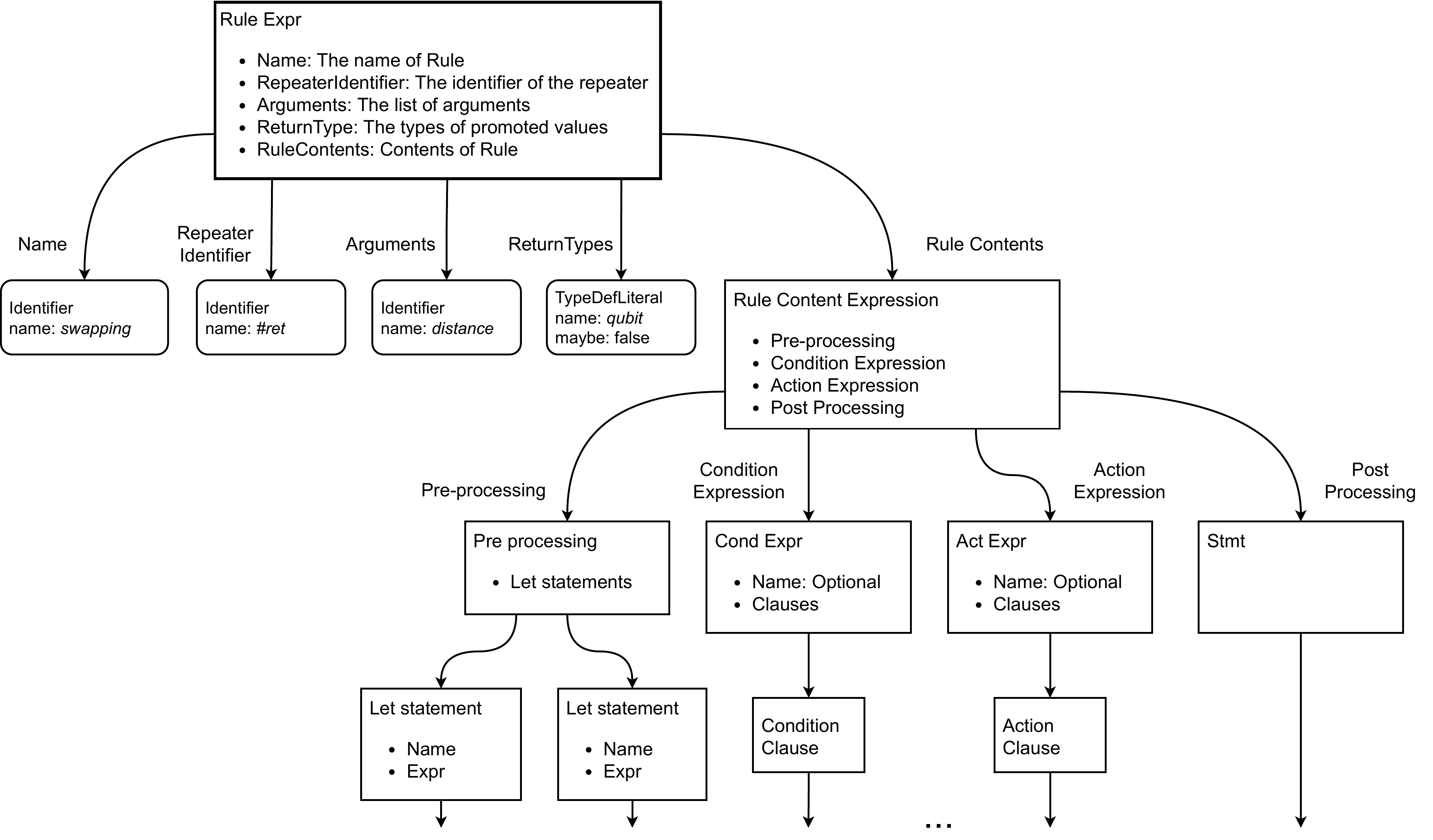}
    \caption[Example AST]{An example AST construction for Rule Expression}
    \label{fig:rule_def_ast}
\end{figure}

Figure~\ref{fig:rule_def_ast} is an example of an AST structure. The definition of the rule starts from \verb|RuleExpr| which consists of four different values.
\begin{itemize}
    \item \textit{Name}: The identifiable name of the rule.
    \item \textit{Repeater Identifier}: The repeater of the identifier starts with \verb|#|.
    \item \textit{Arguments}: Similarly to function arguments, the rule definition can also take some arguments that are passed by when the rule is evaluated. Those arguments are also a series of identifiers.
    \item \textit{RuleContents}: Finally, the actual content of the rule is expressed by \verb|RuleContentExpr| which is made up of several important components for the definition of the rule.
\end{itemize}
The entire RuLa program is translated into a single large tree that is composed of multiple AST nodes. 

\subsection{Generator}
RuLa generator translates the AST given by the parser into the Rust code~\cite{matsakis2014rust}. Each statement and expression is corresponding to some pieces of Rust code that can generate RuleSet instructions. For example, the \verb|if| statement is translated into a set of closures that describes what is the conditions to be met and the actions to be performed. Some built-in functions such as \verb|send()|, \verb|free()| are also converted into a set of RuleSet instructions. 

This generator first generates Rust code and the generated Rust code is compiled by the Rust compiler. Some serialized RuleSets are generated as outputs of the compilation.

%% file: Chapters/4.Proposal/4.6.Example.tex
\section{Examples}
In this section, we introduce several examples of the RuLa programs and configurations.
\subsection{Simple Entanglement Swapping}
The first example is the simple entanglement swapping protocol shown in Listing~\ref{lst:swapping_rule}.
The program starts with repeater vector definition in line~\ref{ln:repeater_def} which stores repeater values. Line~\ref{ln:importing_functions} imports quantum operation functions Pauli Z gate (\verb|z()|), Pauli X gate (\verb|x()|) and Bell State Measurement (\verb|bsm()|).

A \verb|swapping| rule is defined between line~\ref{ln:ruledef_start} and line~\ref{ln:ruledef_end}.
\verb|swapping| rule takes a repeater argument (\verb|#rep|) and an integer argument \verb|distance|.
Line~\ref{ln:left_partner_def} and ~\ref{ln:right_partner_def} define the target partner quantum repeaters with a function called \verb|hop| function which is implemented in the \verb|Repeater| value.

Condition clauses are defined between line~\ref{ln:condition_def_start} and line~\ref{ln:condition_def_end}. A qubit returned from \verb|res| function is stored in a variable \verb|q1| that can be used in action clauses.

Action clauses are defined from line~\ref{ln:action_def_start} to \ref{ln:action_def_end}. Firstly, perform Bell State Measurement in line~\ref{ln:bsm}. According to the measurement result, send different messages to the partner repeaters (line~\ref{ln:result_check_start} to line~\ref{ln:result_check_end}).

\begin{minipage}{\linewidth}
\begin{lstlisting}[language=RuLa, label={lst:swapping_rule}, escapechar=|]
// Abstraction for repeater nodes
#repeaters: vec[Repeater]|\label{ln:repeater_def}|

import std::operation::{z, x, bsm}|\label{ln:importing_functions}|

rule swapping<#rep>(distance: int){|\label{ln:ruledef_start}|
    let left_partner: Repeater = #rep.hop(-distance)|\label{ln:left_partner_def}|
    let right_partner: Repeater = #rep.hop(distance)|\label{ln:right_partner_def}|
    cond {|\label{ln:condition_def_start}|
        @q1: res(1, 0.8, left_partner, 0)  
        @q2: res(1, 0.8, right_partner, 1)
    } |\label{ln:condition_def_end}|=> act {|\label{ln:action_def_start}|
        let result: Result = bsm(q1, q2)|\label{ln:bsm}|
        match result {|\label{ln:result_check_start}|
            "00" => {},
            "01" => {update(q1, z()) -> left_partner},
            "10" => {update(q2, x()) -> right_partner},
            "11" => {update(q1, z()) -> left_partner,
                     update(q2, x()) -> right_partner},
            otherwise => {
                // swapping faild
                free(q1) -> left_partner,
                free(q2) -> right_partner}
        }|\label{ln:result_check_end}|
        // swapping success
        transfer(q1) -> left_partner
        transfer(q2) -> right_partner
    }|\label{ln:action_def_end}|
}|\label{ln:ruledef_end}|

ruleset entanglement_swapping{
    for d in 1..(#repeaters.len()/2){
        for i in 1..#repeaters.len()-1{
            if (i % (2 * d) == d){
                swapping<#repeaters(i)>(d)
            }
        }
    }
}
\end{lstlisting}
\end{minipage}

\begin{figure}[h]
    \centering
    \includegraphics[width=\linewidth]{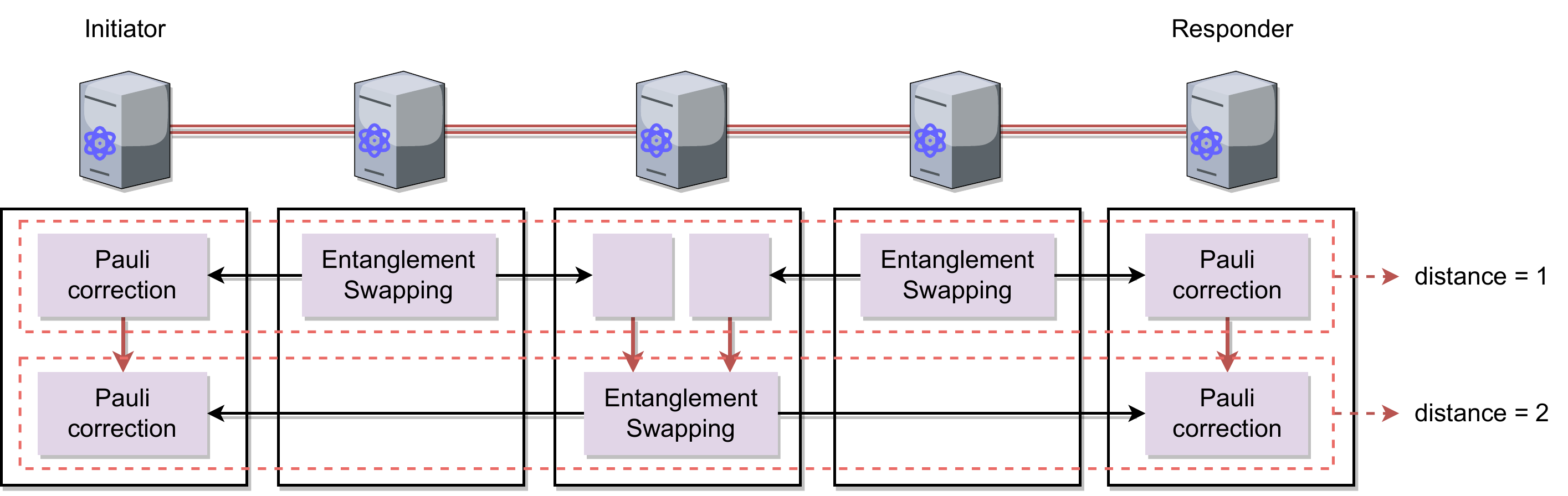}
    \caption[Example Usage]{A diagram of RuleSets generated in Listing~\ref{lst:swapping_rule}.}
    \label{fig:example_rule_generation_1}
\end{figure}

Figure~\ref{fig:example_rule_generation_1} is the RuleSet generated by a RuLa program in Listing~\ref{lst:swapping_rule}.

\subsection{Entanglement Swapping with Purification}
The following example (Listing~\ref{lst:pur_with_swap}) shows an example program that allows an initial purification before the entanglement swapping rule.
\begin{lstlisting}[language=RuLa, caption={An example RuleSet with a single link-level purification}, captionpos=b, label={lst:pur_with_swap}, escapechar=|]
#repeaters: vec[Repeater]|\label{ln:pur_repeater_def}|

import std::operation::{cx, measure}
import (rule) entanglement_swapping::swapping|\label{ln:pur_rule_import}|

rule local_operation<#rep>(distance: int) :-> Qubit {|\label{ln:local_operation_start}|
    let partner: Repeater = #rep.hop(distance)
    cond {
        @q1: res(1, 0.8, partner, 0)
        @q2: res(1, 0.5, partner, 1)
    } => act {
        // do cx
        cx(q1, q2)
        // measure the target qubit
        let result: Result = measure(q2, "Z")
        meas(q2, result) -> partner 
        set result as self_result
        promote q1 
    }
}|\label{ln:local_operation_end}|

rule parity_check<#rep>(distance: int, promoted: Qubit) :-> Qubit? {
    let partner: Repeater = #rep.hop(distance)
    cond {
        @message: recv(partner)
    } => act {
        if(message.result == get self_result){
            // purification success
            promote promoted
        }else{
            // purification failed (This can be an expression)
            free(promoted)
        }
    }
}

ruleset purification {
    // initial purification
    for i in 0..#repeaters.len()-1{
        let promoted_qubit: Qubit = local_operation<#repeaters(i)>(1)
        let promoted1: Qubit = parity_check<#repeaters(i)>(1, promoted_qubit)
        let promoted_qubit2: Qubit = local_operation<#repeaters(i+1)>(-1)
        let promoted2: Qubit = parity_check<#repeaters(i+1)>(-1, promoted_qubit2)
    }
    // entanglement swapping
    for d in 1..(#repeaters.len()/2){
        for i in 1..#repeaters.len()-1{
            if (i % (2 * d) == d){
                swapping<#repeaters(i)>(d)
            }
        }
    }
} |\label{ln:ruleset_pur_def_end}|
\end{lstlisting}

The second example is entanglement swapping with a single round purification.
Same as the previous example, the program starts with a definition of the repeater vector in line~\ref{ln:pur_repeater_def}. 

In this example, the \verb|swapping| rule defined in the different file is imported in line~\ref{ln:pur_rule_import}.

A \verb|local_operation| rule is defined from line~\ref{ln:local_operation_start} to line~\ref{ln:local_operation_end}. Same as the previous example, this rule takes a single repeater argument and an integer argument. However, since this rule contains \verb|promote| statement that promotes a qubit to the next rule, the type annotation is put after the argument definition.

This example only contains one round of purification, but it is also possible to add multiple rounds of purification to the RuleSet before swapping by adding one more loop. 
\begin{figure}
    \centering
    \includegraphics[width=\linewidth]{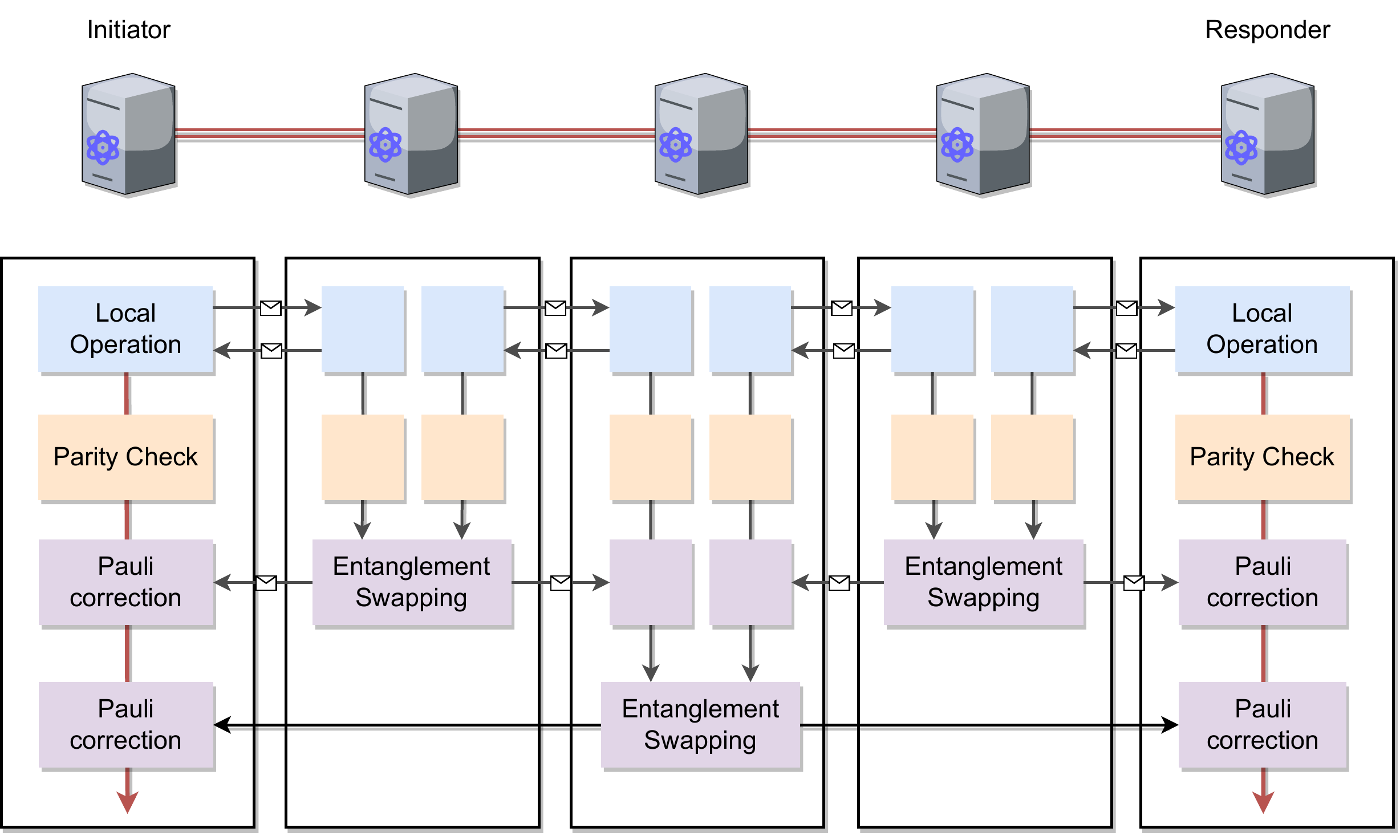}
    \caption[RuleSet with purification and entanglement swapping]{A diagram of the RuleSet generated in Listing~\ref{lst:pur_with_swap}.}
    \label{fig:ruleset_pur_swap}
\end{figure}

\subsection{Entanglement Swapping with Multiple Round Purification}
It is also possible to assume that all purification is done after the end-to-end entanglement is generated.

\begin{minipage}{\linewidth}
\begin{lstlisting}[language=RuLa]
ruleset purification { 
    // swappings
    for d in 1..#repeaters.len()/2 +1 {
        for i in 1..#repeaters.len() -1{
            if (i % 2 == 0){
                swapping<#repeaters(i)>(d)
            }
        }
    }
    // final purification
    let (promoted_qubit: qubit, result_i: str) = local_operation<#repeaters(0)>(#repaters.len()-1)
    parity_check<#repeaters(0)>(promoted_qubit, result_i)
    let (promoted_qubit2: qubit, result_next: str) = local_operation<#repeaters.last()>(#repeaters.len()-1)
    parity_check<#repeaters(i+1)>(promoted_qubit2, result_next)
}
\end{lstlisting}
\end{minipage}

%% file: Chapters/5.Evaluation/5.0.Evaluation.tex
\chapter{Evaluation}
\label{chap:evaluation}
In this chapter, we evaluate our proposal in terms of three aspects. 
First, we evaluate the requirements introduced ~\ref{sec:requirements} and check how RuLa satisfies those requirements in Section~\ref{sec:requirement_satisfaction}.
Second, we explain the limitations of current RuLa in Section \ref{sec:limitations}. Finally, we enumerate several use cases of RuLa in practical situations in Section~\ref{sec:usecase}.
\input{Chapters/5.Evaluation/5.1.ReqSatisfaction.tex}
\input{Chapters/5.Evaluation/5.2.Limitations.tex}
\input{Chapters/5.Evaluation/5.3.Usecases.tex}

\input{Chapters/5.Evaluation/5.4.Validity.tex}

%% file: Chapters/5.Evaluation/5.1.ReqSatisfaction.tex
\section{Requirement Satisfaction}
\label{sec:requirement_satisfaction}
This section discusses whether the requirements listed in Section~\ref{sec:requirements} are satisfied in RuLa and the coverage of the basic instructions shown in Table~\ref{tab:condition_clauses}, \ref{tab:action_clauses} and \ref{tab:proto_message}.

\subsection{Requirement 0: Instruction Coverage}
The primitive instructions introduced in Section~\ref{sec:ruleset_def} must be covered by some elements in RuLa such as functions, expressions, and statements.

\vskip\baselineskip
\noindent\textbf{Condition Clauses}

\noindent\textbf{RES}: 
\verb|res(<# of resource: u_int>, <fidelity: float>,| \\ \verb|<partner repeater: Repeater>, <index: u_int>)| 

\verb|res| function takes four arguments. The first argument is an integer value to count the number of required resources. The second argument is a float value for threshold fidelity value. The third argument is a \verb|Repeater| value that represents a partner repeater node. The final argument is an index value for the resource. This function returns a \verb|Qubit| type reference to the qubit.

\vskip\baselineskip
\noindent\textbf{TIMER}: 
\verb|check_timer(<timer_id: str>)| This function generates the condition \verb|TIMER| to check if the timer has expired or not. Right now, the timer id is given by some string identifiers.

\vskip\baselineskip
\noindent\textbf{CMP}: 
\verb|cmp(<target: Result>, <operation: str>, <value: str>)| 
\verb|cmp| function takes three arguments. The first argument is the \verb|Result| value that can be compared to some values. \verb|operation| is a set of comparison operators (e.g. \verb|==|, \verb|>|).

\vskip\baselineskip
\noindent\textbf{RECV}: 
\verb|recv(<partner repeater: Repeater>)|
\verb|recv| function takes a repeater value and makes an \verb|RECV| instruction based on the repeater information.

\vskip\baselineskip
\noindent\textbf{Action Clauses}

\vskip\baselineskip
\noindent\textbf{SETTIMER}: 
\verb|set_timer(<timer_id: str>, <duration: int>)| 
\verb|set_timer| function set a new timer with an integer duration value. A timer can be identified by a string value \verb|timer_id|.

\vskip\baselineskip
\noindent\textbf{PROMOTE}:
\verb|promote| statement introduced in Section~\ref{sec:promote_stmt} is corresponding to the \verb|PROMOTE| instruction that promotes a qubit to the next rule. 

\vskip\baselineskip
\noindent\textbf{FREE}:
\verb|free(<resource_id: Qubit>)| 
\verb|free| function takes a \verb|Qubit| value as an argument and issues an instruction to free that qubit.

\vskip\baselineskip
\noindent\textbf{SET}:
\verb|set| statement explained in 
SET instruction is supported by the \verb|return| expression. Once a Rule returns the value, the parent RuleSet stores the value and provides it to the other Rules. This operation corresponds to the SET operation in RuleSet.

\vskip\baselineskip
\noindent\textbf{MEAS}:
Measurement operations are defined in the standard library for RuLa.

\verb|measure(<qubit: Qubit>, <measurement_basis: str>)| is a function to make a measurement instruction. The first argument is a \verb|Qubit| value that identifies the target qubit and the second argument is the measurement basis that can be \verb|X|, \verb|Y| or \verb|Z|

\vskip\baselineskip
\noindent\textbf{QCIRC}: 
A quantum gate is also defined as a function in the standard library.
For example, \verb|x(<qubit: Qubit>)| make a \verb|QCIRC| instruction with a qubit identifier.

\vskip\baselineskip
Protocol messages are defined as \verb|Send| expressions with the \verb|->| symbol.
\begin{itemize}
    \item \verb|free(<resource_id>) -> repeater| frees the target resource in the \verb|repeater|.
    \item \verb|update(<resource_id>, <pauli_correction>) -> repeater| notify the result of entanglement swapping to the \verb|repeater| with corresponding Pauli correction operations.
    \item \verb|meas(<resource_id>, <measurement_result>) -> repeater| send the raw measurement results to the \verb|repeater|.
    \item \verb|transfer(<resource_id>) -> repeater| tell the \verb|repeater| the successful entanglement swapping.
\end{itemize}
\subsection{Requirement 1: Visibility Enhancement}

RuLa supports the following functionalities to enhance the visibility of the entire Rule process.

\vskip\baselineskip
\noindent\textbf{Modular Rule Definition}: RuLa supports \verb|import| statement to get Rules defined in different files into the current RuleSet definition. This \verb|import| statement improves the maintainability of the Rule definition. Users do not have to define all the required Rules in a single file that lose the visibility of the entire program.

\vskip\baselineskip
\noindent\textbf{Clear Rule definition process}: The way to define the Rule is clearly divided into four steps as explained in Section~\ref{sec:rule_def}. It is possible to see who is the owner of the Rule and who is the partner of the Rule easily and also possible to check what are the conditions and actions.

\verb|Repeater| type values also provide some useful functions that contribute entire visibility, such as \verb|#rep.hop()| to identify the repeater placed in a specific number of hops from the owner repeater. 

\vskip\baselineskip
\noindent\textbf{RuleSet definition with Rule Call}:
In the \verb|ruleset| statement, the rules defined in the Rule definition process can be called in the RuleSet definition process. This allows us to see the order of the Rules and the relationship between the Rules. It is also possible to see the flow of resources from one Rule to the next Rule.

\subsection{Requirement 2: Consistency Support}
The consistency of the Rule is automatically supported in the RuLa compiler. \verb|send| statement automatically resolves the two dependent Rules.

For example, a piece of code shown in Listing~\ref{lst:send} shows a \verb|send| statement that sends a \verb|free| message to a \verb|repeater| with a qubit identifier. 

\begin{lstlisting}[language=RuLa, label={lst:send}]
// repeater is a Repeater type variable.
free(qubit) -> repeater
\end{lstlisting}

This statement is translated into the two different RuleSet instructions. The first instruction is a \verb|send| instruction for the owner of this Rule. The second instruction is a \verb|wait| instruction for the partner repeater.

\subsection{Requirement 3: Expressibility Enhancement}
RuLa allows us to generate RuleSets clearly and easily in terms of the following four aspects.

\vskip\baselineskip
\noindent\textbf{Condition expantion}
In RuLa, \verb|match| and \verb|if| expressions are automatically expanded to the series of Rules within a single Stage. This allows us to set multiple complex conditions easier than defining each of them. For example, if there are two match expressions with four conditions in a Rule, the number of Rules at the end of two match expressions is $4\times4 = 16$.

\vskip\baselineskip
\noindent\textbf{Dynamic Repeater Selection}
RuLa supports natively the \verb|Repeater| type and useful methods for repeater values. This allows us to specify the partner repeater nodes dynamically and easily. 

\vskip\baselineskip
\noindent\textbf{Type checks for promoted values}
There might be a promotion operation in a Rule. The type of promoted value and given value from the previous rule need to be consistent. RuLa supports type checks for the promoted values and if there are inconsistent promotions between the Rules, RuLa spits out compile errors.

%% file: Chapters/5.Evaluation/5.2.Limitations.tex
\section{Limitations}
\label{sec:limitations}
In this project, we made some assumptions that should be removed in the future. These assumptions result in the limitations of RuLa.

\subsection{RuleSet Generation}

We enumerate several limitations on the RuleSet generation process.

\textbf{Limitation 1: RuleSet is done statically}:
Since all RuleSets are created in the responder node, the responder node must know the intermediate device topologies before the RuleSet creation. Thus, a specific path must be established with some policies, and that path cannot be changed until the entanglement generation is done. This assumption might lose the robustness of the entire process when an intermediate repeater had some trouble with the generation process. The biggest issue with this assumption is that we cannot flexibly change the number of rounds of purification and error correction based on the required fidelity and the state of the fiber.

In terms of the operations inside the RuleSet, it is possible to give some conditions for semi-dynamic operations (e.g. do measurements based on the result). However, the RuleSet needs to describe all possible combinations of operations to be fully functional.

\textbf{Limitation 2: The primary target state is the bipartite state}:
Related to limitation 1, the current target quantum state is all bipartite state because the multipartite state might require a more complex routing table. It is still possible to handle the GHZ state in a straight line without any branching but it is difficult to generate a complex graph state in the current system. Especially in the network-level entangled states, this assumption is fairly acceptable in the near future. However, there will be a demand for multipartite state communication in the future.

\subsection{Optimizations}

We consider possible future optimizations.

\begin{figure}[ht]
    \centering
    \includegraphics[width=\textwidth]{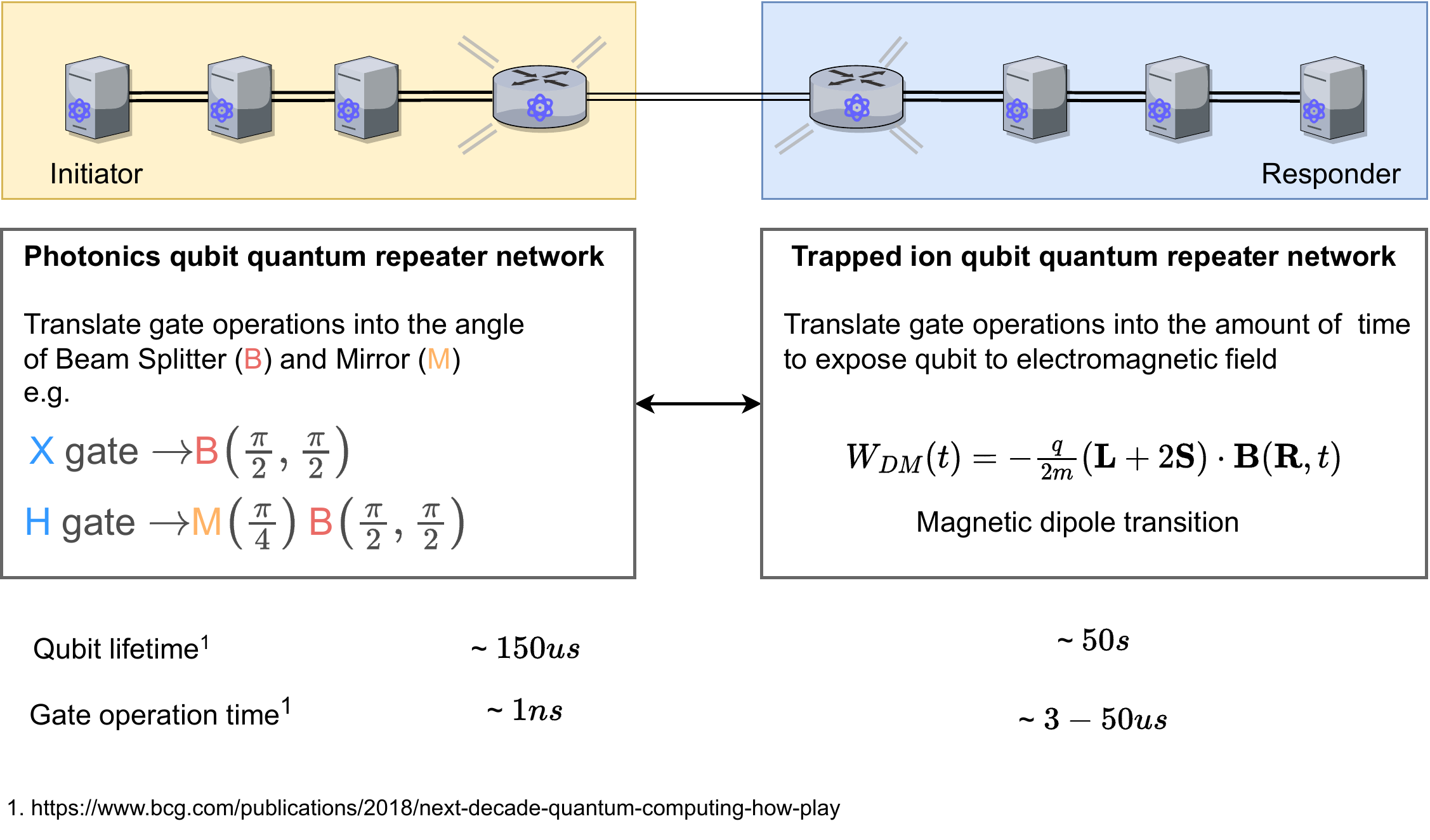}
    \caption[RuleSet in a Heterogeneous Network]{Translation of instructions over the different physical systems. Abstract gate-level instructions are translated into lower-level instructions. For example, in the case of linear optical quantum repeaters, X gate operation is translated into Beam splitter angles. }
    \label{fig:ruleset_in_heterogeneous}
\end{figure}

\textbf{Optimization 1: RuleSet generation for heterogeneous quantum network}
When it comes to heterogeneous quantum networks as shown in Figure~\ref{fig:ruleset_in_heterogeneous}, there might be operation time differences between the different physical systems. For example, the gate time of the linear optical quantum repeater network tends to be faster than that of any other physical system. When we think of the combination of linear optical quantum repeaters and trapped ion quantum repeaters, the linear optical quantum repeaters have to wait for trapped ion repeaters to finish entanglement generation.
The possible direction of optimization to solve this problem is to put some \verb|TIMER| operations to align the operations. 

\textbf{Optimization 2: Link property adaptive RuleSet generation}
In the current RuleSet scheme, it is not possible to generate dynamically by considering the link cost between repeaters. For example, if there is a bad link whose fidelity of the link-level entanglement is 0.8 and there is another link whose fidelity of the link-level entanglement is 0.95. 
In this case, the number of required entanglement purification would be different (e.g. the first link requires three rounds of purification and the second link requires one round of purification). 
Their error characteristics are also different (e.g. the first link-level entanglement tends to have X error but the second link-level entanglement does not). 

One possible solution to deal with this situation is to use some variables that represent link-specific parameters and process them in the RuLa compiler. 

%% file: Chapters/5.Evaluation/5.3.Usecases.tex
\section{Use Cases}
\label{sec:usecase}
In this section, we discuss several use cases of RuLa. 

\begin{figure}[ht]
    \centering
    \includegraphics[width=\textwidth]{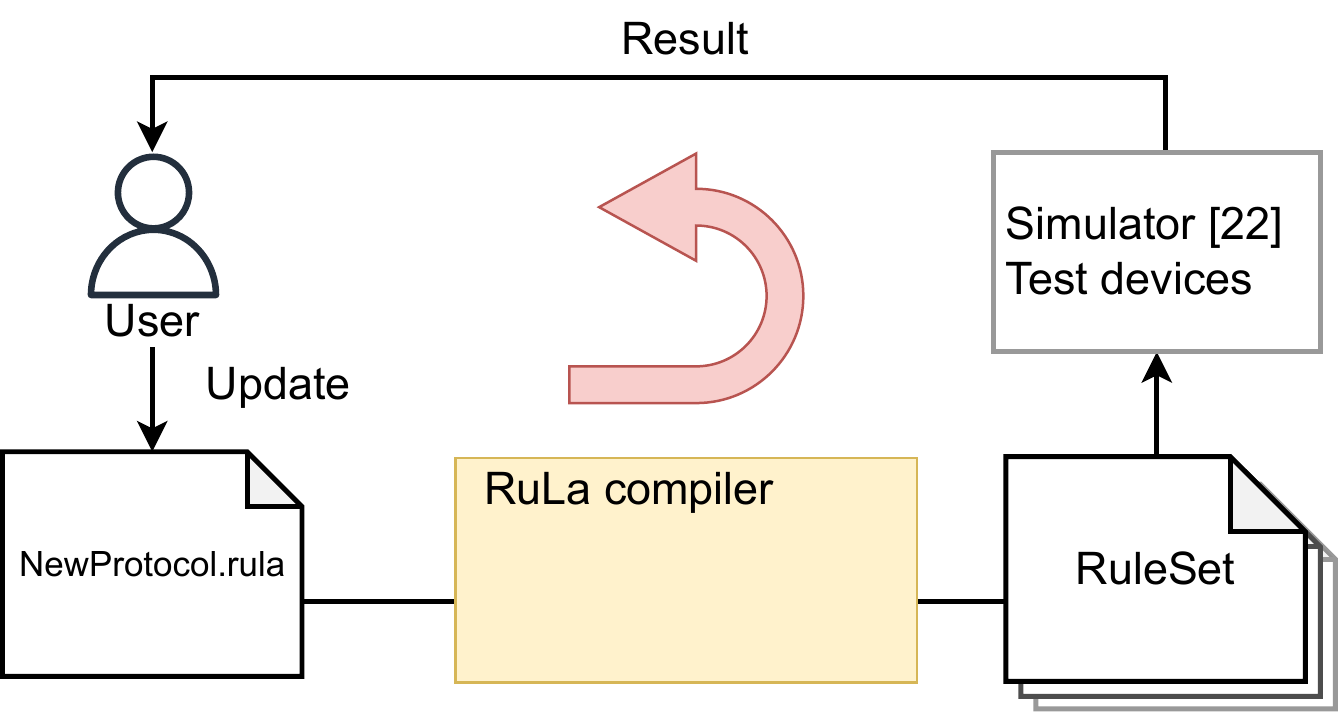}
    \caption[Testing a New Protocol with RuLa]{Testing a new protocol with RuLa. 1. Users create a RuLa program that describes a new protocol. 2. Generate RuleSets with RuLa compiler. 3. Put the generated RuleSet into the simulator or testing devices. 4. Get the result and update the protocol.}
    \label{fig:protocol_testing}
\end{figure}

\textbf{Use case 1: Protocol testing}: The first use case is to accelerate protocol development for quantum networking by making the RuleSet creation process easier (Figure~\ref{fig:protocol_testing} shows a cycle of protocol creation with RuLa).
The first step of protocol testing is to create a RuLa program that describes how a new protocol works. After writing the program, users prepare a configuration and give it to the RuLa compiler with the written program. RuLa compiler gives users the ability to deserialize RuleSets that can be tested on the RuleSet-based simulator (e.g. QuISP~\cite{quisp2022}) or actual devices that support RuleSet execution. As a result of the execution, users could obtain the final fidelity of the generated quantum states or other performance indices that can be used to evaluate the performance of the protocol.
Based on those performance tests, users might come up with a better structure of protocols and test again without touching the raw RuleSets.

\begin{figure}[h]
    \centering
    \includegraphics[width=\textwidth]{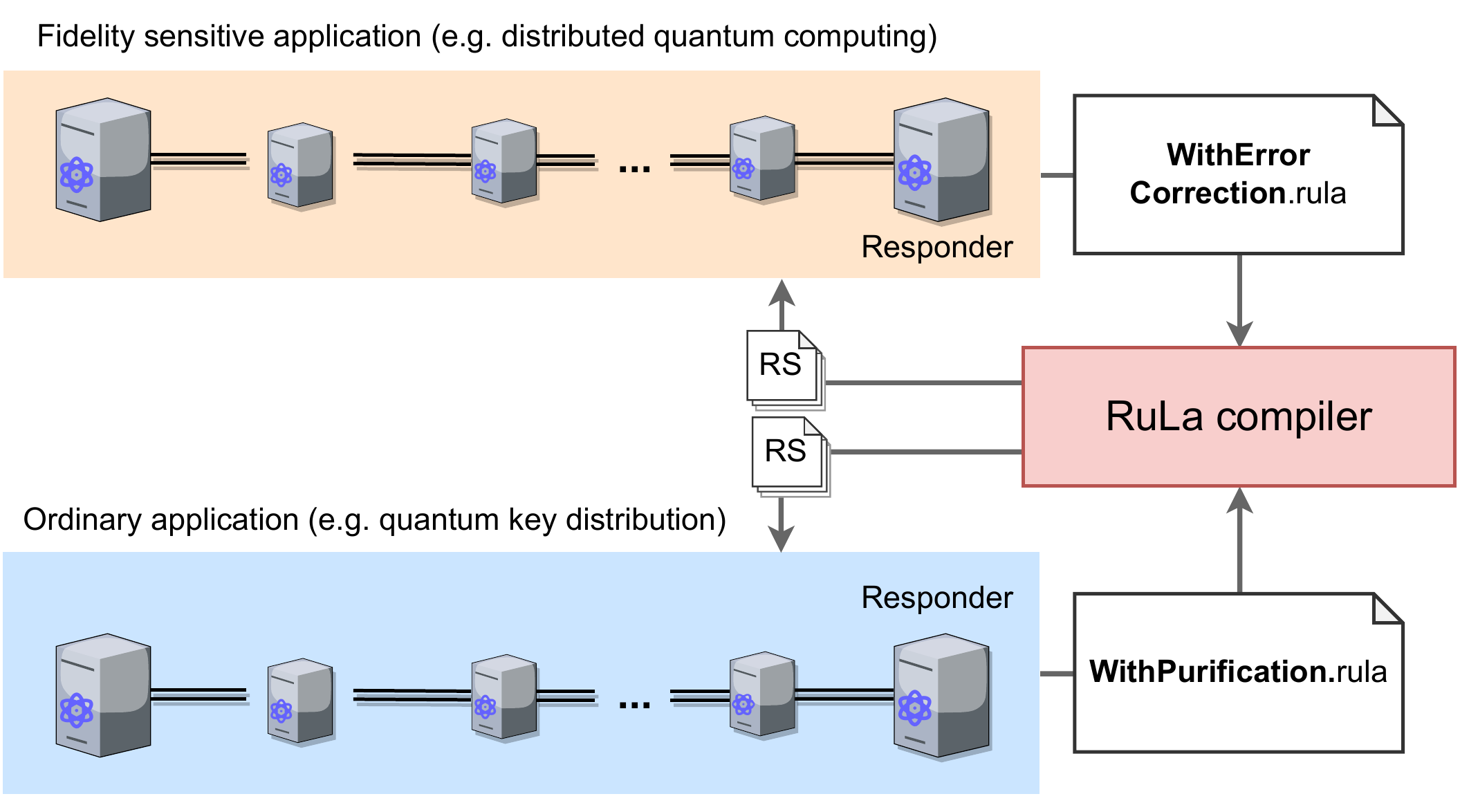}
    \caption[Network Management]{Network Management with RuLa}
    \label{fig:network_management}
\end{figure}

\textbf{Use case 2: Network management}: 
The second use case is network management, shown in Figure~\ref{fig:network_management}.
Depending on the type of application, we might prepare for entanglement with different fidelities. For example, QKD might require 98\% fidelity Bell pairs to increase the key generation rate, but DQC might require Bell pairs with 99.999\% fidelity to ensure greater reliability. This strategy differs depending on the applications. In that case, it takes time to generate the corresponding RuleSets from scratch. 

If the responder node has multiple RuLa programs according to the applications or the fidelity of the states, the RuLa program works as if the policy of the entanglement generation. Responder receives the intermediate repeater information and recompiles the prepared RuLa programs with the configuration and distributes generated RuleSets to the intermediate repeaters.

%% file: Chapters/5.Evaluation/5.4.Validity.tex

%% file: Chapters/6.Conclusion/6.0.Conclusion.tex
\chapter{Conclusion}
\label{chap:conclusion}
\section{Conclusion}
In this thesis, we presented a new programming language RuLa for RuleSet-based quantum repeaters. RuLa is the first programming language for RuleSet-based quantum repeaters, which allows us to create RuleSets in a very straightforward way. We formulate the requirements for the RuLa as these three things. 

The first requirement is reliable and scalable RuleSet generation. RuLa supports higher-level conditioning and looping which allow us to generate many Rules at the same time. There is also a way to specify a partner repeater dynamically. These functionalities give us the ability towards scalable RuleSet generation.

The second requirement is the coverage of the RuleSet instructions. RuLa covers all the instructions supported in the RuleSet with statements, expressions, and built-in functions.

The third requirement is that RuLa should make RuleSet definition clearer and easier. In terms of clearance, RuLa follows the order of the Rule execution. This means that it is possible to understand the programs in the way that is executed.

We also mentioned the limitation of the current RuLa. In RuLa, since all the information for the repeaters is given before the execution, no dynamic RuleSet generation can be done such as deciding operations based on the current fidelity of the link. However, it is still possible to prepare several RuLa programs in one place and choose one of them based on the requirements for the application. 
In addition, RuLa does not care about the lower-level physical architectures which would be another source of errors and increase the waiting time. These limitations are the tasks that should be solved in a future version.

Finally, we also enumerated two use cases of RuLa. The first use case is to accelerate the testing of a new protocol. Since writing RuleSet became easier, it is possible to test and update a RuleSet for a new protocol easier and faster. The second use case is the selective RuleSet generation based on the applications.

This thesis explained the motivations to use RuLa and how RuLa is implemented. We believe that RuLa advances quantum networking research.

\section{Future Work}
In future works of this project, this approach should be more general to support internetworking. Currently, RuleSet generation is assumed to be done in a single network. However, we should be able to handle internetworking RuleSet generation. 

It is also important for RuLa to show more examples and test them in an actual simulator which would give us more confidence that the entire system works properly.

In terms of language design, we might need more theoretical discussion such as Turing completeness, livelock, and deadlock.

\section{Availability}
The source code is available on GitHub (\url{https://github.com/Chibikuri/rula})

%% file: Chapters/B.Appendix.tex
\chapter{Full Specification of RuLa}
\label{chap:full_specification}
The following Listing shows the whole PEG expressions of RuLa.
\begin{lstlisting}[language=PEG]
// Top AST node
rula <- SOI COMMENT? program COMMENT? EOI

program <- repeaters? import_stmt* rule_stmt* ruleset_stmt?

// Import expression to get functions and rules. (import test::func)
import_stmt <- ^"import" rule_import? ident ("::" ident )* 
                ( "::" "{"ident_list"}")? !"::"
rule_import <- "(rule)"

// Defining Rules
rule_stmt <- ^"rule" ident "<"repeater_ident">" argument_def 
            (":->"ret_type_annotation)? "{"rule_contents"}"
ret_type_annotation <- typedef_lit maybe? / 
                    "("(typedef_lit maybe?)("," typedef_lit maybe?)*")"
maybe <- "?"
argument_def <- "(" ((ident_typed/ident)? (","(ident_typed/ident))*)")"
rule_contents <- let_stmt* cond_expr "=>" act_expr stmt*

cond_expr <- ^"cond" "{"cond_clauses* "}"
cond_clauses <- res_assign / fn_call_expr / variable_call_expr
res_assign <- "@" ident ":" fn_call_expr

act_expr <- ^"act" "{" stmt* "}"

// RuleSet definition
ruleset_stmt <- ^"ruleset" ident "{" stmt* "}" 

stmt <- let_stmt     / 
        if_stmt      / 
        for_stmt     / 
        match_stmt   / 
        promote_stmt / 
        set_stmt     / 
        send_stmt    / 
        expr

repeaters <- "#repeaters: vec[Repeater]"

let_stmt <- ^"let" (ident_typed/"("ident_typed 
            ("," ident_typed)*")" ) "=" expr

// If expression (e.g. if (block > 0) {expreesion;};)
if_stmt <- ^"if" "("if_block")" "{"stmt*"}"else_if_stmt* else_stmt?
if_block <- get_expr / comp_expr / literal_expr
else_if_stmt <- ^"else" ^"if" "("if_block")""{"stmt*"}"
else_stmt <- ^"else" "{" stmt* "}"

// For expression for looping
for_stmt <- ^"for" (ident/for_multi_block) "in" for_generator "{"stmt* "}"
for_multi_block <- "("ident_list")"
for_generator <- series / expr
series <- int ".." expr 

// match expression for conditioning
// match expression {
//  00 => {do something},
//  11 => {do something else},
// otherwise => {hey}
// }
match_stmt <- ^"match" expr "{"(match_arm ",")* 
                (^"otherwise" "=>"match_action)? "}"
match_arm <- match_condition "=>" match_action 
match_condition <- satisfiable 
satisfiable <- literal_expr
match_action <- "{"stmt? ("," stmt)*"}"

// promote values 
promote_stmt <- ^"promote" promotable ("," promotable)*}
promotable <- comp_expr          /  
              term_expr          / 
              vector             / 
              tuple              / 
              variable_call_expr / 
              literal_expr 

// Set expression that set ruleset variable
set_stmt <- ^"set" ident ("as" ident)?

// Send mesage to the repeaters
send_stmt <- fn_call_expr "->" expr

expr <- rule_call_expr     /
        get_expr           /
        comp_expr          /
        term_expr          /
        vector             /
        tuple              /
        fn_call_expr       /
        variable_call_expr /
        literal_expr

// Call the defined Rules
rule_call_expr <- ident "<"repeater_call">" 
                "("fn_call_args? ("," fn_call_args)*")"
repeater_call <- "#repeaters" "(" (term_expr/ident/int)")"

// Get expression that get ruleset variable
get_expr <- ^"get" ident

// Comparison (no chain comparison allowed right now)
comp_expr <- comparable comp_op comparable
comparable <- get_expr           / 
              term_expr          / 
              variable_call_expr / 
              fn_call_expr       /  
              literal_expr
comp_op <- "<" / ">" / "<=" / ">=" / "==" / "!="

// Term expression for calculation
term_expr <- inner_term (op inner_term)+ 
inner_term <- terms / "("term_expr")"
terms <- get_expr           / 
        variable_call_expr  / 
        fn_call_expr        / 
        literal_expr 
        
// A set of arithmetic operations for tems
op <-  plus / minus / asterisk / slash / percent / caret
plus <- "+"
minus <- "-" 
asterisk <- "*" 
slash <- "/" 
percent <- "%" 
caret <- "^" 

// vector expresion
vector <- "["literal_expr? ("," literal_expr)* ","? "]"

// tuple expression
tuple <- "("expr? ("," expr)*","? ")"

fn_call_expr <- ident "("fn_call_args? ("," fn_call_args)*")"
fn_call_args <- term_expr           / 
                fn_call_expr        / 
                variable_call_expr  / 
                literal_expr 
                
variable_call_expr <- callable "." callable ("." callable)*
callable <- fn_call_expr   / 
            repeater_ident / 
            ident
repeater_ident <- "#" ident

ident_list <- ( ident/ident_typed ) 
            (","( ident/ident_typed))*
ident_typed <- ident ":" typedef_lit
ident <- ASCII_ALPHA (ASCII_ALPHA/ASCII_DIGIT/"_")*

literal_expr <- bool    /
                string  /
                ident   /
                number  / 
                binary  /
                hex     / 
                unicord

bool <- true_lit / false_lit 
true_lit <- "true" 
false_lit <- "false"

// number literal (int, float)
// "-some_number" is also fine
number <- (plus/minus)? (float/int/ident) ("e"("+"/"-")? ASCII_DIGIT+)? }

int <- (ASCII_NONZERO_DIGIT ASCII_DIGIT+/ASCII_DIGIT) !ASCII_ALPHA
float <- int "." int

binary <- "0b" bin_num
bin_num <- ("0"/"1")*

// Hex digit literals (e.g. x109ab2)
hex <- "0x" hex_num
hex_num <- (ASCII_DIGIT / "a" / "b" / "c" / "d" / "e" / "f" / "A" / "B" / "C" / "D" / "E" / "F" )*

// Unicord
unicord <- "0u" hex_num

// String literals 
string <- "\"" raw_string* "\""
raw_string <- (!("\\"/"\"") ANY )+

// Type definition literal
typedef_lit <- vector_type              / 
                integer_type            /
                unsigned_integer_type   /
                float_type              / 
                boolean_type            /
                string_type             /
                qubit_type              /
                repeater_type           /
                message_type            /
                result_type             /

vector_type <- "vec" "[" typedef_lit "]"
integer_type <- "int" 
unsigned_integer_type <- "u_int"
float_type <- "float"
boolean_type <- "bool" 
string_type <- "str"
qubit_type <- "Qubit"
repeater_type <- "Repeater" 
message_type <- "Message" 
result_type <- "Result" 

COMMENT <- ("/*"(!"*/"ANY)* "*/" / "//" (!"\n" ANY)* ("\n" / EOI))
WHITESPACE <- " "/"\n"/"\t"
\end{lstlisting}

\section{Notation}
The syntax is defined by Parsing Expression Grammar (PEG).
Listing~\ref{lst:match_peg} is an example expression with PEG.
\begin{lstlisting}[caption={An example PEG expression}, captionpos=b, label={lst:match_peg}, language=PEG]
rula <- SOI COMMENT? program COMMENT? EOI 
program <- repeaters? import_stmt* rule_stmt* ruleset_stmt?
\end{lstlisting}
To define a new set of non-terminated rules, \verb|<-| is the operator that defines a new rule with a name.
The rules to generate PEG expressions are following.
\begin{itemize}
    \item $E_1 E_2$: $E2$ follows immediately after $E_1$.
    \item $E_1/E_2$: If $E_1$ is not evaluated, $E_2$ is evaluated.
    \item $E_1$?: $E_1$ can be omitted
    \item $E_1$*: Zero or more $E_1$
    \item $E_1$+: One or more $E_1$
\end{itemize}

The tool we used to define the PEG parser is \verb|pest|~\cite{pest_rust}which provides the following extended syntax. 
\begin{itemize}
    \item $!E_1$: $E_1$ is not allowed here.
    \item $@E_1$: Explicit definition that removes all white spaces inside
    \item $\_E_1$: Inplicit definition that is ignored from the upper rules
    \item $^\wedge\text{""}$: Case-sensitive character.
\end{itemize}

\section{Termination}
Two types of values can be terminated values.

\textbf{Letters}:
The lower-case ASCII letters ('a'.. 'z') and upper-case ASCII letters ('A'..'Z') are the terminated values in letters.
\begin{lstlisting}[label={lst:ascii_alpha}, language=PEG]
ASCII_ALPHA <- 'a' / 'b' / ... / 'z' / 'A' / 'B' / ... / 'Z'
\end{lstlisting}

\textbf{Digits}
ASCII digits can also be terminated values. 
\begin{lstlisting}[label={lst:ascii_digit}, language=PEG]
ASCII_DIGIT <- '1'/'2'/'3'/'4'/'5'/'6'/'7'/'8'/'9'/'0'
\end{lstlisting}

\textbf{EOI}
The end of the input is defined as the \verb|EOI| symbol.
\begin{lstlisting}[label={lst:eoi}, language=PEG]
EOI
\end{lstlisting}

\textbf{ANY}
There is a useful rule called \verb|ANY| that contains \verb|ASCII_ALPHA|, \verb|ASCII_DIGIS| and also special characters such as \verb|@|, \verb|_|, etc. 
\begin{lstlisting}[label={lst:any}, language=PEG]
ANY
\end{lstlisting}

\section{Lexical Elements}
\subsection{Comments}
Comments allow us to provide some instructions or leave notes in the program.
There are two types of comments.
\begin{itemize}
    \item Line comments: The notes follow two slash lines \verb|//|.
    \item Cross-line comments: Comments quoted by \verb|/*| and \verb|*/|.
\end{itemize}
\begin{lstlisting}[label={lst:comments}, language=RuLa]
// This is line comment
/* 
Also possible to comment 
over the different lines
*/
\end{lstlisting}

\subsection{Identifier}
The identifier gives a name to variables, functions, rules, etc.
\begin{lstlisting}[label={lst:identifier_def}, language=PEG]
ident <- ASCII_ALPHA (ASCII_ALPHA / ASCII_DIGIT / "_" )*
\end{lstlisting}
Some identifiers listed in Section~\ref{sec:keywords} are reserved for predefined types and other expressions.

\subsection{Keywords}
\label{sec:keywords}

The keywords which are used in other expressions are as follows. 
\begin{lstlisting}[label={lst:keywords}, language=RuLa]
let, import, if, else, for, in, match, ruleset, rule, cond, act, set, get, true, false, promote
\end{lstlisting}

There are also type values and literals that cannot be an identifier.
\begin{lstlisting}[label={lst:type_def}, language=RuLa]
u_int, int, str, float, bool, qubit, Repeater, Result, Message
\end{lstlisting}

\subsection{Operators}
In order to compare the values, and do some mathematical operations, the following operators can be used. 
\begin{lstlisting}[label={lst:operator_def}, language=RuLa]
+, -, *, /, ^, >, >=, <, <=, ==, !=
\end{lstlisting}

\subsection{Punctuation}
The following braces are used as punctuations in some expressions
\begin{lstlisting}[language=RuLa]
[, ], {, }, (, )
\end{lstlisting}

\subsection{Integer Literals}
In RuLa, integer values are expressed by a set of digits. There are several types of integer value definitions.
\begin{lstlisting}[language=PEG]
int <- (ASCII_NONZERO_DIGIT ASCII_DIGIT+ / ASCII_DIGIT ) !ASCII_ALPHA 
binary <- "0b" bin_num
bin_num <- ("0" / "1")*
hex <- "0x" hex_num
hex_num <- ( ASCII_DIGIT / "a" / "b" / "c" / "d" / "e" / "f" / "A" / "B" / "C" / "D" / "E" / "F" )*
\end{lstlisting}
Internally, any integer values are converted to 64-bit integer values. 

\begin{lstlisting}[language=RuLa]
120
-12343
0b1001011
0x13ed232
\end{lstlisting}

\subsection{Floating Literals}
Floating point values are composed of an integer part, decimal point, and fraction values. 
\begin{lstlisting}
float <- int  "."  int
\end{lstlisting}
The floating point value is also converted to a 64-bit floating point value.

\subsection{String Literals}
Define the string values, it can be done with some values quoted by double quotations \verb|"|.
Since the PEG expression uses double quotations, double quotation
\begin{lstlisting}
string <- "\"" ( raw_string )* "\"" 
raw_string <- (ASCII_ALPHA/ ASCII_DIGIT)+
\end{lstlisting}

\section{Variables}
A variable stores a value that can be used in later expressions. A variable can be defined in the following way. 
\begin{lstlisting}[language=PEG]
let_stmt <- ^"let" ( ident_typed / "(" ident_typed ("," ident_typed)* ")" ) "=" expr
ident_typed <- ident ":" typedef_lit
\end{lstlisting}
where the \verb|ident_typed| is an identifier with type annotations shown in \ref{sec:types}.
\begin{lstlisting}[language=RuLa]
let x: int = -10
let y: u_int = 120
let hello: str = "world"
\end{lstlisting}
Putting the identifier refers to the predefined identifier.
\begin{lstlisting}[language=RuLa]
x  //-10
hello // "world"
\end{lstlisting}

\section{Types}
\label{sec:types}
In RuLa, there are five primitive data types and five composite data types.
\subsection{Boolean Type}
The boolean type can represent boolean values \verb|ture| or \verb|false|.
\begin{lstlisting}[language=RuLa]
true
false
\end{lstlisting}
A variable of boolean type can be either \verb|true| or \verb|false|.
\begin{lstlisting}[language=RuLa]
let boolean_true: bool = true
\end{lstlisting}

\subsection{Numeric Type}
There are two kinds of integer types \verb|int| and \verb|u_int| and one type for floating value \verb|float|.
\begin{lstlisting}[language=RuLa]
int  // 64 bit signed integer type
u_int  // 64 bit unsigned integer type
float
\end{lstlisting}

\subsection{String Type}
A string type is a set of string values that are surrounded by double quotations \verb|"|.
String type value is annotated by \verb|str|.
\begin{lstlisting}[language=RuLa]
let rula: str = "RuLa"
let hello_world: str = "Hello World"
\end{lstlisting}

\subsection{Vector Type}
A vector type contains a set of values that is annotated by \verb|vec[<inner_type>]| where \verb|<inner_type>| can be types values such as \verb|int|, \verb|bool|. 

\begin{lstlisting}[language=RuLa]
let int_vector: vec[int] = [1, 2, 3, 4, 5]
let bool_vector: vec[bool] = [true, false, true, true, true]
\end{lstlisting}

\subsection{Repeater Type}
In RuLa, the repeater can be represented as a \verb|Repeater| type value.
Repeaters are special values in RuLa. All RuLa programs start with \verb|#repeaters: vec[Repeater]| which stores the target repeater values generated from the configuration.
The repeater value contains the name and index of the repeaters. Figure~\ref{fig:repeater_indexing} shows how repeater indices are assigned.

The target repeater can be called from the repeater vector \verb|#repeater(<index>)| where the \verb|<index>| is the index of the repeater in the path.

\subsection{Qubit Type}
\verb|Qubit| type is just an alias of the target qubit. Since the actual qubit value can only be resolved when the RuleSet is executed, the qubit data type can only hold the index or identifier of the qubit within a single Rule.

For example, if one Rule consumes two qubits, these qubits are necessary to be distinguishable. When a qubit value is initialized, an index value can be assigned.


\subsection{Result Type}
\verb|Result| type is a composite data type that is used to store measurement results and to make comparisons with specific values with the target measurement result. Since many measurement operations happen in quantum computing, values with result type can be used in \verb|match| or \verb|if| as a condition.

\section{Expressions and Statements}
RuLa supports the following expressions.
\begin{lstlisting}[language=PEG]
expr <-  fn_call_expr / rule_call_expr / get_expr / comp_expr / term_expr / vector / tuple / variable_call_expr / literal_expr
\end{lstlisting}

\subsection{Function Call Expression}
Functions can be called by their names.
\begin{lstlisting}[language=PEG]
fn_call_expr <- ident "(" fn_call_args? ("," fn_call_args)* ")" 
fn_call_args <- term_expr / fn_call_expr / variable_call_expr / literal_expr 
\end{lstlisting}
The arguments are surrounded by the \verb|(| and \verb|)| with commas \verb|,|.

\subsection{Rule Call Expression}
Defined rules can be called by a rule-call expression. 
\begin{lstlisting}[language=PEG]
rule_call_expr <- ident "<" repeater_call ">" "(" fn_call_args? ("," fn_call_args)* ")"
repeater_call <- "#repeaters" "(" ( term_expr / ident / int ) ")"
\end{lstlisting}
The rule can be identified by an \verb|ident| and a repeater argument \verb|repeater_call| and a set of arguments \verb|fn_call_args|.
\begin{lstlisting}
test_rule<#repeaters(0)>(1, 2)  
// call `test_rule` rule with a repeater at index 0 and integer arguments (1, 2)
\end{lstlisting}

\subsection{Get Expression}
\verb|get| expression is to get the value registered by the \verb|set| statement.
\begin{lstlisting}
get_expr <- ^"get" ident
\end{lstlisting}
The identifier must be set properly before the get expression is called. Otherwise, the 
\subsection{Comp Expression}
The value can be compared with comparison operators, and the comparison value returns a Boolean value.
\begin{lstlisting}[language=PEG]
comp_expr <- comparable comp_op comparable
comparable <- get_expr / term_expr / variable_call_expr / fn_call_expr / literal_expr 
comp_op <- "<" / ">" / "<=" / ">=" / "==" / "!=" 
\end{lstlisting}

\subsection{Term expression}
The numbers can be added, subtracted, multiplied, or divided from the other numbers.
\begin{lstlisting}
term_expr <- inner_term ( op inner_term )+ 
inner_term <- terms / "(" ~ term_expr ~ ")"
terms <- get_expr / variable_call_expr / fn_call_expr / literal_expr 
\end{lstlisting}

\subsection{Variable Call Expression}
When the functions or variables in a composite-type variable are called, those variables and functions can be referred to with \verb|.|.
\begin{lstlisting}
variable_call_expr <- callable "." callable ("." callable)*
callable <- fn_call_expr / repeater_ident / ident
\end{lstlisting}

\subsection{Vector}
A vector can be defined as the following expressions.
\begin{lstlisting}
vector <- "[" literal_expr? ("," literal_expr)* ","? "]"
\end{lstlisting}
A vector can store a set of literal expressions. However, the type of literal value must be a single literal type.

\subsection{Tuple}
A tuple works similarly to the vector value. However, a tuple value can contain arbitrary expressions.
\begin{lstlisting}
tuple <- "(" expr? ("," expr)* ","? ")"
\end{lstlisting}

\subsection{Literal Expression}
The literal expressions can be defined as follows.
\begin{lstlisting}
literal_expr <-  bool / string / ident / number / binary / hex / unicord 
bool <- true_lit / false_lit 
true_lit <- "true"
false_lit <- "false"
string <- "\"" ( raw_string )* "\""
raw_string <- (!( "\\" / "\"" ) ANY )+
ident <- ASCII_ALPHA ( ASCII_ALPHA / ASCII_DIGIT / "_" )*
number <- ( "+" / "-" )? ( float / int / ident ) ( "e" ( "+" / "-" )? ASCII_DIGIT+ )? 
binary < "0b" bin_num
bin_num <- ( "0" / "1" )*
hex <- "0x" hex_num
hex_num <- ( ASCII_DIGIT / "a" / "b" / "c" / "d" / "e" / "f" / "A" / "B" / "C" / "D" / "E" / "F" )*
unicord <- "0u" hex_num 
\end{lstlisting}

\subsection{Statement}
\begin{lstlisting}[language=PEG]
program <- repeaters? import_stmt* rule_stmt* ruleset_stmt?
stmt <-  let_stmt / if_stmt / for_stmt / match_stmt / promote_stmt / set_stmt / send_stmt / expr 
\end{lstlisting}

\subsection{Import Statement}
To keep the modularity of the rules and functions, RuLa supports importing other rules and functions.
\begin{lstlisting}[language=PEG]
import_stmt <- ^"import" ident ("::" ident)* ( "::" "{"ident_list"}")?  !"::"
rule_annotation <- "(rule)"
ident_list <- ( ident / ident_typed )  ( "," ( ident / ident_typed ) )*
\end{lstlisting}
A set of identifiers and delimiter \verb|::| represent the position of the file, function, and rules.
\subsection{Rule Statement}
Rule statement allows us to define independent rules.

\begin{lstlisting}[language=PEG]
rule_stmt <- ^"rule" ident "<" repeater_ident ">" argument_def 
            (":->" ret_type_annotation)? "{" rule_contents "}"
repeater_ident <- "#" ident
ret_type_annotation <- typedef_lit maybe? / "(" (typedef_lit maybe?) 
                    ("," typedef_lit maybe?)* ")"
maybe <- "?" 
argument_def <- "(" ( ( ident_typed / ident )? 
                (","( ident_typed / ident))*)")"
rule_contents <- ( let_stmt )* cond_expr "=>" act_expr ( stmt )*
\end{lstlisting}

The rule statement is composed of five different components. The first \verb|ident| is the name of the target rule and the \verb|repeater_ident| is an identifier for the target rule.
In the rule definition, the repeaters can be referred to with a special identifier starting with \verb|#|.
Then the argument definition and type annotations are following after that.
In the rule contents, there are four different components.
The first series of let statements are used to define values that live till the end of the rule. 
\verb|cond_expr| defines a set of conditions to be satisfied.
\begin{lstlisting}
cond_expr <- ^"cond" "{" (cond_clauses)* "}"
cond_clauses <- res_assign / fn_call_expr / variable_call_expr
res_assign <- "@" ident ":" fn_call_expr
\end{lstlisting}
 There are three possible candidates for the condition clauses. The first \verb|res_assign| assigns a generated value by a function and passes it to the action clauses. The other two are \verb|fn_call_expr| and \verb|variable_call_expr|.

\verb|action+expr| defines a set of actions to be performed when the conditions are met.
\begin{lstlisting}[language=PEG]
act_expr <- ^"act" "{" stmt* "}"
\end{lstlisting}

\subsection{RuleSet Statement}
At the end of the program, the ruleset can be defined with \verb|ruleset| statement.
\begin{lstlisting}[language=PEG]
ruleset_stmt <- ^"ruleset" ident "{" stmt* "}" 
\end{lstlisting}
When the defined rule is called in this statement, a created rule is stored in the ruleset. 
The \verb|ident| follows \verb|"ruleset"| is the name of the ruleset.
\subsection{Let Statement}
A \verb|let| statement is a statement that defines a new variable.
\begin{lstlisting}
let_stmt <- ^"let"(ident_typed/"("ident_typed ("," ident_typed)* ")" ) "=" expr
\end{lstlisting}
Let statement requires a typed identifier as a left-hand side value. It is also possible to put multiple typed identifiers surrounded by \verb|(| and \verb|)|.

\subsection{Match Statement}
The \verb|match| statement supports higher-level conditioning for generating complex rules.
\begin{lstlisting}
match_stmt <- ^"match" expr "{"(match_arm ",")* 
                (^"otherwise" "=>" match_action)? "}"
match_arm <- match_condition "=>" match_action
match_condition <- satisfiable
satisfiable <- literal_expr
match_action <- "{" stmt? ("," stmt)* "}"
\end{lstlisting}
\verb|match| statement expands multiple conditions to multiple different rules in a single stage. Figure~\ref{fig:auto_expantion} shows how the \verb|match| is expanded to the different rules in a single stage. The top expression for matching needs to be comparable with literal values.  

\subsection{If Statement}
If the statement works similarly to the match statement but it can explain more complex conditioning.
\begin{lstlisting}
if_stmt <- ^"if" "(" if_block ")" "{" stmt* "}" else_if_stmt* else_stmt?
if_block <- get_expr / comp_expr / literal_expr
else_if_stmt <- ^"else" ^"if" "(" if_block ~ ")" "{" (stmt)* "}"
else_stmt <- ^"else" "{" stmt* "}"
\end{lstlisting}
In the rule statement, this if statement is also expanded to the multiple different rules in a stage.

\subsection{For Statement}
\verb|for| statement takes one by one from the given expression.
\begin{lstlisting}
for_stmt <- ^"for" (ident/for_multi_block) "in" 
            for_generator "{" stmt* "}" 
for_multi_block <- "(" ident_list ")"
for_generator <- series / expr
series <- int ".." expr
\end{lstlisting}
One or multiple temporary variables follow after \verb|"for"|. After these identifiers, \verb|"in"| and the corresponding expression follow.
In the statements in \verb|for|, it is possible to use defined identifiers as ordinary identifiers. 

\subsection{Promote Statement}
In order to promote a resource from one rule to the next rule, \verb|promote| statement returns a value to a parent ruleset and makes it available for later rules.
\begin{lstlisting}
promote_stmt <- ^"promote" promotable ("," promotable)*}
promotable <- comp_expr / term_expr / vector / 
             tuple / variable_call_expr / literal_expr
\end{lstlisting}
Currently, the promotable value must be a \verb|Qubit| type value.

\subsection{Set Statement}
In order to share the values other than the \verb|Qubit| type value, \verb|set| statement allows us to share the value among the rules in the same ruleset.
\begin{lstlisting}
set_stmt <- ^"set" ident ("as" ident)?
\end{lstlisting}
The first identifier is the identifier defined in the rule and the second identifier is an alias to that value.

\subsection{Send Statement}
\verb|send| expression is used to create Send operation to send the measurement result, promote notification, etc.
\begin{lstlisting}
send_stmt <- fn_call_expr "->" expr
\end{lstlisting}
At the syntax level, any function call can be in the send expression. However, there are four valid functions to be in the send statement. 

\section{Built-in Functions}
RuLa supports several built-in functions that are operations frequently used in the Rule definition.
\subsection{Free}
Free the qubit that is no longer used.
\begin{lstlisting}[language=RuLa]
free(<qubit>)
\end{lstlisting}

\subsection{Res}
Check the available resource.
This returns a set of qubit resources.
\begin{lstlisting}[language=RuLa]
res(<# of required resource>, <required fidelity>, <partner repeater>, <qubit index>)
\end{lstlisting}

\subsection{Recv}
Check if there is a message from the target repeater or end node.
\begin{lstlisting}[language=RuLa]
recv(<partner repeater>)
\end{lstlisting}
